\documentclass[10pt]{article}

\usepackage{geometry}
\usepackage{setspace}
\usepackage{graphicx}
\usepackage[numbers,square]{natbib}
\usepackage{amsmath,latexsym}

\title{Optimal perturbations of gravitationally unstable, transient, boundary layers in porous media}

\author{Don Daniel\thanks{dond@lanl.gov}, Nils Tilton\thanks{ntilton@mines.edu}, Amir Riaz\thanks{{ariaz@umd.edu, Department of Mechanical Engineering, University of Maryland,
             College Park, MD 20742, USA}}}

\date{}

\newcommand{\wip}{w_{\mathrm{p}}}
\newcommand{\cip}{c_{\mathrm{p}}}

\newcommand{\tp}{{t}_{\mathrm{p}}}
\newcommand{\tc}{{t}_{\mathrm{c}}}

\newcommand{\cb}{c_{\mathrm{b}}}

\newcommand{\Phic}{\Phi_{\mathrm{c}}}
\newcommand{\Phiw}{\Phi_{\mathrm{w}}}
\newcommand{\Phie}{\Phi_{\mathrm{e}}}
\newcommand{\Phiq}{\Phi_{\mathrm{q}}}
\newcommand{\ton}{{t}_{\mathrm{o}}}

\newcommand{\ch}{\widehat{c}}
\newcommand{\wh}{\widehat{w}}
\newcommand{\uh}{\widehat{u}}

\newcommand{\cs}{c^*}
\newcommand{\ws}{w^*}

\newcommand{\tf}{{t}_{\mathrm{f}}}
\newcommand{\ti}{{t}_{\mathrm{p}}}
\newcommand{\D}{\mathcal{D}}
\newcommand\Ra{\mbox{\textit{Ra}}}  
\newcommand{\kw}{{k}}
\newcommand{\M}{ A_\infty }
\newcommand{\opa}{ \mathrm{COP} }
\newcommand{\opb}{ \mathrm{MOP} }
\newcommand{\Ep}{ E_\Psi (\ti)}

\begin{document}

\maketitle

\begin{abstract}
We study gravitationally unstable, `transient', diffusive boundary
 layers in porous media using modal and nonmodal stability methods. 
Using nonmodal stability theory, we demonstrate that both the onset of 
instabilities and the shape of  optimal 
perturbations are highly sensitive to perturbation amplification 
measures and also the time at which the boundary layer is perturbed. 
This behavior is in
contrast to traditional studies of steady or non-transient  diffusive 
boundary layers
where perturbation dynamics are independent of perturbation measure or time. 
We demonstrate that any analysis of transient layers
produced through  
classical methods can result in
physically unrealizable perturbation structures. To resolve the issue, 
we propose a non-modal stability procedure which additionally 
constrains the perturbation 
dynamics to physically admissible fields. 
The proposed procedure predicts that instabilities grow primarily due to 
unstable perturbations featuring much larger spanwise wavenumbers
(modes) and smaller amplifications 
compared to perturbations predicted using classical methods. 
We validate our predictions using  direct numerical 
 simulations that emulate the onset of convection in physical systems. 
\end{abstract}

\section{Introduction}
Gravitationally unstable, transient, solute boundary layers in 
porous media have been studied extensively due to their importance in 
carbon dioxide sequestration in porous, brine-saturated, subsurface aquifers. After injection in an  aquifer, buoyant $\mathrm{CO}_2$ rises and forms a horizontal layer beneath an impermeable caprock.
With time, the $\mathrm{CO}_2$ dissolves into the underlying brine and forms a downwardly growing diffusive boundary layer, as illustrated in figure \ref{figure1}. 
As $\mathrm{CO}_2$ diffuses downwards, the solute boundary
layer is naturally perturbed by local aquifer heterogeneities.
 Because the $\mathrm{CO}_2$-rich brine in the boundary layer is denser than the underlying brine, a gravitational instability eventually causes perturbations to grow and form finger-like structures that break away from the boundary layer and convect $\mathrm{CO}_2$ downwards into the aquifer.
A clear understanding of the physical mechanisms and
dominant perturbation structures that cause finger-formation is vital to
modelling $\mathrm{CO}_2$ sequestration. Furthermore, similar transient boundary layers occur in purely fluid media \cite[]{Elder1968} and are important for heat transfer devices \cite[]{Goldstein1959} and geophysical flows \cite[]{Wooding1997}.

In comparison to classical Rayleigh-B\'{e}nard convection \cite[]{DrazinReid1981},
the stability of transient diffusive boundary layers is complicated by the
transient base-state that renders the linear stability
operator non-autonomous. 
At small times when the boundary layer is beginning to form, perturbations to the layer are damped by stabilizing effects of diffusion. Eventually, a
critical time for linear instability, $t=\tc$, is reached after which perturbations begin to grow. For small initial perturbations, linear mechanisms can dominate for
considerable time beyond $\tc$ \cite[]{Riaz2006JFM, Rouhi2007, Selim2007b, Rapaka2008}. Within this linear regime, the flux of CO$_2$ into the brine, $J$, decreases monotonically. Eventually nonlinear mechanisms cause the flux of CO$_2$ to increase from that predicted by linear theory such that there is a turning point where $dJ/dt=0$. Motivated by experimental studies \cite[]{Blair1969}, we define the time at which this turning point occurs as the onset time of nonlinear convection, $t=\ton$. 
  
Various methods have been used to study the linear regime preceding onset of convection, for detailed review,
 see Refs. \cite[]{Rapaka2008, Rees2008, Kim2012}. One approach invokes the quasi-steady-state-assumption (QSSA) and performs a modal analysis. 
The QSSA eigenmodes, however, are non-orthogonal, and there is potential for nonmodal growth \cite[]{Schmid2007}. 
Thus motivated, recent studies (for example, see Ref. \cite[]{Rapaka2008})  use  traditional nonmodal methods \cite[]{Farrell1996a, Farrell1996b, Schmid2007} to the
nonautonomous linear initial value problem (IVP) to determine optimal perturbations
with maximum amplification at a later time.
However, in this study, we show that optimal perturbations predicted through such classical means
 cannot lead to onset of convection in finite time. Rather, onset of convection results from 
the growth of ``suboptimal'' perturbations localized within the diffusive boundary layer.
To determine such perturbations, we propose a non-modal method based on an adjoint Lagrangian formulation. 

This study is organized as follows. The governing equations are presented in \S 2. The classical optimization procedure, without application of the physical constraint, is described in \S 3. The classical optimization results are presented in \S 4.   The proposed modifications to the classical optimization procedure and associated results  are presented in \S 5.    
DNS results are reported in \S6. The main findings are summarized in \S7. 

\section{Geometry and governing equations}
\begin{figure}
\begin{center}
  \hspace{0.0cm}
    (\emph{a})
    \hspace{6.5cm}
    (\emph{b})
    \\
    \includegraphics[width=7.3cm]{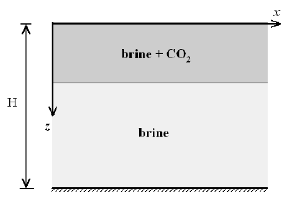}
    \includegraphics[width=6.7cm]{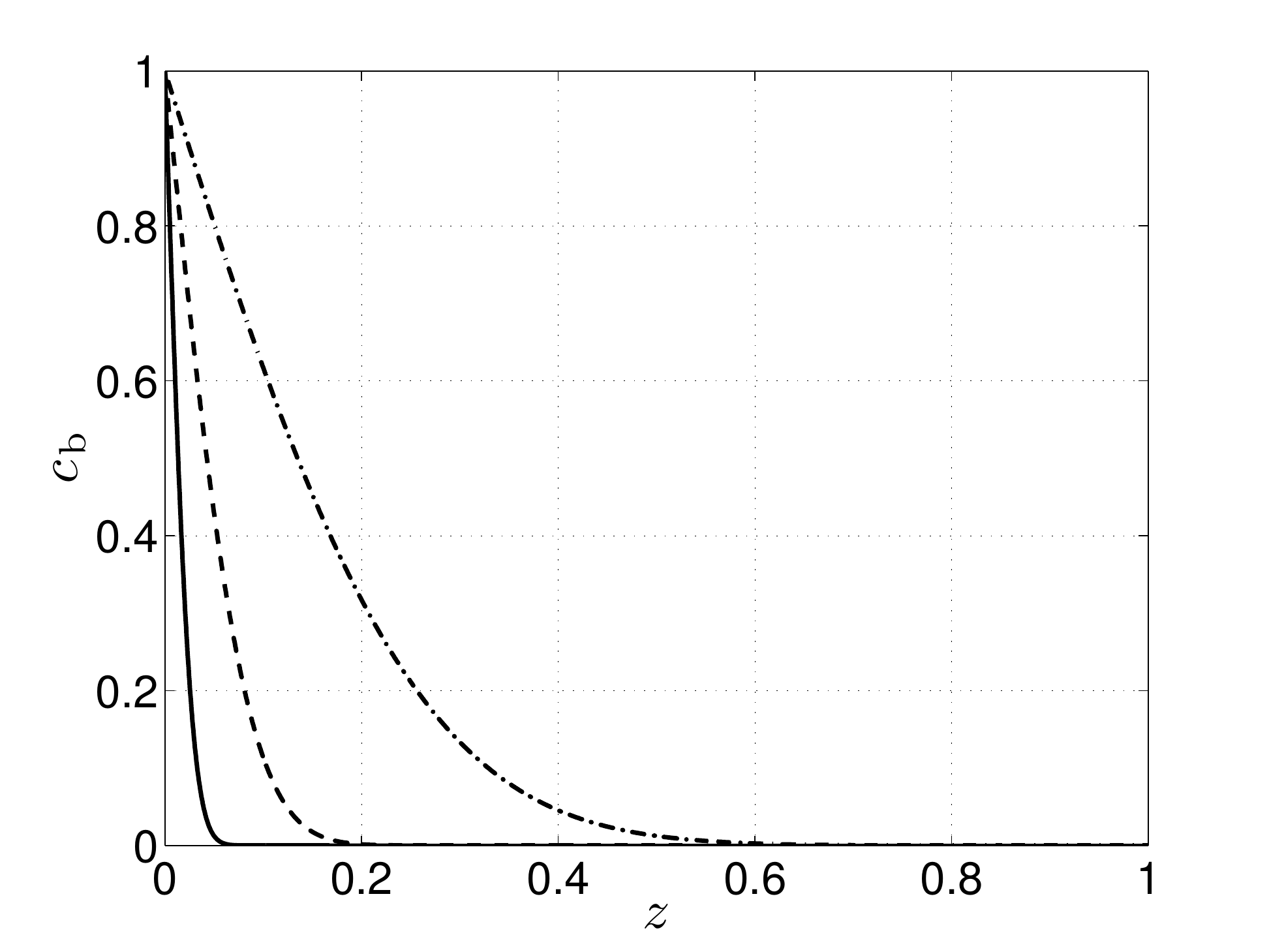}
        \caption{ (\emph{a}) Sketch, not to scale, of the geometry considered in this study. (\emph{b}) Base-state (\ref{eq:bs}) for $\Ra=500$ and $t= 0.1$ (solid line), $t=1$ (dashed line), and $t=10$ (dash-dotted line). } 
    \label{figure1}
\end{center}
\end{figure}

Due to the fundamental nature of the current study, we consider an isotropic, homogeneous, fluid-saturated porous medium
of infinite horizontal extent in the $x$ and $y$ directions, and of finite
depth $H$ in the vertical $z$ direction, see figure \ref{figure1}(\emph{a}). This facilitates comparison with the previous nonmodal analysis of \cite{Rapaka2008}. The analysis considered in the current study can be extended to anisotropic heterogeneous media. We define the vertical $z$
direction as positive in the downward direction of gravity,
$g$. The domain is bounded by an impermeable wall at $z = H$. The porous
medium is characterized by its permeability, $K$, dispersivity, $D$, and
porosity,  $\phi$, respectively. Initially, the brine is quiescent with
zero $\mathrm{CO}_2$ concentration, $c=0$, and constant density, $\rho=\rho_0$. At time
$t=0$, saturated brine is supplied at $z=0$ with a constant concentration
$c= C_1$ and density $\rho_1$. The fluid viscosity, $\mu$, is assumed to be
constant. The density difference $\Delta \rho = \rho_1-\rho_0$ is assumed
to be much less than $\rho_0$, i.e. $\Delta \rho \ll \rho_0$. 

Fluid flow and mass transport in the porous medium are governed by Darcy's law and
volume averaged forms of the continuity and advection-diffusion equations \cite[]{Whitaker1999}. The governing
equations are written in nondimensional form as,

\begin{equation}
\boldsymbol{v}   +   \boldsymbol{\nabla} p -  c\boldsymbol{e}_z  = 0,  \qquad  \boldsymbol{\nabla} \boldsymbol{\cdot} \boldsymbol{v}  = 0,  \qquad 
 \frac{\partial c}{\partial t} + \boldsymbol{v} \boldsymbol{\cdot} \boldsymbol{\nabla} c - \frac{1}{\Ra} \nabla^2 c=0,
 \label{eq:ge}
\end{equation}
using the characteristic length $L=H$, time $T=\phi H/U$, buoyancy velocity $U = K \Delta \rho g / \mu$, pressure $P=\Delta \rho g H$, and concentration $C=C_1.$ The dimensionless equations (\ref{eq:ge}) have been obtained using the Boussinesq approximation and a linear fluid density profile, $\rho = \rho_0 + \Delta \rho (c / C_1).$ The Rayleigh number is defined as $\Ra = U H/  (\phi D).$  
The symbol $\boldsymbol{v} =[u,v,w]$ is the nondimensional velocity vector, $c$ is the nondimensional concentration and $p$ is the nondimensional pressure obtained from the dimensional pressure $\hat{p}$ through the relation $ p = (\hat{p} - \rho_0 g z)/P.$ The symbol $\boldsymbol{e}_z$ is the unit vector in the $z$ direction. 
Equations (\ref{eq:ge}) must satisfy the following boundary conditions, 
\begin{equation}
c \Big|_{z=0} = 1, \qquad \frac{\partial c}{\partial z} \Big|_{z=1} = 0, \qquad w \Big|_{z=0} = w \Big|_{z=1} = 0, \qquad t \ge 0. 
\label{eq:ge-bc}
\end{equation}

Equations (\ref{eq:ge}) admit the transient base state,
\begin{equation}
\boldsymbol{v}_\mathrm{b} = \textbf{0}, \qquad \cb(z,t) = 1 - \frac{4}{\pi} \sum_{\mathrm{n}=1}^{\infty}  \frac{1}{2\mathrm{n} \! - \! 1} \mathrm{sin} \! \Bigg[ \! \left( \mathrm{n}-\frac{1}{2} \right) \pi z \Bigg] \mathrm{exp} \! \Bigg[ \! - \! \left( \mathrm{n} - \frac{1}{2} \right)^{\! 2} \frac{\pi^2 t}{Ra}  \Bigg],
\label{eq:bs}
\end{equation}
We study the linear stability of base-state (\ref{eq:bs})  with respect to small wavelike perturbations of the form,
\begin{equation}\label{eq:normalmodes}
\widetilde{c}=\widehat{c}(z,t)\mathrm{e}^{\mathrm{i}(\alpha x + \beta y)},
\quad
\widetilde{\boldsymbol{v}}=\widehat{\boldsymbol{v}}(z,t)\mathrm{e}^{\mathrm{i}(\alpha x + \beta y)},
\quad
\widetilde{p}=\widehat{p}(z,t)\mathrm{e}^{\mathrm{i}(\alpha x + \beta y)},
\end{equation} 
where $\mathrm{i}=\sqrt{-1}$, $\alpha$ and $\beta$ are wavenumbers in the $x$ and $y$ directions respectively, and $\widehat{c}(z,t)$, $\widehat{\boldsymbol{v}}(z,t)$ and $\widehat{p}(z,t)$ are time-dependent perturbation profiles in the $z$ direction. Following the standard procedure \cite[see][]{Riaz2006JFM}, the linear stability problem can be written as the following initial value problem for $\ch$ and $\wh$,
\begin{equation}
\frac{\partial \ch}{\partial t}   + \wh \frac{\partial \cb}{\partial z} -  \frac{1}{Ra}  \D \ch  = 0, \qquad
\D \wh + \kw^2 \ch = 0,
\label{eq:linear}
\end{equation}
\begin{equation}
\ch \Big|_{z=0}=0,  \qquad   \frac{\partial \ch}{\partial z} \Big|_{z=1}=0,  \qquad   \wh \Big|_{z=0}=  \wh \Big|_{z=1}=0, 
\label{eq:bcs}
\end{equation}
where $\D=  \partial^2 / \partial z^2- \kw^2$ and $\kw = \sqrt{ \alpha^2 + \beta^2}.$ Because the base-state is transient, the boundary layer is sensitive to the time at which it is perturbed. 
We assume the layer is perturbed at time $t=\ti$ with the following initial perturbation profiles,
\begin{equation}
\ch \, \Big|_{t = \ti} = \cip(z), \qquad  \wh \, \Big|_{t = \ti} =  \wip(z),
 \label{eq:ic}
 \end{equation}
where $\cip$ and $\wip$ must satisfy equations (\ref{eq:linear})--(\ref{eq:bcs}).  

\section{Classical Optimization}
The initial perturbation profiles, $\cip$ and $\wip$, are determined so that the perturbation amplification is maximized at some prescribed final time $t=\tf$. Previous studies  \cite[]{Tan1986, Doumenc2010} have observed that the perturbation amplification is sensitive to the perturbation flow field used to measure the perturbation magnitude. 
To investigate how different measures of perturbation magnitude influence nonmodal results, we define the perturbation magnitude at time $t$ as,
\begin{equation}
{{E}(t)} = \int_0^1 \left[ {\mathrm{A}_1 \ch(z,{t})^2} +  {\mathrm{A}_2 \wh(z,{t})^2} +  {\mathrm{A}_3\uh(z,{t})^2} \right] \,\mathrm{d}z,
\end{equation} 
where $\mathrm{A}_1,\mathrm{A}_2$ and $\mathrm{A}_3$ are constants to be defined shortly.
We introduce the following measures of perturbation amplification, $\Phi(t)$,  
\begin{subequations}
\label{pert_measure}
\begin{equation}
\quad \, \, \,  \Phic(t) =  \Bigg[  \frac{E(t)}{E(\ti)} \Bigg] ^{\frac{1}{2}},\quad \mathrm{A}_1 = 1, \quad \mathrm{A}_2 = \mathrm{A}_3=0,
\end{equation}
\begin{equation}
\quad \, \, \,  \Phiw(t) =  \Bigg[  \frac{E(t)}{E(\ti)} \Bigg] ^{\frac{1}{2}},\quad  \mathrm{A}_2= 1, \quad \mathrm{A}_1= \mathrm{A}_3=0,
\end{equation}
\begin{equation}
\Phie(t) =  \Bigg[  \frac{E(t)}{E(\ti)} \Bigg] ^{\frac{1}{2}},\quad  \mathrm{A}_1 =  \mathrm{A}_2 = \mathrm{A}_3=1.
\end{equation}
\end{subequations}
Most previous studies of transient boundary layers measure amplification with respect to the perturbation's concentration field, $\Phic$ \cite[]{Caltagirone1980,Tan1986,Ennis-King2003,Kim2005,Rapaka2008}, or the vertical velocity field, $\Phiw$ \cite[]{Foster1965a,Gresho1971,Kaviany1984,Tan1986}. In addition, we introduce $\Phie$ as a measure of perturbation energy that includes both the velocity and concentration fields.
We  optimize $\Phi(\tf)$ using an adjoint procedure described by \cite{Doumenc2010} in which $E(\tf)$ is maximized
subject to the constraint that $E(\tp)=1$. For this purpose, we define the Lagrangian,
\begin{eqnarray}
\nonumber \mathcal{L}(\ch,\cs ,\wh,\ws,\uh,\mathrm{s}) &=& E(\tf) -  \mathrm{s} \big[ E(\ti) - 1 \big]  - \int_{\ti}^{\tf} \int_0^1 \ws \left( \D \wh + \kw^2 \ch \right)  \,\mathrm  {d}z \, \mathrm{d}t  \\
&& -    \int_{\ti}^{\tf} \int_0^1 \cs \left( \frac{\partial \ch}{\partial t}   -  \frac{1}{\Ra}  \D \ch + \wh \frac{\partial \cb}{\partial z}  \right) \,\mathrm{d}z \,\mathrm{d}t ,  
\label{eq:L}
\end{eqnarray}
where $\mathrm{s}$ is a scalar Lagrange multiplier and the adjoint variables $\cs(z,t)$ and $\ws(z,t)$ are Lagrange multipliers dependent on $z$ and $t$. The double integrals on the right-hand-side of (\ref{eq:L}) assure satisfaction of the governing IVP (\ref{eq:linear})--(\ref{eq:ic}). 
To obtain first-order optimality conditions, the variational of
the Lagrangian, $\delta \mathcal{L}$, is set to zero.
Integrating by parts, $\delta \mathcal{L}$ can be written as,
\begin{displaymath}
 \delta \mathcal{L} =  \int_0^1 \Big[ 2 \left( {\mathrm{A}_1 \ch \,  \delta \ch + \mathrm{A}_2 \wh \,  \delta \wh + \mathrm{A}_3 \uh \,  \delta \uh }  \right) - \cs  \,  \delta \ch  \, \Big]_{t=\tf}  \mathrm{d}z \textrm{\hspace{4cm}}
 \end{displaymath}
 \begin{displaymath}
\hspace{0.45cm} -  \int_0^1 \Big[ 2 \mathrm{s} \left( {\mathrm{A}_1 \ch \,  \delta \ch + \mathrm{A}_2 \wh \,  \delta \wh + \mathrm{A}_3 \uh \,  \delta \uh } \right)  - \cs \, \delta \ch  \, \Big]_{t=\ti} \mathrm{d}z  \textrm{\hspace{4cm}} \\
 \end{displaymath}
  \begin{displaymath}
\hspace{0.45cm} -  \int_{\ti}^{\tf} \int_0^1 \left[ \delta \ch \,\ \left( - \frac{\partial \cs}{\partial t}  -  \frac{1}{\Ra}  \D \cs + \kw^2 \ws \right)   + \delta \wh \,\  \left( \D \ws   + \frac{\partial \cb}{\partial z}  \cs \right) \right] \mathrm{d}z \, \mathrm{d}t  \textrm{\hspace{0.9cm}}\\
 \end{displaymath}
\begin{eqnarray} 
+    {\int_{\ti}^{\tf} \left[ \frac{1}{\Ra} \left( \cs  \frac{\partial \delta \ch}{\partial z}  - \delta \ch  \frac{\partial \cs}{\partial z}  \right)  - \ws  \frac{\partial \delta \wh}{\partial z} + \delta \wh  \frac{\partial \ws}{\partial z}  \right]_{z=0}^{z=1}  \mathrm{d}t }  = 0. \textrm{\hspace{2.5cm}}
\label{eq:adj}
\end{eqnarray}
The optimality conditions are met when $\cs$ and $\ws$ satisfy the following adjoint problem,
\begin{equation}
-\frac{\partial \cs}{\partial t}  -  \frac{1}{\Ra}  \D \cs  + \kw^2  \ws= 0, \qquad
\D \ws   =  -\frac{\partial \cb}{\partial z}  \cs,
\label{eq:adj1}
\end{equation}
\begin{equation}
\cs \Big|_{z=0}=0,  \qquad   \frac{\partial \cs}{\partial z} \Big|_{z=1}=0,  \qquad   \ws \Big|_{z=0}=  \ws \Big|_{z=1}=0, 
\label{eq:bc_adj}
\end{equation}
along with the following coupling conditions between the adjoint and
physical variables,
\begin{equation}
2 \left( \mathrm{A}_1 \ch \,  \delta \ch + \mathrm{A}_2 \wh \,  \delta \wh + \mathrm{A}_3 \uh \,  \delta \uh  \right)\, \Big|_{t=\tf}   = \cs  \,  \delta \ch  \, \Big|_{t=\tf},   
\label{eq:ref1}
\end{equation}
\begin{equation}
2 \mathrm{s} \left( \mathrm{A}_1 \ch \,  \delta \ch + \mathrm{A}_2 \wh \,  \delta \wh + \mathrm{A}_3 \uh \,  \delta \uh  \right)\, \Big|_{t=\ti}   = \cs  \,  \delta \ch  \, \Big|_{t=\ti} .  
\label{eq:ref2}
\end{equation}
The optimal initial perturbations are found using an iterative
procedure. Given an initial guess for $\cip$ and
$\wip$, we integrate the IVP (\ref{eq:linear})--(\ref{eq:ic}) forward in time to $t=\tf$. We then
apply the condition (\ref{eq:ref1}) to obtain a final condition for the
adjoint IVP (\ref{eq:adj1})--(\ref{eq:bc_adj}). The adjoint IVP is then integrated backwards in
time to $t=\ti$. We then use condition (\ref{eq:ref2}) to obtain
improved initial profiles $\cip$ and $\wip$. This procedure is repeated
until satisfaction of the convergence criteria, $\| \cip^\mathrm{n} - \cip^{\mathrm{n}-1} \|_\infty  / \| \cip^{\mathrm{n}-1} \|_\infty   \le 10^{-4},$ where $\mathrm{n}$ is the iteration number. The iterative procedure is insensitive to the initial guess; however, the number of iterations is reduced using $\cip^0= \xi \mathrm{exp}(-\xi^2)$ where $\xi=z \sqrt{\Ra/(4 t)}$.  The IVPs are solved using standard second-order finite-difference methods.

The application of the  coupling conditions (\ref{eq:ref1})--(\ref{eq:ref2}) depends on the definition of the perturbation amplification.
When maximizing $\Phic$, conditions (\ref{eq:ref1})--(\ref{eq:ref2}) are satisfied for,
\begin{equation}
2 \ch \, \Big|_{t=\tf}  = \cs \Big|_{t=\tf}  , \qquad  2  \mathrm{s} \ch \, \Big|_{t=\ti}  = \cs\Big|_{t=\ti} .   
\label{eq:rel1}
\end{equation}
The derivation of (\ref{eq:rel1}) is described in Ref. \cite{Doumenc2010}. When maximizing $\Phiw$ or $\Phie$, however, the application of the coupling conditions is less straightforward than in the case of \cite{Doumenc2010} because in the current study, the momentum equation lacks a temporal derivative. For those cases, we found it necessary to replace the Neumann boundary conditions for $\ch$ and $\cs$ at the lower wall with the following Dirichlet condition,
\begin{equation}
\ch \, \Big|_{z=1}= \cs \, \Big|_{z=1}=0.
\label{eq:bc_new}
\end{equation}
Consequently, when maximizing $\Phiw$, coupling conditions (\ref{eq:ref1})--(\ref{eq:ref2}) are satisfied for,
\begin{equation}
-2 \kw^2  \wh \Big|_{t=\tf}  =  \D  \cs \Big|_{t=\tf} , \qquad -2 \kw^2  \mathrm{s} \wh \Big|_{t=\ti}   =  \D  \cs \Big|_{t=\ti}.  
\label{eq:rel2}
\end{equation}
When maximizing $\Phie$, conditions (\ref{eq:ref1})--(\ref{eq:ref2}) are satisfied for,
\begin{equation}
\left( \kw^2  \frac{\partial^2 }{\partial z^2} - \kw^4 - \D^2 \right) \wh \Big|_{t=\tf}      =  \frac{\kw^2}{2} \D  \cs \Big|_{t=\tf} ,
\label{eq:rel3}
\end{equation}
\begin{equation}
\mathrm{s} \left( \kw^2  \frac{\partial^2 }{\partial z^2} - \kw^4 -\D^2 \right) \wh \Big|_{t=\ti}      =  \frac{\kw^2}{2} \D  \cs \Big|_{t=\ti} .
\label{eq:rel4}
\end{equation}

For the parameter space $(\ti,\tf,\Ra,\kw)$ considered in the current study, the Dirichlet condition (\ref{eq:bc_new}) is valid because the optimal perturbations are concentrated near $z=0$ and decay to zero outside the boundary layer such that they are not influenced by the lower wall \cite[]{Rapaka2008,Slim2010}.  
We validated our results by directly optimizing the IVP (\ref{eq:linear})--(\ref{eq:ic}), subject to the standard boundary conditions (\ref{eq:bcs}), using MATLAB routines. The adjoint-based method shows excellent agreement with direct optimization but is an order-of-magnitude faster.    

\section{Classical optimization results} 
Previously, Ref. \cite{Rapaka2008} reported optimal
perturbations that maximize $\Phic$ for a fixed initial perturbation
time, $\ti$.
We extend their work in the following manner.
First, we explore how the amplification measure (\ref{pert_measure}) affects the optimization results. Second, we investigate the role of the initial perturbation time. Third, we study the influence of the final time on the initial optimal profiles. Fourth, we compare the optimal perturbations with quasi-steady eigenmodes. Finally, in \S5, we assess the relevance of the optimal perturbations to physical experiments.

\subsection{Effect of amplification measure}

\begin{figure}
\begin{center}
  \hspace{0.0cm}
    (\emph{a})
    \hspace{6.5cm}
    (\emph{b})
    \\
    \includegraphics[trim = 0 0 0 10, clip=true,width=7cm]{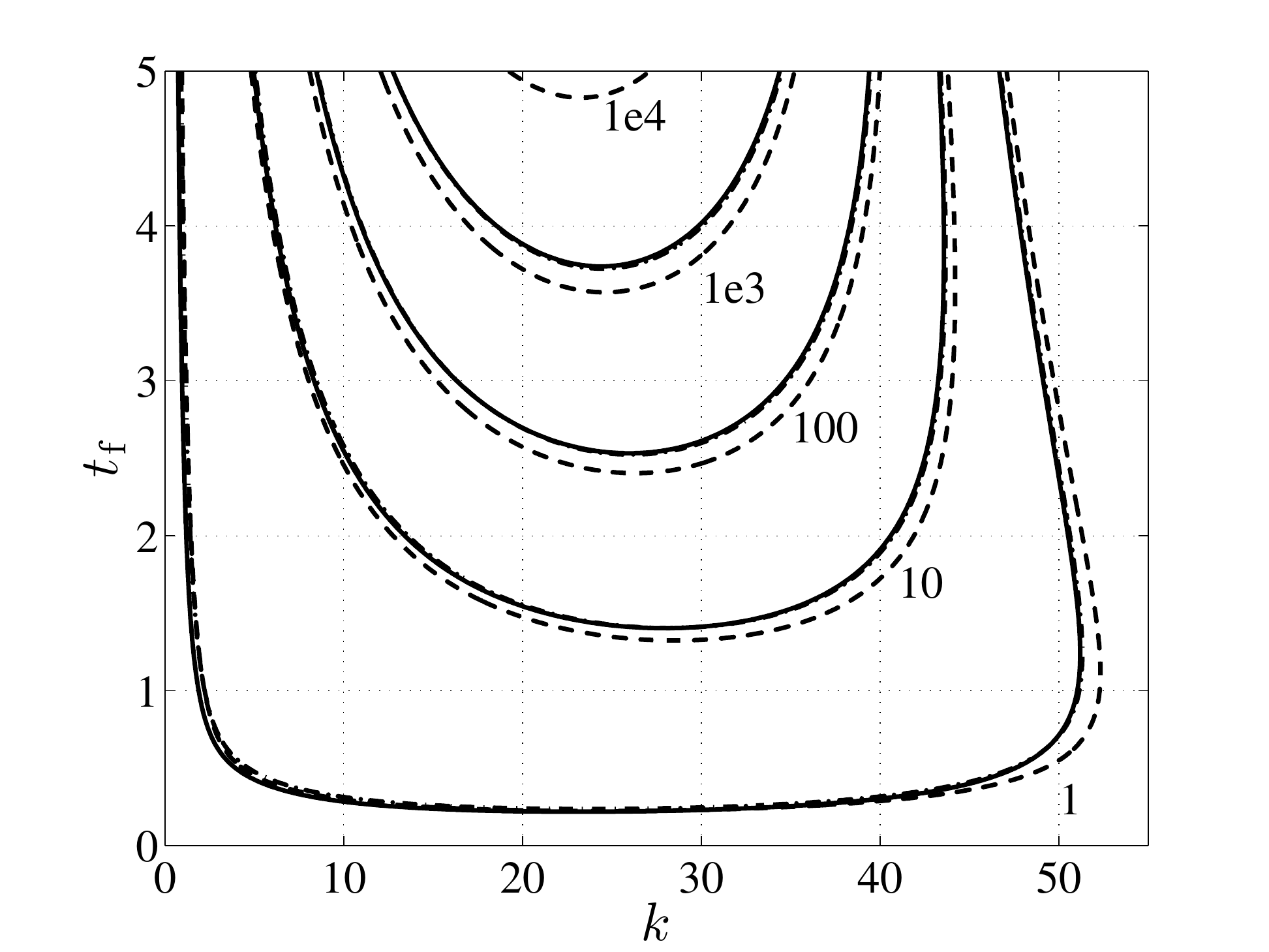}
    \includegraphics[trim = 0 0 0 10, clip=true,width=7cm]{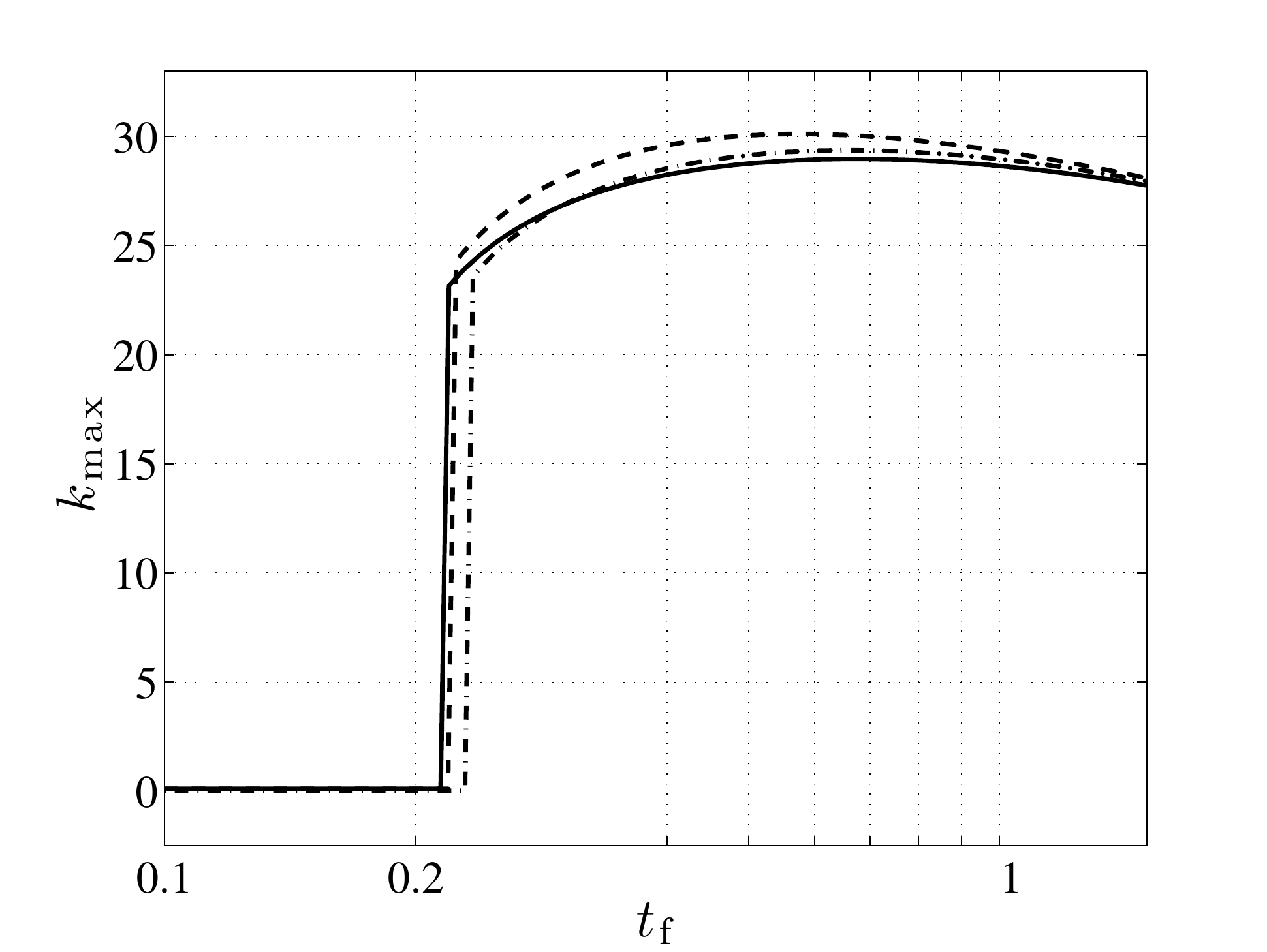} 
     \\
     \hspace{0.0cm}
     (\emph{c})
     \hspace{6.5cm}
     (\emph{d})
     \\
     \includegraphics[trim = 0 0 0 10, clip=true,width=7cm]{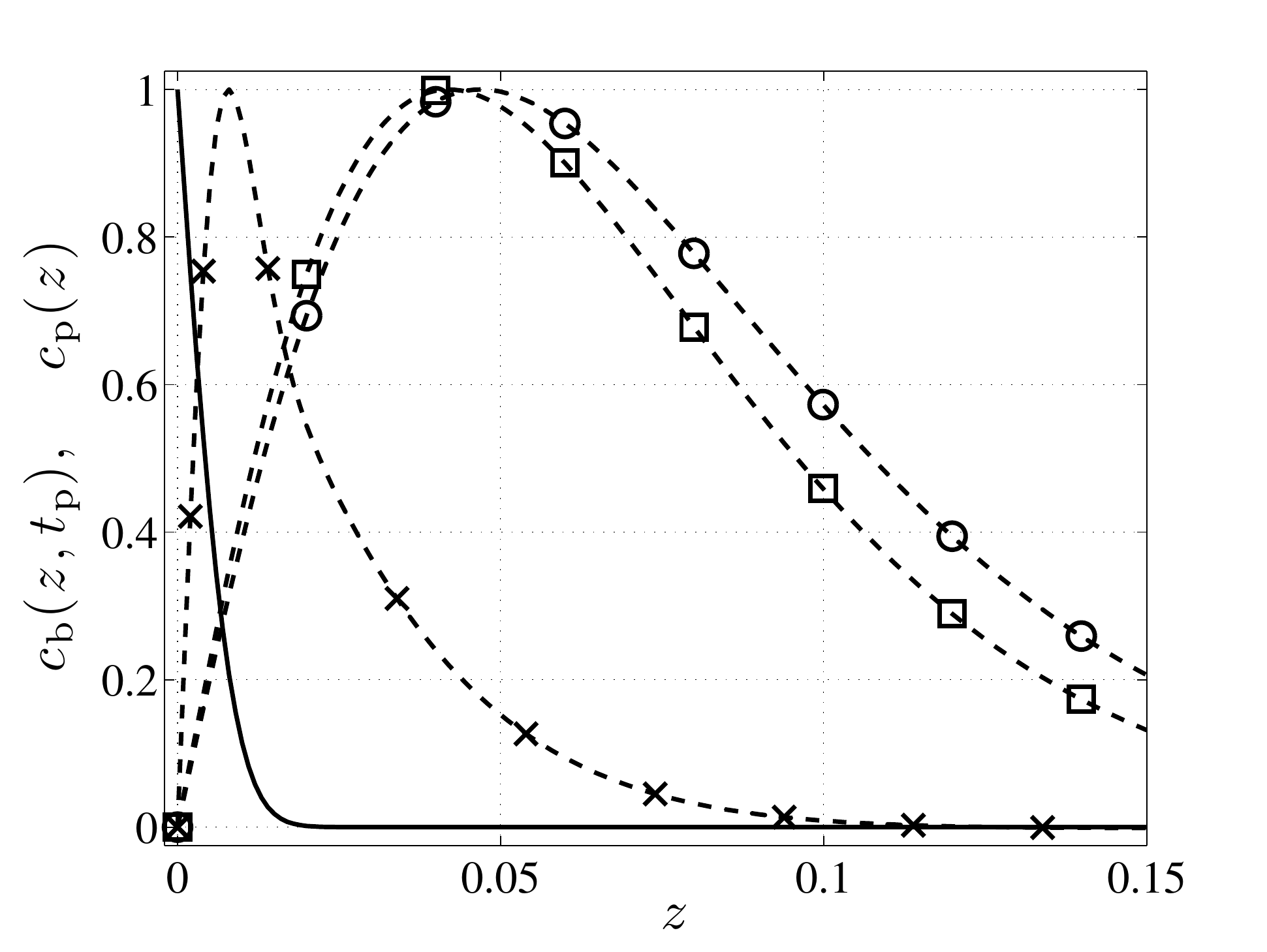}
    \includegraphics[trim = 0 0 0 10, clip=true,width=7cm]{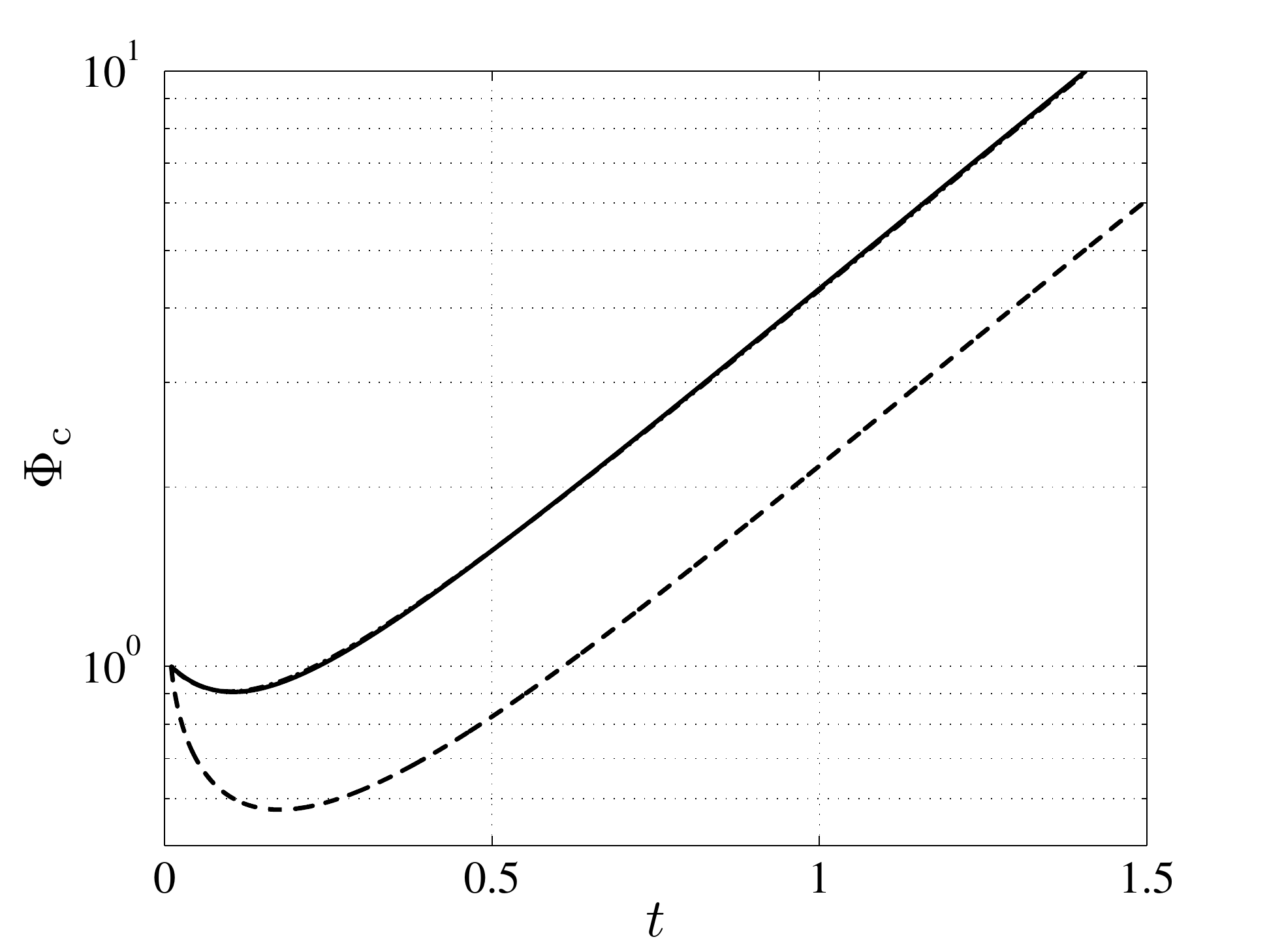}
      \\
     \hspace{0.0cm}
     (\emph{e})
     \hspace{6.5cm}
     (\emph{f})
     \\
     \includegraphics[trim = 0 0 0 10, clip=true,width=7cm]{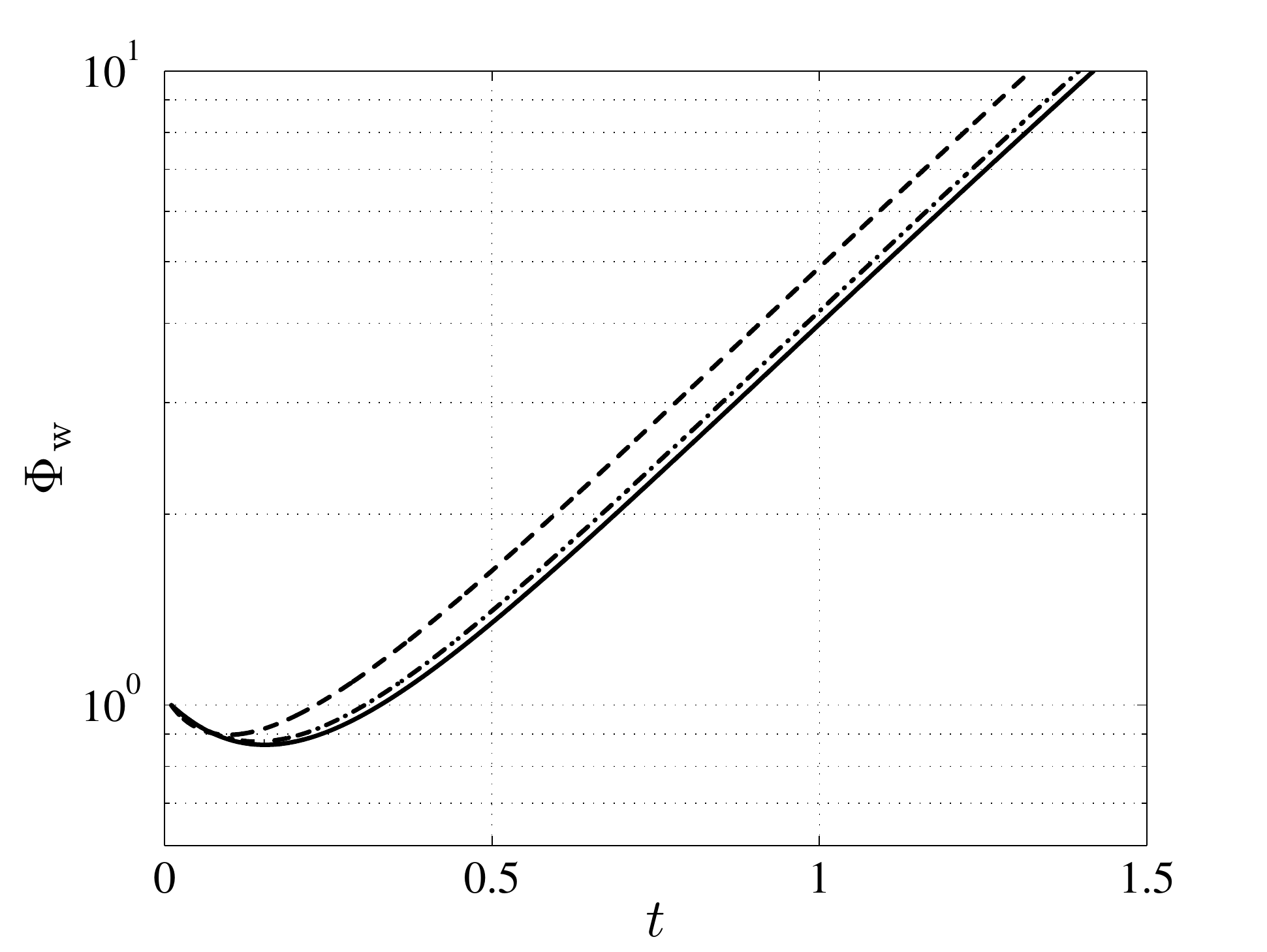}
    \includegraphics[trim = 0 0 0 10, clip=true,width=7cm]{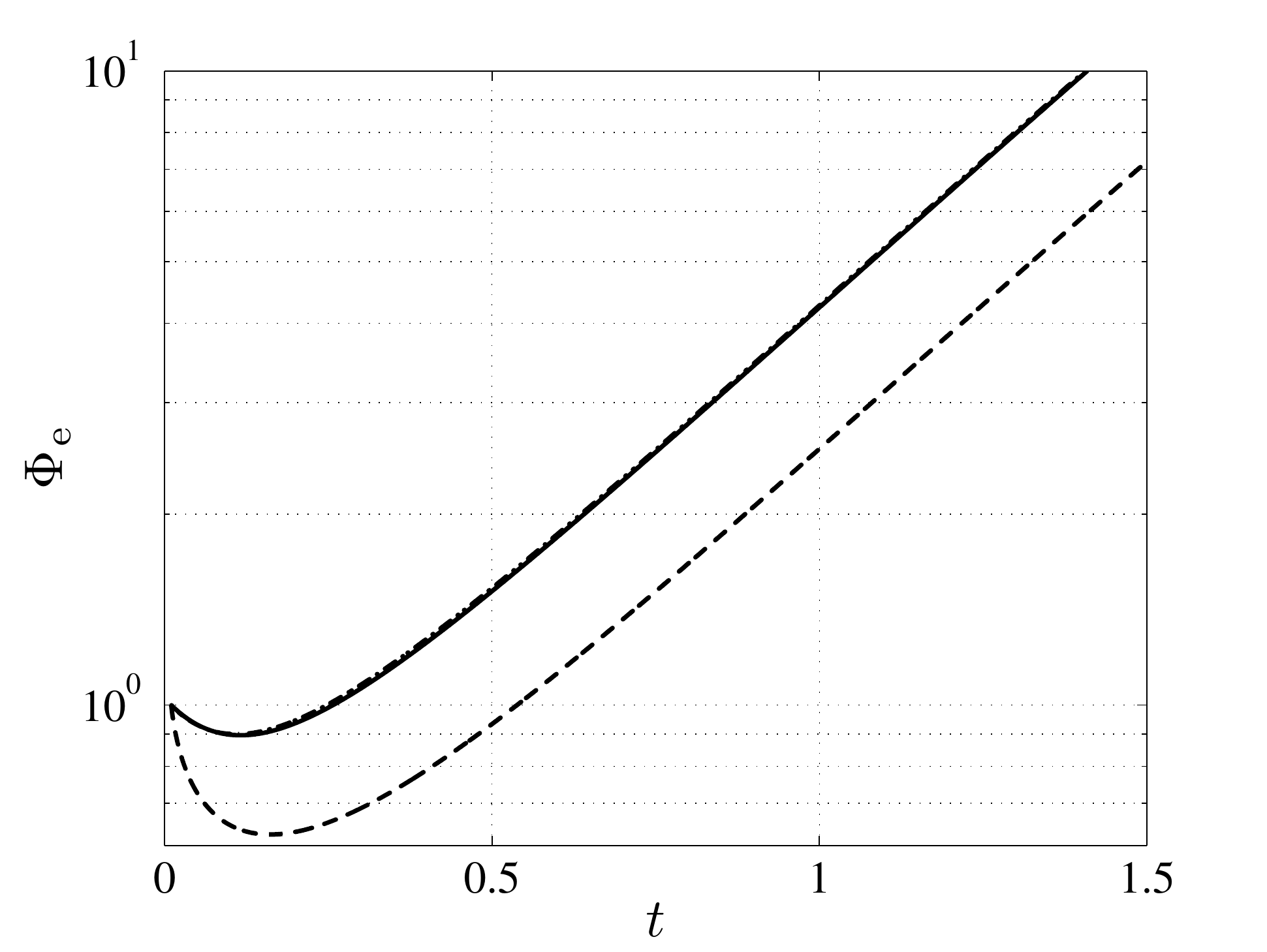}
        \caption{ Optimization results for $Ra=500$ and $\ti=0.01$ when maximizing $\Phic$, $\Phiw$ and $\Phie$, respectively. (\emph{a}) Isocontours of $\Phic$ (solid line), $\Phiw$ (dashed line), and $\Phie$ (dash-dotted line) in the ($\kw$, $\tf$) plane. The $\Phic$ and $\Phie$ lines are visually indistinguishable. (\emph{b}) Dominant wavenumbers, $\kw_\mathrm{max}$, vs. $\tf$, when maximizing $\Phic$ (solid line), $\Phiw$ (dashed line), and $\Phie$ (dash-dotted line). A log scale has been used for $\tf$. (\emph{c}) The optimal $\cip$ profiles when maximizing $\Phic$ (circles), $\Phiw$ (crosses), and $\Phie$ (squares) for $\kw=30$ and $\tf=5$. The base state $\cb(z,\ti)$ is shown as a solid line. (\emph{d})--(\emph{f}) Amplifications $\Phic$, $\Phiw$, and $\Phie$ vs. $t$ when integrating the forward IVP (\ref{eq:linear})--(\ref{eq:ic}) in time using the optimal initial $\cip$ profiles shown in panel (\emph{c}), that maximize $\Phic$ (solid line), $\Phiw$ (dashed line), and $\Phie$ (dash-dotted line). }
    \label{figure2}
    \end{center}
\end{figure}

Figure \ref{figure2} presents optimization results for $Ra=500$ and $\ti=0.01$ when maximizing $\Phic$, $\Phiw$, and $\Phie$. The Rayleigh number is set to a  typical value for $\mathrm{CO}_2$ sequestration \cite[]{Ennis-King2005}. The initial perturbation time is chosen to be one order-of-magnitude smaller than the critical time for instability, $\tc$, where the critical time is the time at which $d \Phi/ dt =0$, after which $\Phi$ begins to increase. The critical time depends on the initial condition and choice of the amplification measure; however, previous analyses find the minimum critical time is on the order of $\tc \sim O(0.1)$ for $\Ra=500$ \cite[]{Riaz2006JFM, Selim2007a, Slim2010}. 
Panel (\emph{a}) illustrates optimal isocontours of $\Phic$ (solid lines), $\Phiw$ (dashed lines), and $\Phie$ (dash-dotted lines) in the ($\kw$, $\tf$) plane. 
The three amplification measures produce qualitatively similar behavior. 
The isocontours for $\Phic$ and $\Phie$ are visually indistinguishable. For much of the ($\kw$, $\tf$) plane, $\Phiw$ is marginally larger than $\Phic$ or $\Phie$. 

We define the dominant wavenumber, $\kw_\mathrm{max}$, as the wavenumber for which the amplification is maximized at $\tf$,
\begin{equation}
\Phi_\mathrm{max}(\tf) = \sup_{0 \le \kw < \infty} {\Phi(\tf, \kw)}.
\label{phimax}
\end{equation} 
Figure \ref{figure2}(\emph{b}) illustrates the dominant wavenumbers that maximize $\Phic$ (solid line), $\Phiw$ (dashed line), and $\Phie$ (dash-dotted line) for the final times $0.1 \le \tf \le 2$. 
The dominant wavenumbers for the three amplification measures are qualitatively similar. When $\tf \le 0.21 $, the dominant wavenumbers are zero.  When $\tf > 0.21$, the dominant wavenumbers jump discontinuously to values around $\kw_\mathrm{max} \approx 25$.
The dominant zero-wavenumber perturbations were not reported by \cite{Rapaka2008} because they considered late values of $\tf$ for which $\kw_\mathrm{max}$ is non-zero. When comparing results of \cite{Rapaka2008} with the current study, one must note that \cite{Rapaka2008} nondimensionalized the problem with a diffusive time scale, while we use an advective time scale. Consequently, the nondimensional times, $t$, in this study are related to those of \cite{Rapaka2008}, $t^{(\mathrm{R})}$, through the relation $t^{(\mathrm{R})}=t/\Ra$. 
 
Though maximizing different perturbation fields produces  similar dominant  wavenumbers, $\kw_\mathrm{max}$, the corresponding optimal initial profiles, $\cip$ and $\wip$, are sensitive to the amplification measure. 
Figure \ref{figure2}(\emph{c}) illustrates the optimal $\cip$ profiles that maximize $\Phic$ (circles), $\Phiw$ (crosses), and $\Phie$ (squares) at $\tf=5$ for $\kw=30$. For visualization, the profiles have been scaled so $\| \cip \|_{\infty}=1$. The solid line shows the base-state at $\ti=0.01$.  Figure \ref{figure2}(\emph{c}) shows results for $0 \le z \le 0.15$ because the profiles are concentrated near $z=0$ and decay to zero before interacting with the lower wall $z=1$. 
The profiles for $\Phie$ and $\Phic$ have maxima occurring around $z=0.05$, while the profile for $\Phiw$ has a maximum occurring around $z=0.01$. 

Figure \ref{figure2}(\emph{d}) illustrates the temporal evolution of $\Phic$  when the forward IVP (\ref{eq:linear})--(\ref{eq:ic}) is integrated from $\ti=0.01$ to $t=2$ using the three initial $\cip$ profiles illustrated in figure \ref{figure2}(\emph{c}). The $\cip$ profiles that maximize $\Phic$ (solid line) and $\Phie$ (dash-dotted line) produce indistinguishable results in figure  \ref{figure2}(\emph{d}).  
The $\cip$ profile that maximizes $\Phiw$ (dashed line), however, produces much lower values of $\Phic$. This suggests that maximization of $\Phiw$ occurs at the expense of $\Phic$.
Figure  \ref{figure2}(\emph{e}) illustrates the corresponding results for the evolution of $\Phiw$. The $\cip$ profiles that maximize $\Phic$ (solid line) and $\Phie$ (dash-dotted line) produce nearly indistinguishable results, while the profile that maximizes $\Phiw$ (dashed line) produces marginally larger $\Phiw$.
Finally, figure \ref{figure2}(\emph{f}) illustrates the corresponding results for the evolution of $\Phie$. The initial profiles that maximize $\Phic$ and $\Phie$ again produce indistinguishable results. This indicates that maximizing the perturbation's concentration field naturally maximizes $\Phie$, while maximizing $\Phiw$ does so at the expense of $\Phic$ and $\Phie$. Because $\Phic$ naturally maximizes $\Phie$, hereinafter we focus on maximizing $\Phic$. We choose $\Phic$ over $\Phie$ because the application of the coupling conditions (\ref{eq:ref1})--(\ref{eq:ref2}) is much simpler for $\Phic$.

\subsection{Sensitivity to wavenumber $\kw$}

\begin{figure}  
   \begin{center}  
    \hspace{0.0cm}
    (\emph{a})
    \hspace{6.5cm}
    (\emph{b})
    \\
    \includegraphics[trim = 0 0 0 10, clip=true,width=7cm]{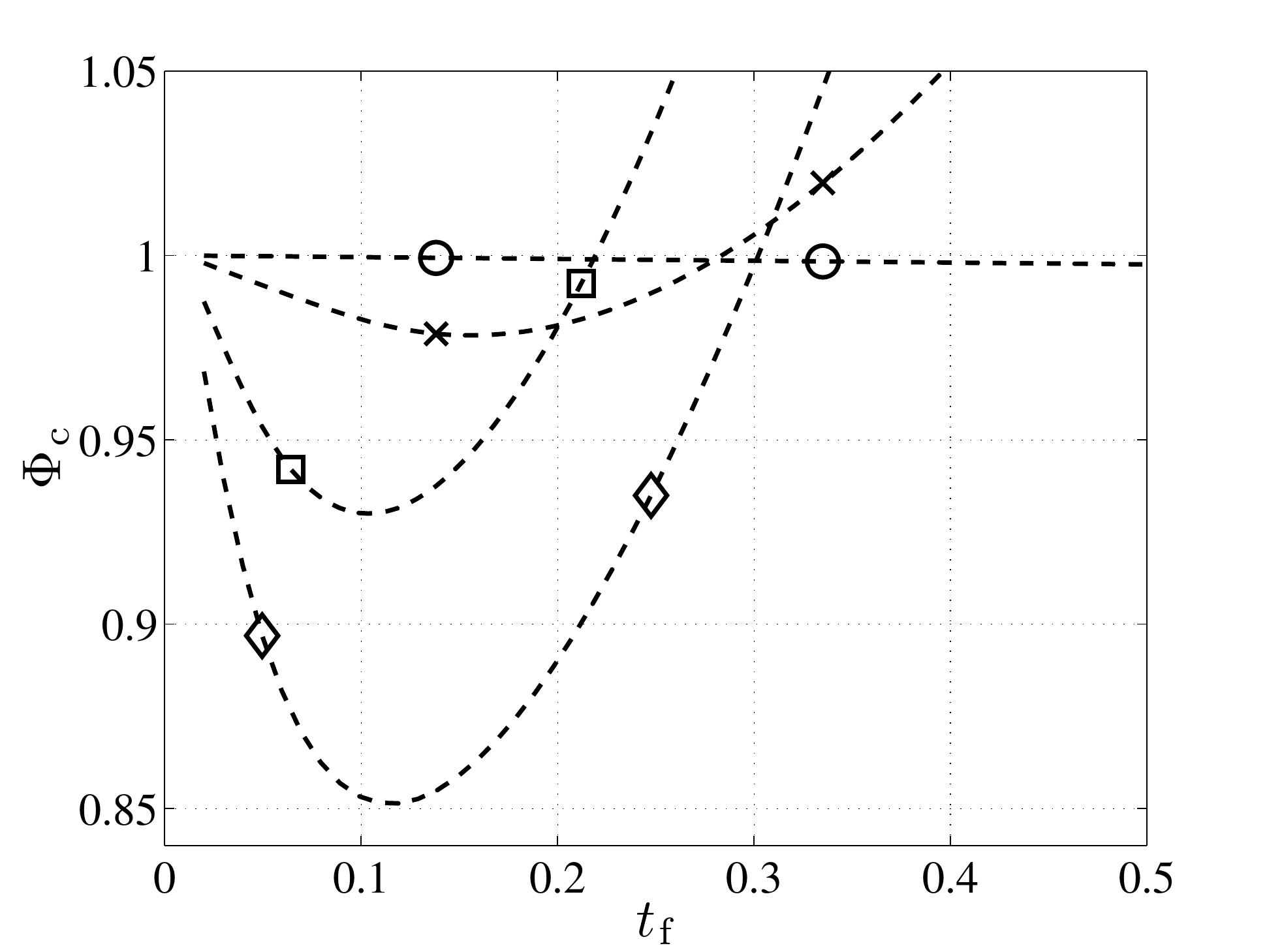}
     \includegraphics[trim = 0 0 0 10, clip=true,width=7cm]{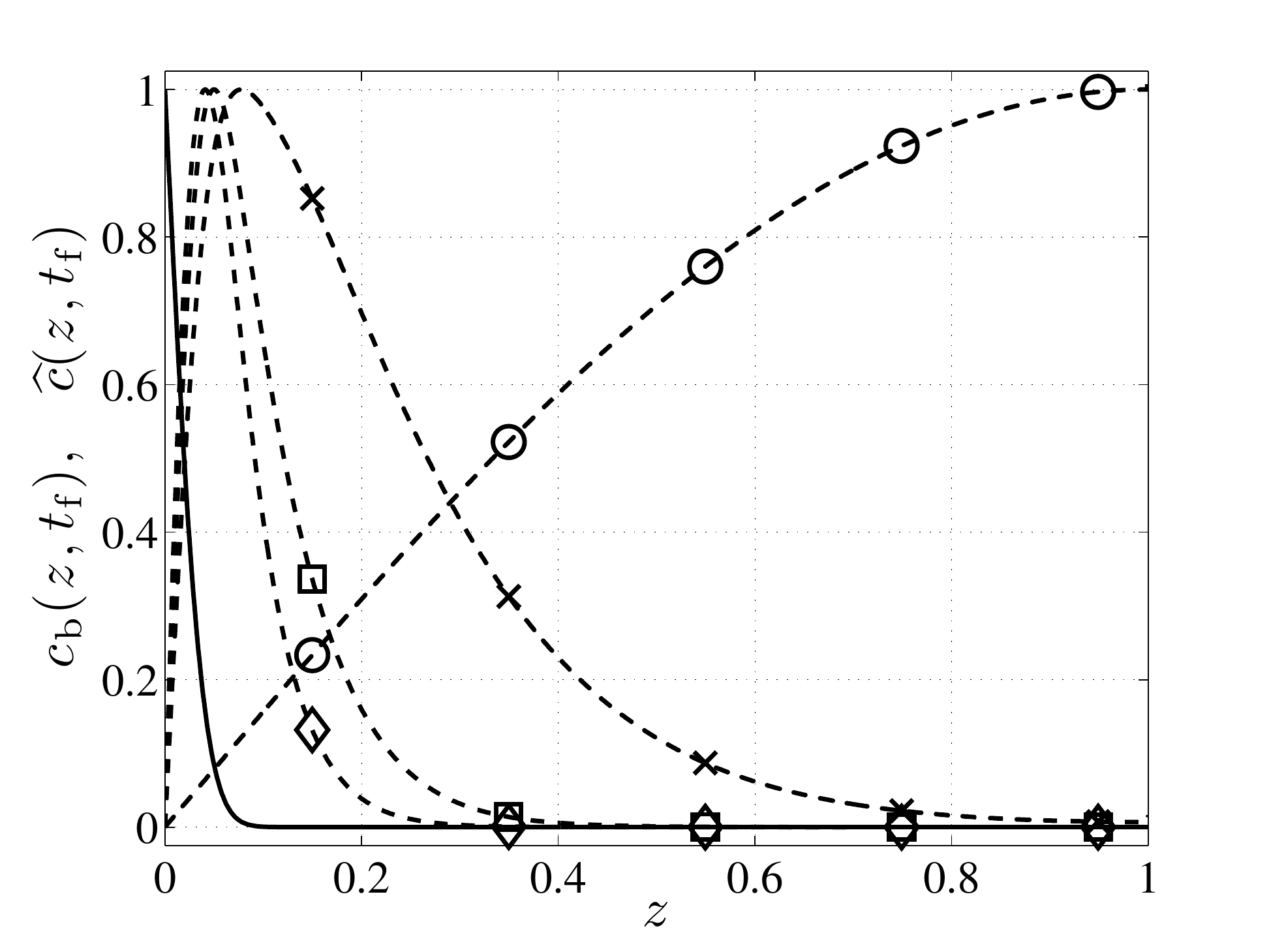}
    \caption{Dominant perturbations for $Ra=500$ and $\ti=0.01$. (\emph{a}) $\Phic$ vs. $\tf$, for $\kw=0$ (circles), $\kw=10$ (crosses), $\kw=25$ (squares), and $\kw=40$ (diamonds).  (\emph{b}) $\cb(z, \tf)$ (solid line) and $\ch(z, \tf)$ for $\tf=0.21$, and $\kw=0$ (circles), $\kw=10$ (crosses), $\kw=20$ (squares), and $\kw=30$ (diamonds). 
    }
   \label{figure3}
   \end{center}
\end{figure}

Figure \ref{figure3}(\emph{a}) illustrates the optimal amplifications $\Phic$ versus $\tf$ for $\ti=0.01$, and $\kw=0$ (circles), $\kw = 10$ (crosses), $\kw = 25$ (squares), and $\kw=40$ (diamonds). For small final times, $\tf < 0.1$, all perturbations decay; however, the $\kw=25$  perturbations are more damped than the $\kw=0$ and $\kw=10$ perturbations. Note that the $k=0$ perturbations have a small constant damping rate. This occurs because the IVP (\ref{eq:linear})--(\ref{eq:ic}) for $\kw=0$ reduces to
\begin{equation}
    \frac{\partial \ch}{\partial t} - \frac{1}{Ra}\frac{\partial^2 \ch }{\partial z^2}=0, \qquad  \widehat{w}=0.
\label{eq:diff}
\end{equation}
Equation (\ref{eq:diff}) can be solved analytically to show that the optimal perturbation is given by 
\begin{equation*}
\ch = \sin{(\pi z /2)}\exp{(-\pi^2 \Ra^{-1} t/4)}.
\end{equation*}
In contrast to the $\kw=0$ perturbations, finite wavenumber perturbations do not have constant growth rates. The $\kw=25$ perturbations begin to grow around $\tf=0.1$ and eventually overtake the $\kw=10$ and $\kw=0$ perturbations.  This explains the discontinuous jump in the dominant wavenumbers from $\kw_\mathrm{max}=0$ to $\kw_\mathrm{max}\approx 25$ illustrated in figure \ref{figure2}(\emph{b}). The $\kw=40$ perturbations experience greater damping, and consequently, never overtake the $\kw=25$ perturbations.   

Figure \ref{figure3}(\emph{b}) illustrates the base state, $\cb(z,\tf)$ (solid line), and optimal profiles, $\ch(z, \tf)$, at $\tf=0.21$ for $\kw=0$ (circles), $\kw=10$ (crosses), $\kw=20$ (squares), and $\kw=30$ (diamonds). The final time is chosen to be near to the discontinuous jump in $\kw_\mathrm{max}$ illustrated in figure \ref{figure2}(\emph{b}). The optimal profile for $\kw=0$ has a maximum at the lower boundary at $z=1$, while the profiles for $\kw=10$, 20, and 30  have maxima near $z=0$. With increasing $\kw$, the optimal profiles become increasingly concentrated within the boundary layer. 

The results for $\Phic$ and $\cip$ illustrated in figure \ref{figure3} can be explained physically by examining the competing effects of the stabilizing diffusive term, $\D \ch / Ra $, and the destabilizing convective term, $\wh \partial \cb / \partial z$, in equation (\ref{eq:linear}). 
At small times, $t \ll \tc$, the convective term $\wh \partial \cb / \partial z$ has only a small effect because $\partial \cb / \partial z$ is nonzero only within the thin boundary layer where  $\wh$ necessarily tends
to zero due to the no-penetration condition at $z=0$.
This explains why the boundary layer is stable at small times. The dominant wavenumber is initially zero because finite wavenumber perturbations have additional damping due to the transverse diffusive term $(\kw^2/\Ra)\ch$ in equation (\ref{eq:linear}).
At later times, the growing boundary layer increases the influence of the destabilizing term $\wh \partial \cb / \partial z$  such that non-zero wavenumber perturbations become unstable. This explains why dominant perturbations at late times tend to be increasingly concentrated in the boundary layer. 

\subsection{Sensitivity to initial perturbation time $\ti$}

\begin{figure}  
 \begin{center}  
 \hspace{0.0cm}
    (\emph{a})
    \hspace{6.5cm}
    (\emph{b})
    \\
        \includegraphics[trim = 0 0 0 10, clip=true,width=7cm]{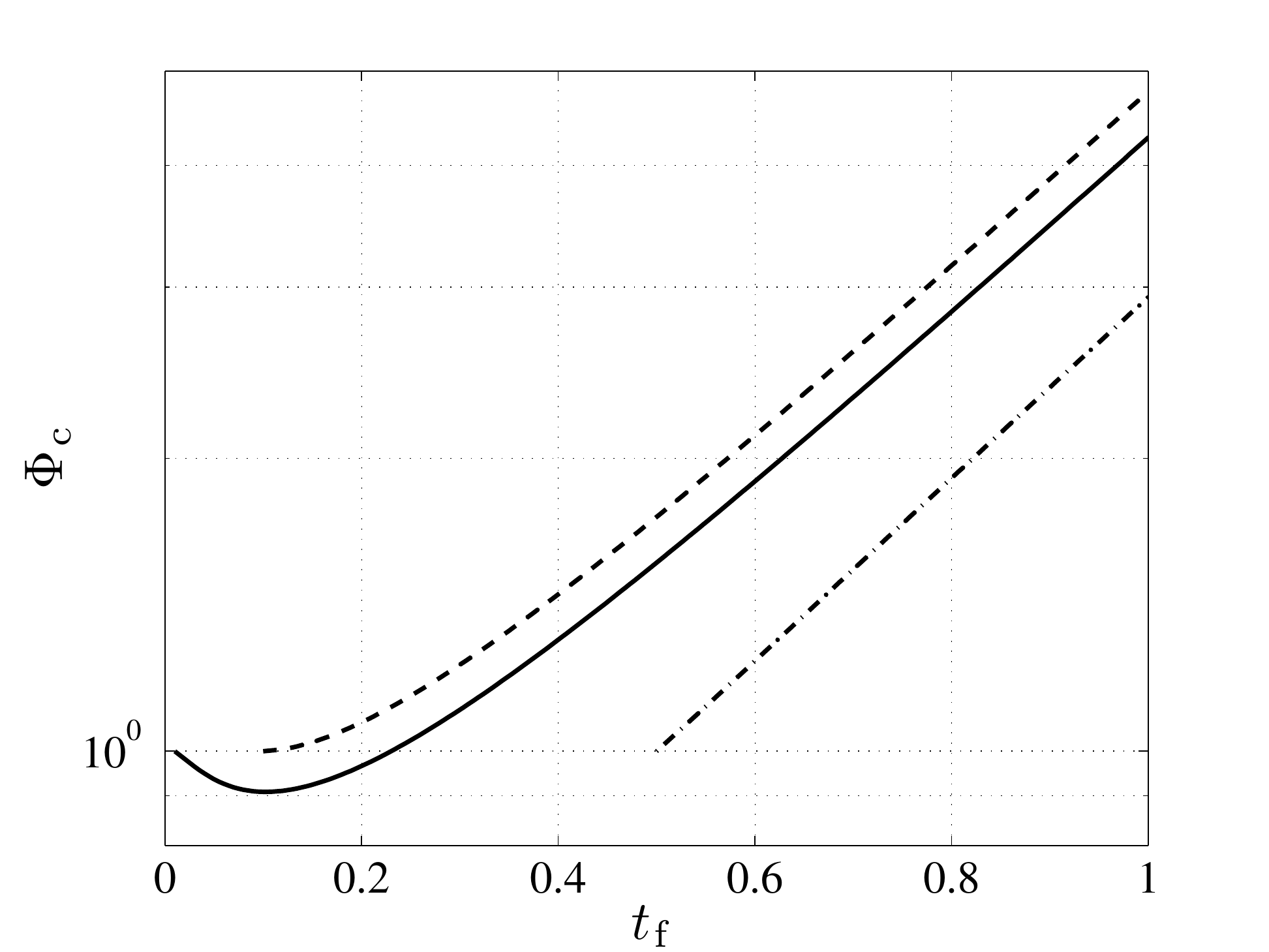}
    \includegraphics[trim = 0 0 0 10, clip=true,width=7cm]{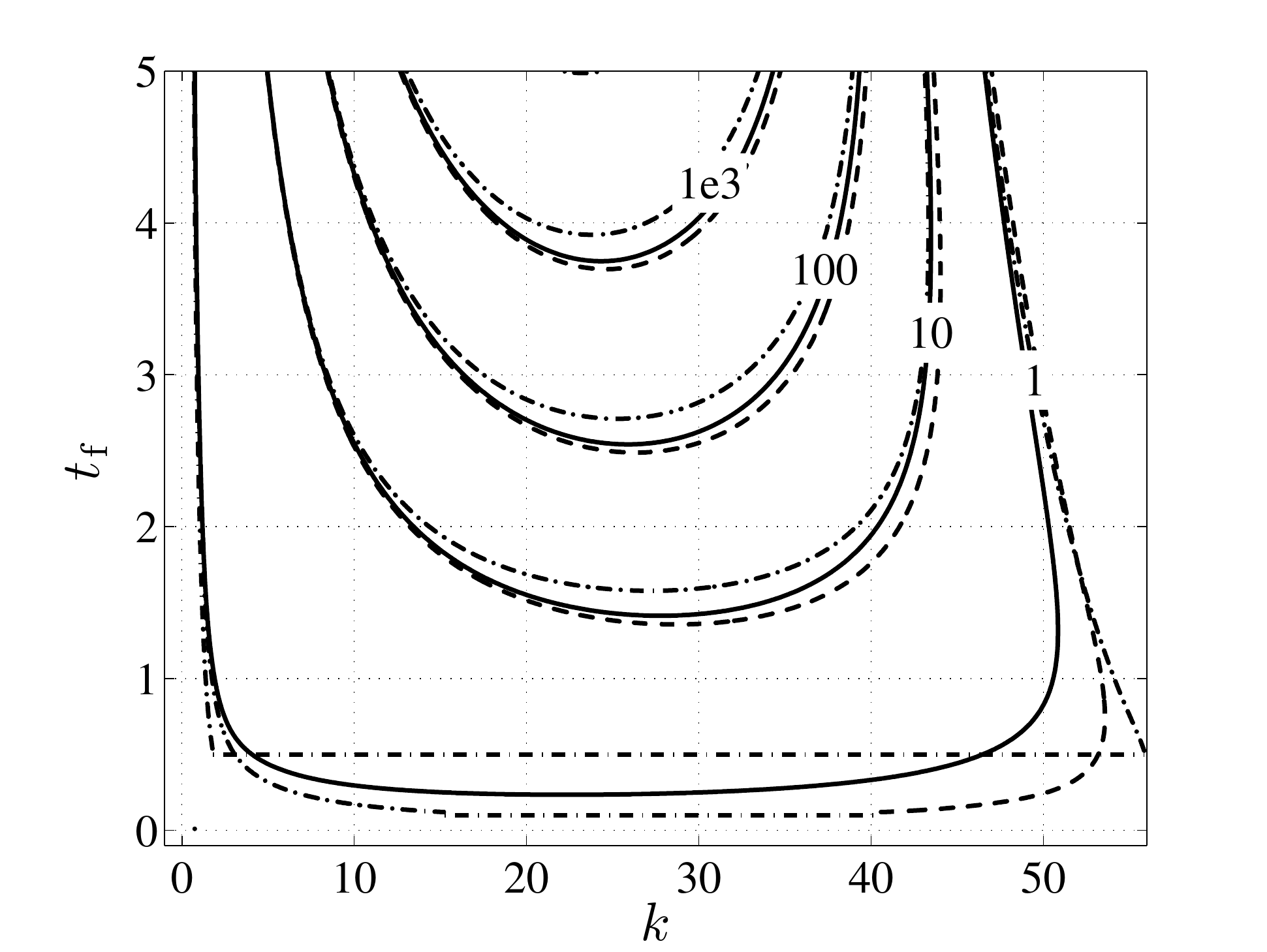}
     \\
    \hspace{0.0cm}
    (\emph{c})
    \hspace{6.5cm}
    (\emph{d})
    \\
    \includegraphics[trim = 0 0 0 10, clip=true,width=7cm]{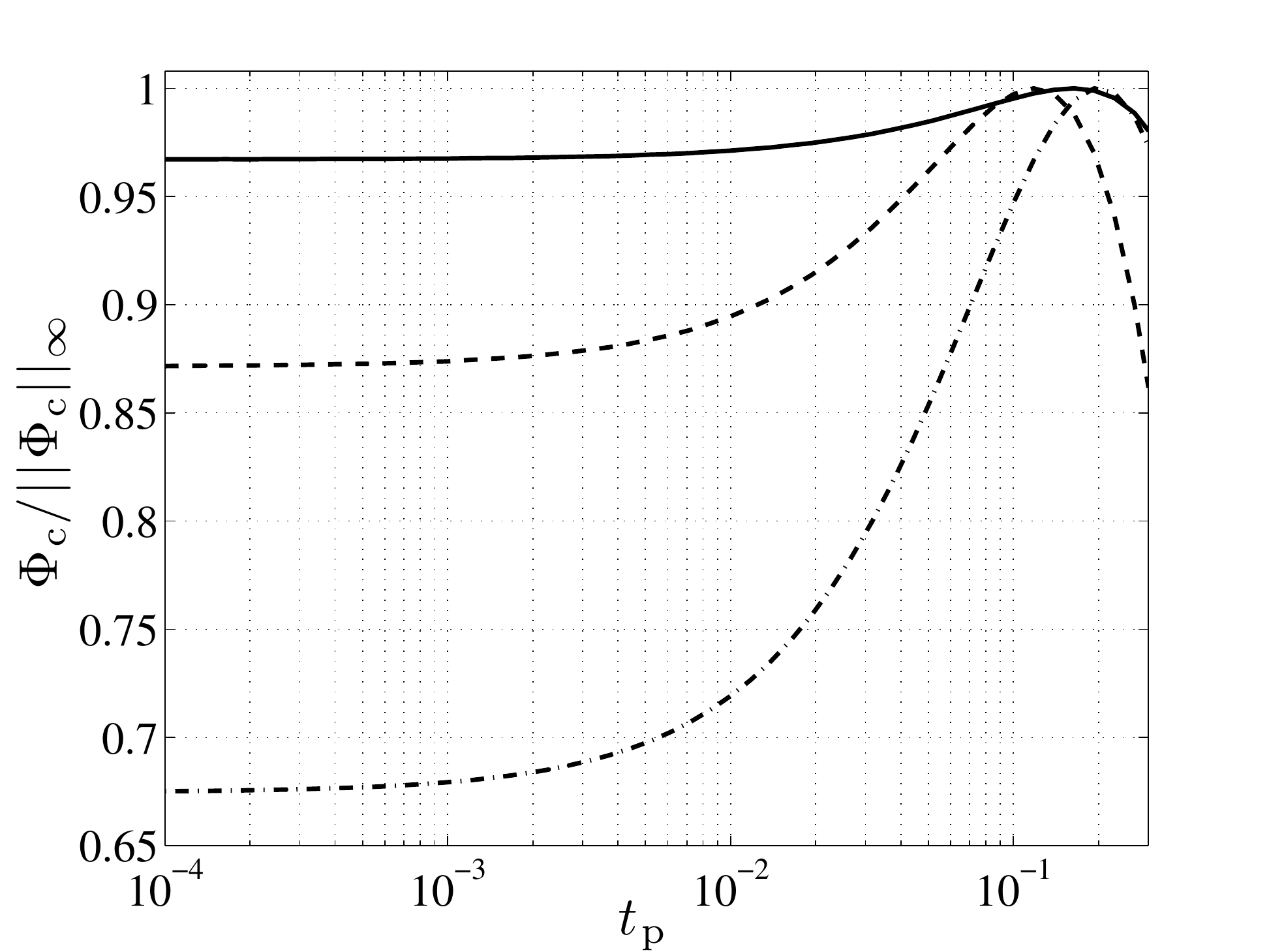}
    \includegraphics[trim = 0 0 0 10, clip=true,width=7cm]{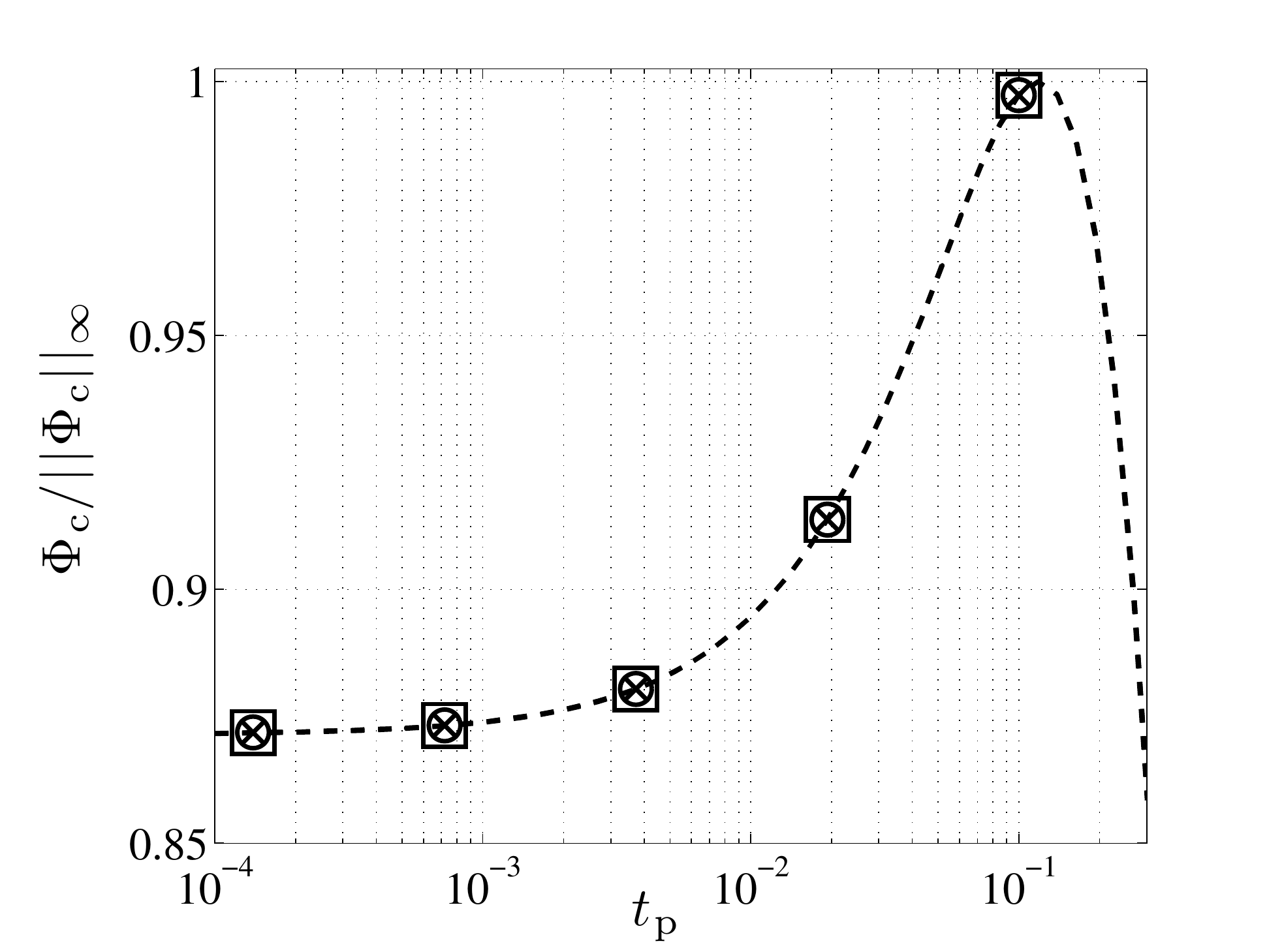}
    \caption{ Effect of initial perturbation time for $Ra=500$. (\emph{a}) $\Phic$ vs. $\tf$ for $k=30$, and $\ti=0.001$ (solid line), $\ti = 0.1$ (dashed line), and $\ti = 0.5$ (dash-dotted line). (\emph{b}) Isocontours of $\Phic$ in the ($k$,$\, \tf$) plane for $\ti=0.001$ (solid line), $\ti = 0.1$ (dashed line), and $\ti = 0.5$ (dash-dotted line). (\emph{c}) $\Phic / ||\Phic||_\infty$, vs. $\tp$ for $\tf=1$, and $\kw=10$ (solid line), $\kw=30$ (dashed line), $\kw=50$ (dash-dotted line). (\emph{d}) $\Phic / ||\Phic||_\infty$ vs. $\tp$ for $\kw=30$ and $\tf=1$ (circles), $\tf=2$ (crosses), and $\tf=3$ (squares). }
   \label{figure4}
  \end{center}
\end{figure}

Due to the transient nature of the base-state, the optimal perturbations also depend on the time, $\ti$, at which the boundary layer is perturbed. 
Figure \ref{figure4} explores the sensitivity of the optimal amplifications $\Phic$ to the initial perturbation time $\ti$ for Ra=500. 
Panel (\emph{a}) illustrates $\Phic$ versus $\tf$ for $\kw = 30$ and $\ti = 0.001$ (solid line), $\ti = 0.1$ (dashed line), and $\ti = 0.5$ (dash-dotted line). Perturbations originating at $\ti=0.001$ have a long initial damping period and consequently have smaller amplifications than perturbations originating at $\ti=0.1$. Perturbations originating at the late time $\ti=0.5$ experience no damping, but have smaller amplifications than perturbations originating at $\ti=0.001$ and $\ti=0.1$ because those perturbations begin growing much earlier. At later times, $\tf>0.5$, the three curves have identical slopes, indicating that the perturbations have identical temporal growth rates.
Figure  \ref{figure4}(\emph{b}) illustrates isocontours of $\Phic$ in the ($\kw$, $\tf$) parameter plane for $\ti = 0.001$ (solid line), $\ti = 0.1$ (dashed line), and $\ti = 0.5$ (dash-dotted line). As expected, perturbations originating at $\tp=0.1$ produce larger amplifications. The horizontal dash-dotted line  indicates that perturbations originating at $\ti=0.5$ grow immediately for $2<\kw<56$. 

\begin{figure}
 \begin{center}
     \hspace{0.0cm}
     (\emph{a})
     \hspace{6.5cm}
     (\emph{b})
     \\
     \includegraphics[trim = 0 0 0 10, clip=true,width=7cm]{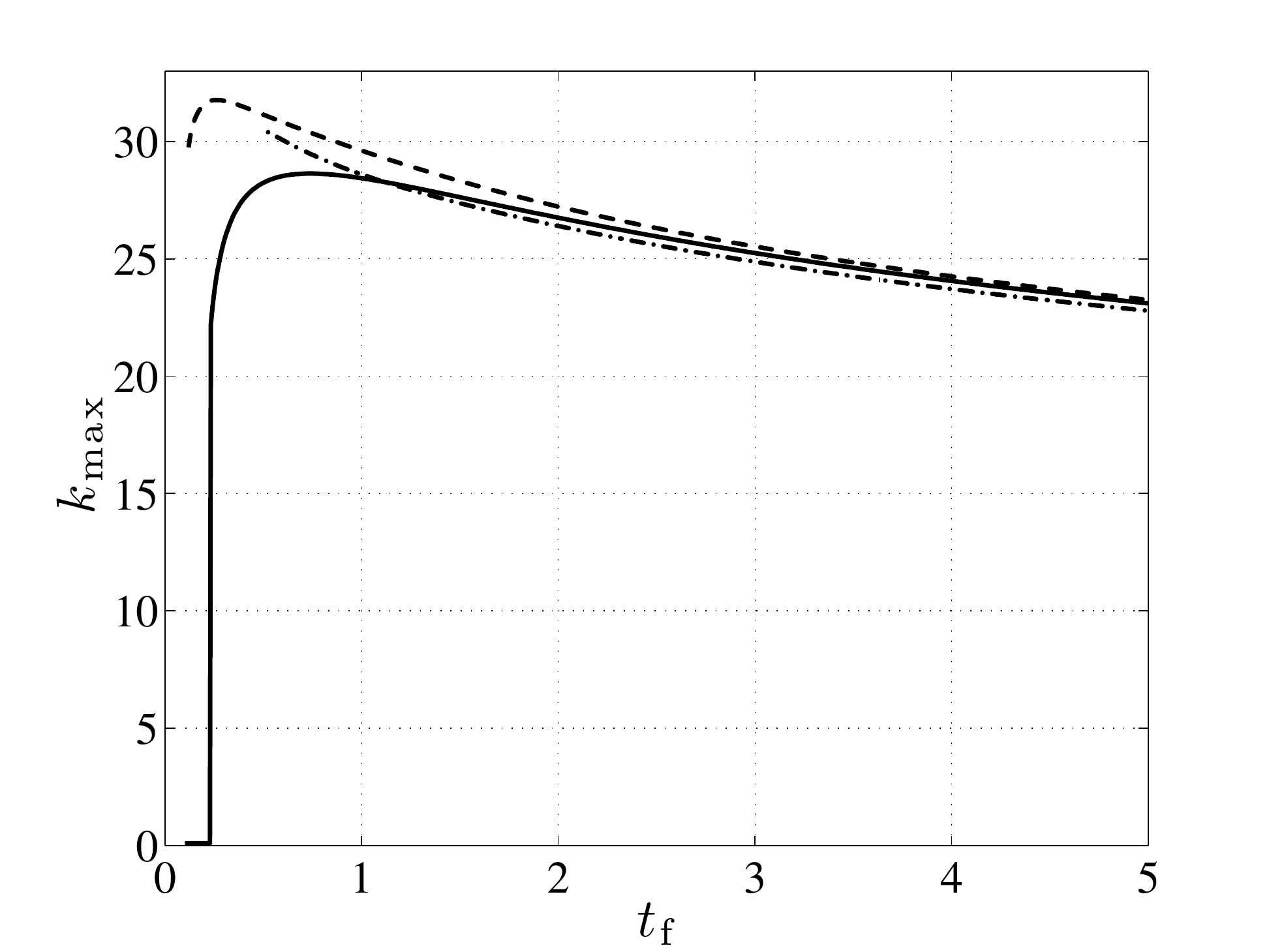}
     \includegraphics[trim = 0 0 0 10, clip=true,width=7cm]{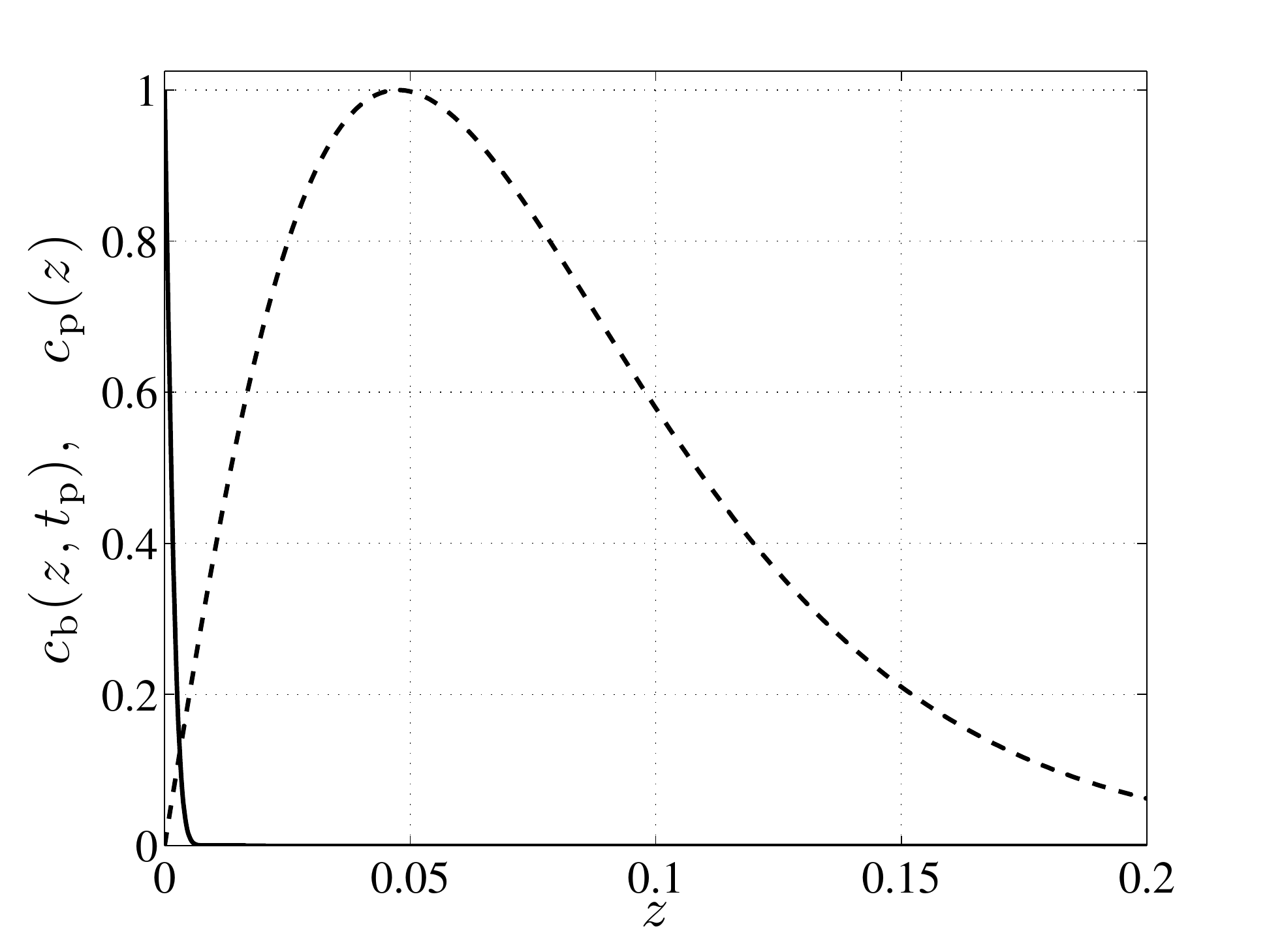}
      \\
     \hspace{0.0cm}
     (\emph{c})
     \hspace{6.5cm}
     (\emph{d})
     \\
     \includegraphics[trim = 0 0 0 10, clip=true,width=7cm]{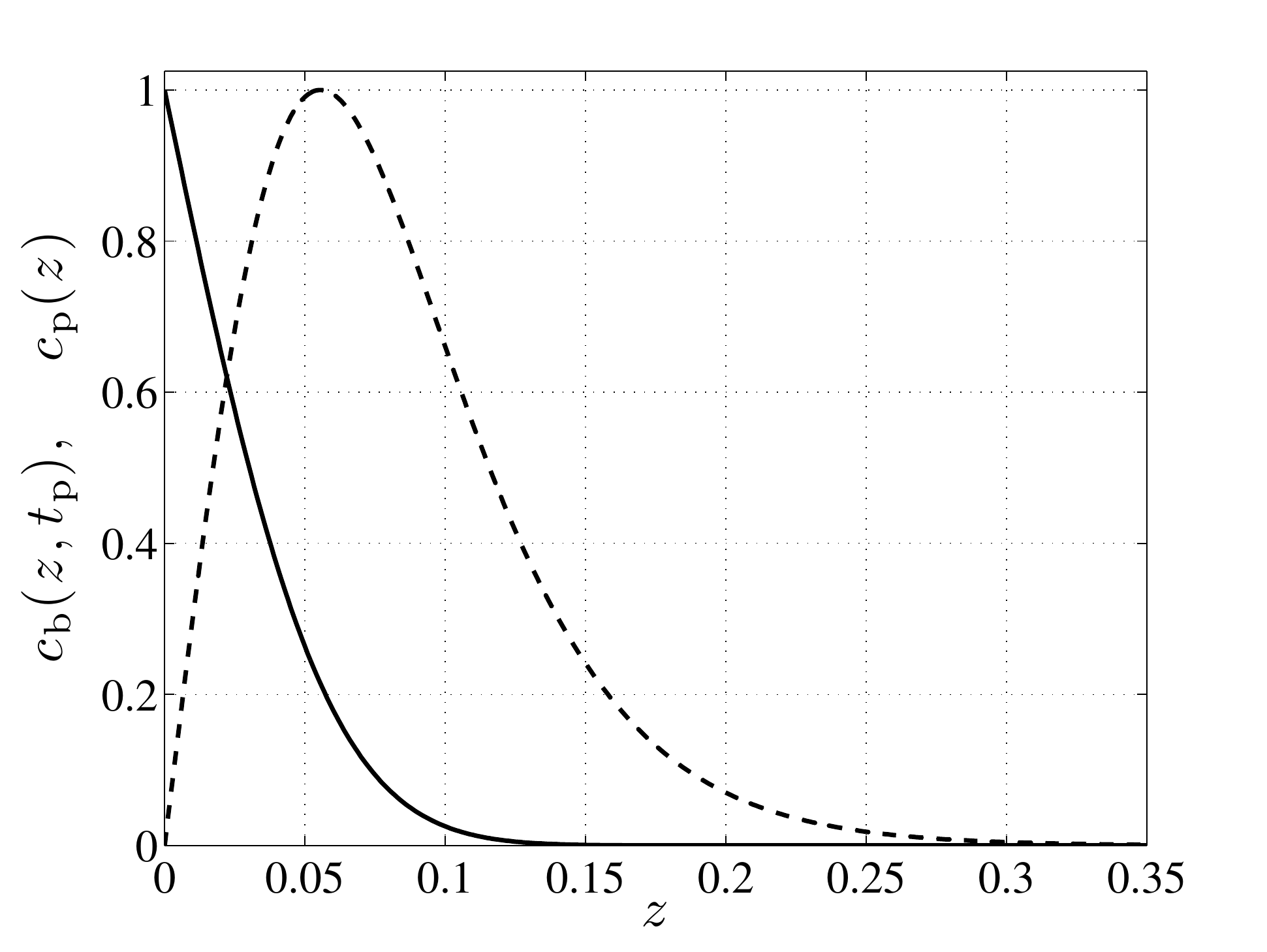}
     \includegraphics[trim = 0 0 0 10, clip=true,width=7cm]{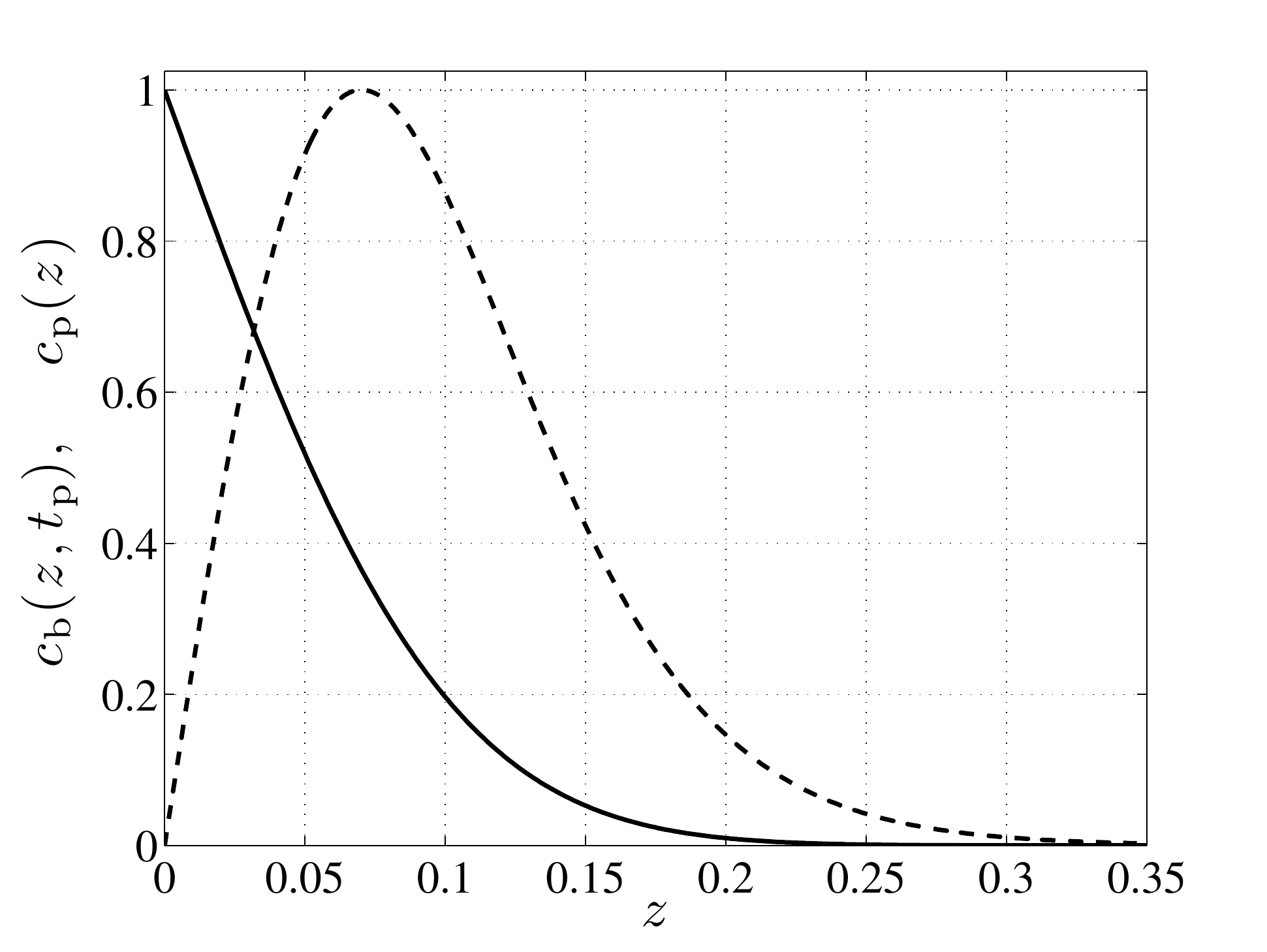}
        \caption{  (\emph{a}) Dominant wavenumbers, $k_\mathrm{max}$ vs. $\tf$ for $\ti=0.001$ (solid line), $\ti = 0.1$ (dashed line), and $\ti = 0.5$ (dash-dotted line). (\emph{b}) Base-state, $c_\mathrm{b}$ (solid line), and optimal $\cip$ profiles (dashed line) for $\ti=0.001$, $\tf=5$, and $\kw=30$. (\emph{c}) Same as panel (\emph{b}) for $\ti=0.5$. (\emph{d}) Same as panel (\emph{b}) for $\ti=1.5$.  With increasing $\ti$, the $\cip$ profiles become increasingly concentrated in the boundary layer.       } 
    \label{figure5}
\end{center}
\end{figure}

Figures \ref{figure4}(\emph{a}) and \ref{figure4}(\emph{b}) suggest that there exists an optimal initial perturbation time, $\ti^\mathrm{o}$, that maximizes $\Phic$. Perturbations originating prior to $\ti^\mathrm{o}$ cannot outgrow the optimal perturbation originating at $\ti^\mathrm{o}$ due to the initial damping period. From figure \ref{figure4}(\emph{a}), we expect $\ti^\mathrm{o}$ to occur near the critical time, $t=\tc$, because this minimizes the damping period. Note that for $\Ra=500$, \cite{Slim2010} report that the minimum critical time is $\tc \approx 0.096$. The notion of an optimal initial perturbation time may appear counterintuitive because in physical systems the boundary layer is continuously perturbed beginning at $\ti=0$. Within the framework of a linear stability analysis, however, the response to this continuous forcing can be expressed as the infinite sum of many impulse responses to forcing at discrete initial times, $\ti$. The optimal perturbation originating at $\ti^\mathrm{o}$ gives a theoretical upper bound for the amplification.    

Figure \ref{figure4}(\emph{c}) illustrates the normalized amplifications, $\Phic / ||\Phic||_\infty$, versus $\tp$ for $\tf=1$, and $\kw=10$ (solid line), $\kw=30$ (dashed line), and $\kw=50$ (dash-dotted line). The amplifications have been normalized with respect to their maximum values to facilitate comparison between the results for different wavenumbers. As $\ti \rightarrow 0$, the amplifications asymptote to constant values. With increasing $\ti$, the amplifications attain maxima near $\ti=\tc$ and then decrease. We observe stronger sensitivity of $\Phic$ to $\ti$ with increasing wavenumber. 
This behavior is similar to that observed in figure \ref{figure3}(\emph{a}) for the sensitivity of $\Phic$ to the final time $\tf$.  The increasing sensitivity of $\Phic$ to both $\ti$ and $\tf$ at higher wavenumbers is likely due to the increase in transverse diffusive damping as noted in the previous section.
Figure \ref{figure4}(\emph{d}) illustrates  $\Phic / ||\Phic||_\infty$ versus $\ti$ for $\kw=30$ and $\tf=1$ (circles), $\tf=2$ (crosses), and $\tf=3$ (squares). The results for different $\tf$ are indistinguishable from each other. This occurs because, as demonstrated in figure \ref{figure4}(\emph{a}), the perturbations have identical growth rates for $\tf>1$.

Figure  \ref{figure5}(\emph{a}) illustrates the temporal evolution of the dominant wavenumbers, $\kw_\mathrm{max}$, when $\ti = 0.001$, (solid line), $\ti = 0.1$ (dashed line), and $\ti = 0.5$ (dash-dotted line). 
As expected from the discussion in \S 4.2, the dominant wavenumbers are initially zero when $\ti=0.001$.
When $\ti=0.1$, however, the dominant wavenumber is initially $k_\mathrm{max}= 29.74$ for $\tf=0.12$ and reaches a maximum at $\tf=0.26$ after which it decays monotonically. 
When $\ti=0.5$, $k_\mathrm{max}$ decreases monotonically with $\tf$. Previously, \cite{Rapaka2008} only reported cases with a monotonic decay of $k_\mathrm{max}$ with $\tf$.
Figures \ref{figure5}(\emph{b})--\ref{figure5}(\emph{d}) illustrate the base state (solid lines) and optimal $\cip$ profiles (dashed lines), for $\ti=0.001$ (panel \emph{b}), $\ti=0.5$ (panel \emph{c}), and $\ti=1.5$ (panel \emph{d}) for $\Ra=500$, $\kw=30$, and $\tf=5$. 
As expected from the discussion in \S 4.2, the optimal profiles become increasingly concentrated within the boundary layer with increasing $\ti$ due to the destabilizing convective term.  

\begin{figure}   
 \begin{center} 
    \hspace{0.0cm}
    (\emph{a})
    \hspace{6.5cm}
    (\emph{b})
    \\
    \includegraphics[trim = 0 0 0 10, clip=true,width=7cm]{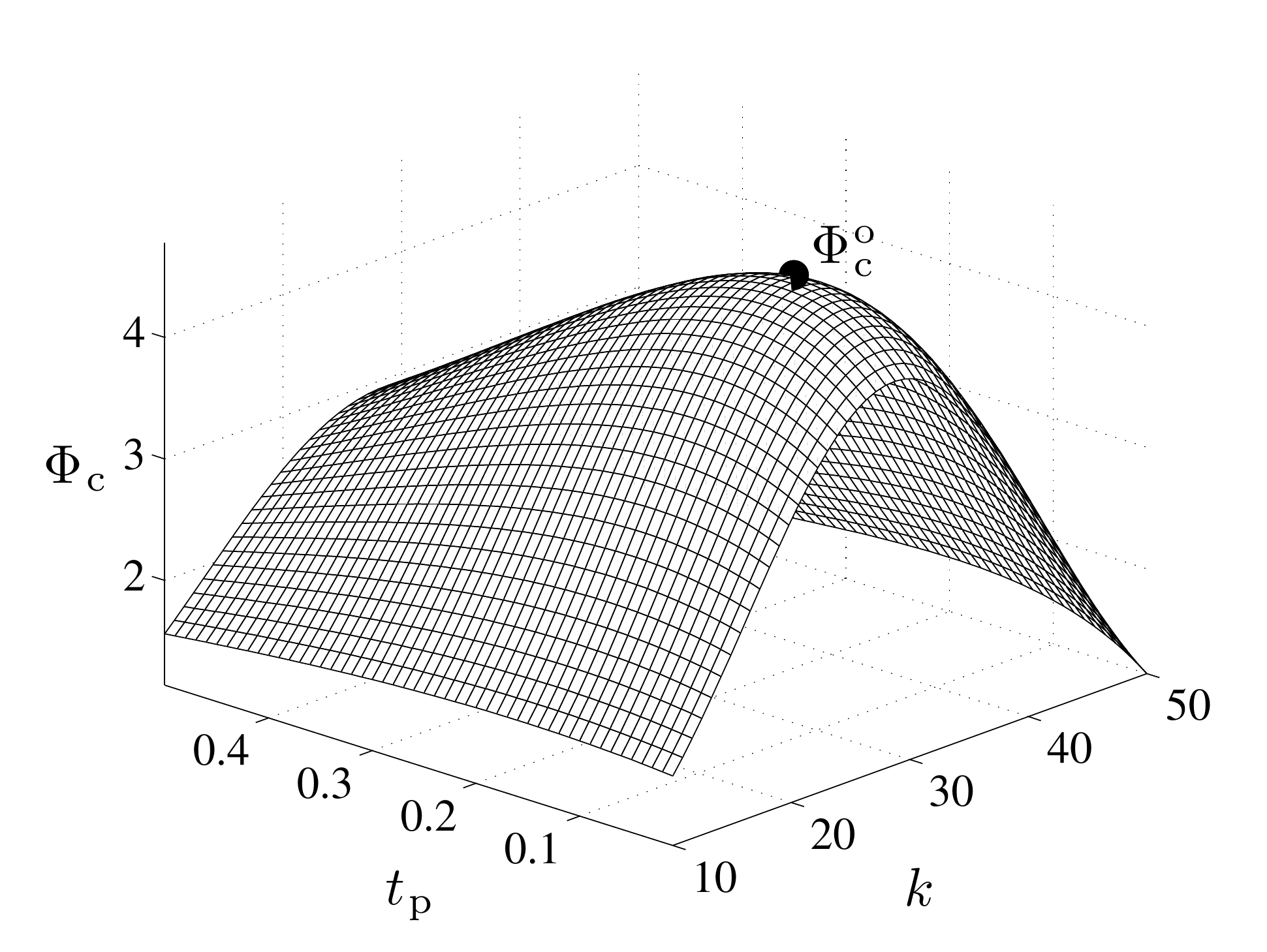}
    \includegraphics[trim = 0 0 0 10, clip=true,width=7cm]{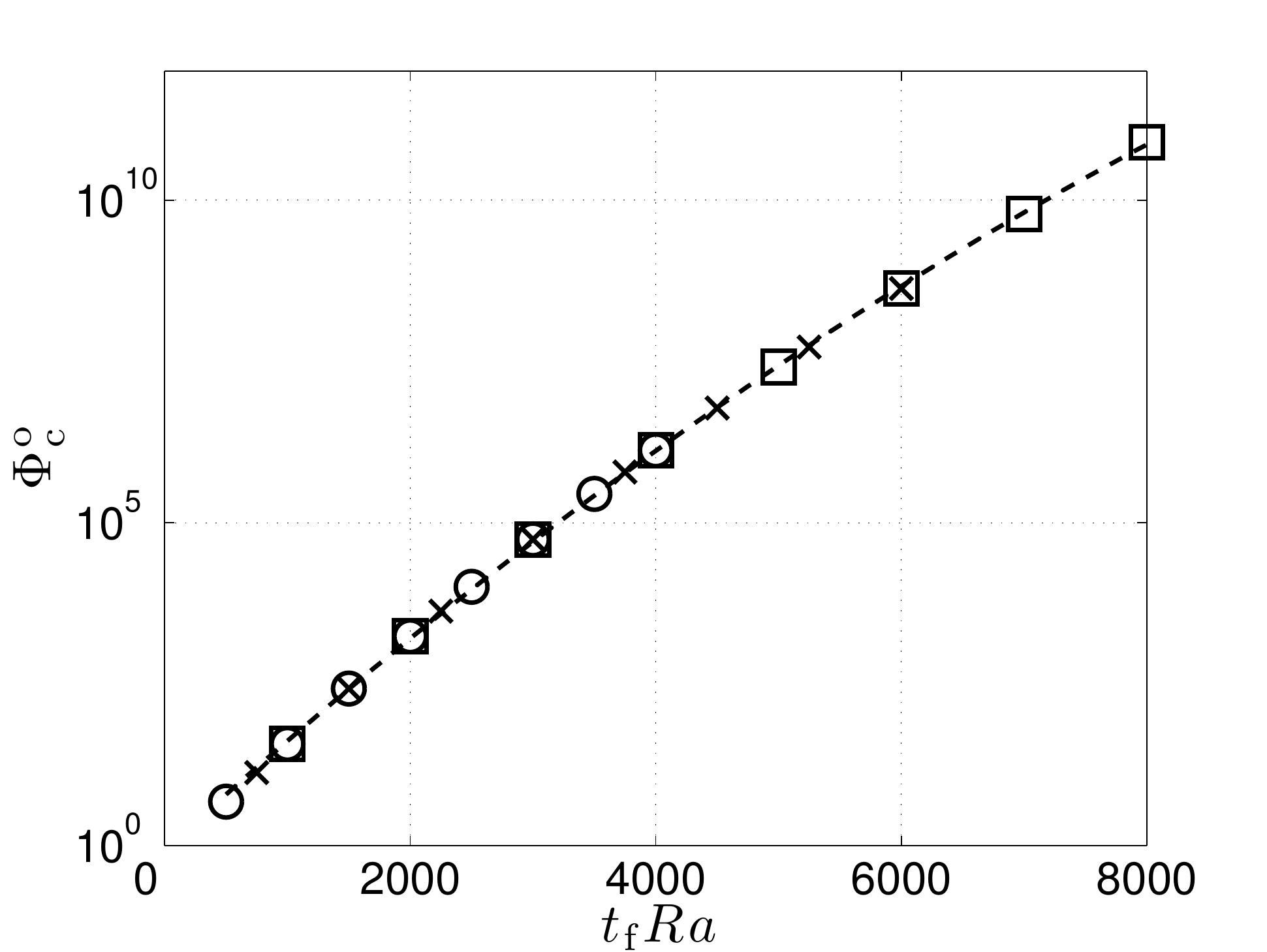}
    \\
     \hspace{0.0cm}
    (\emph{c})
    \hspace{6.5cm}
    (\emph{d})
    \\
    \includegraphics[trim = 0 0 0 10, clip=true,width=7cm]{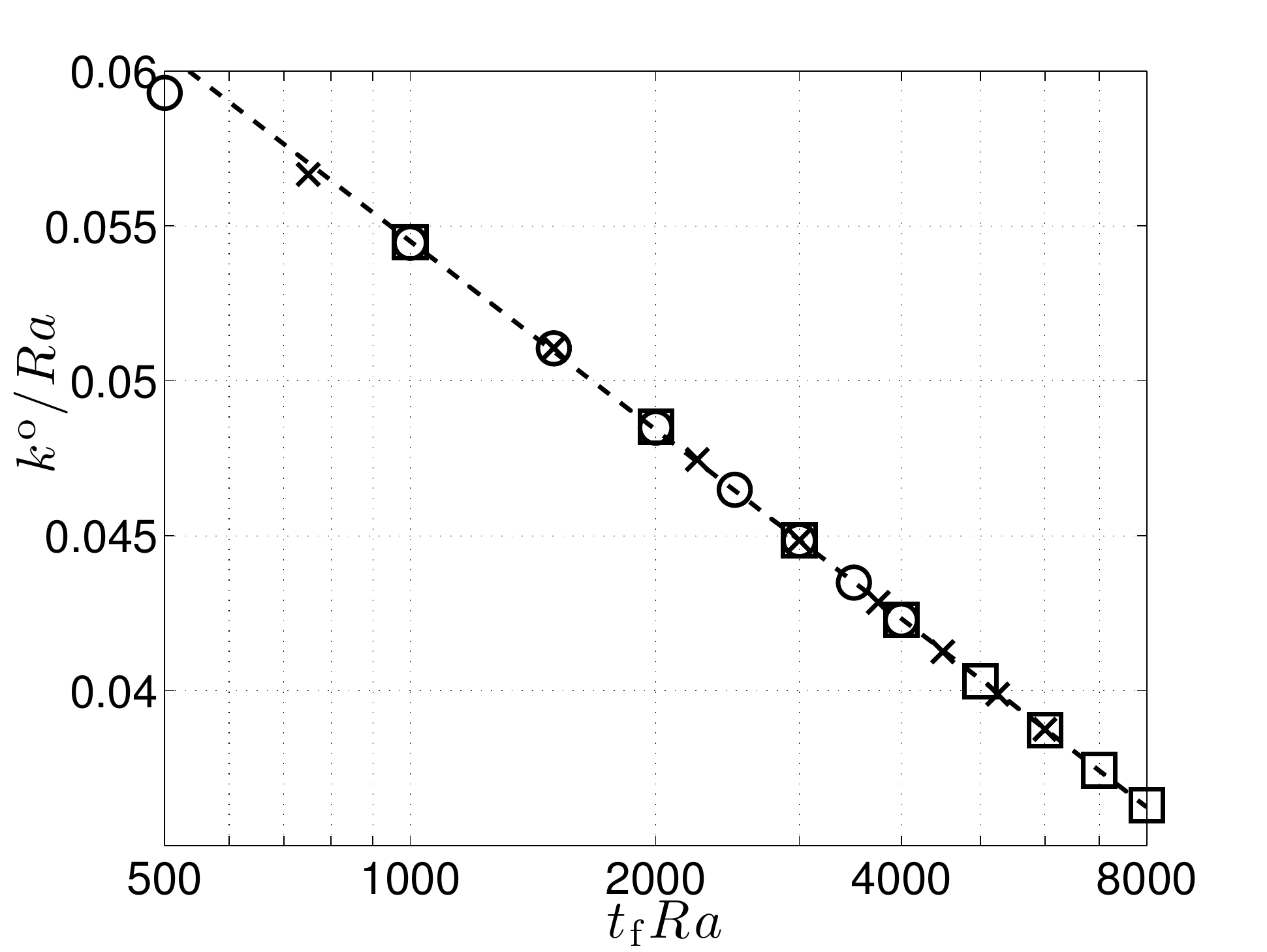}
    \includegraphics[trim = 0 0 0 10, clip=true,width=7cm]{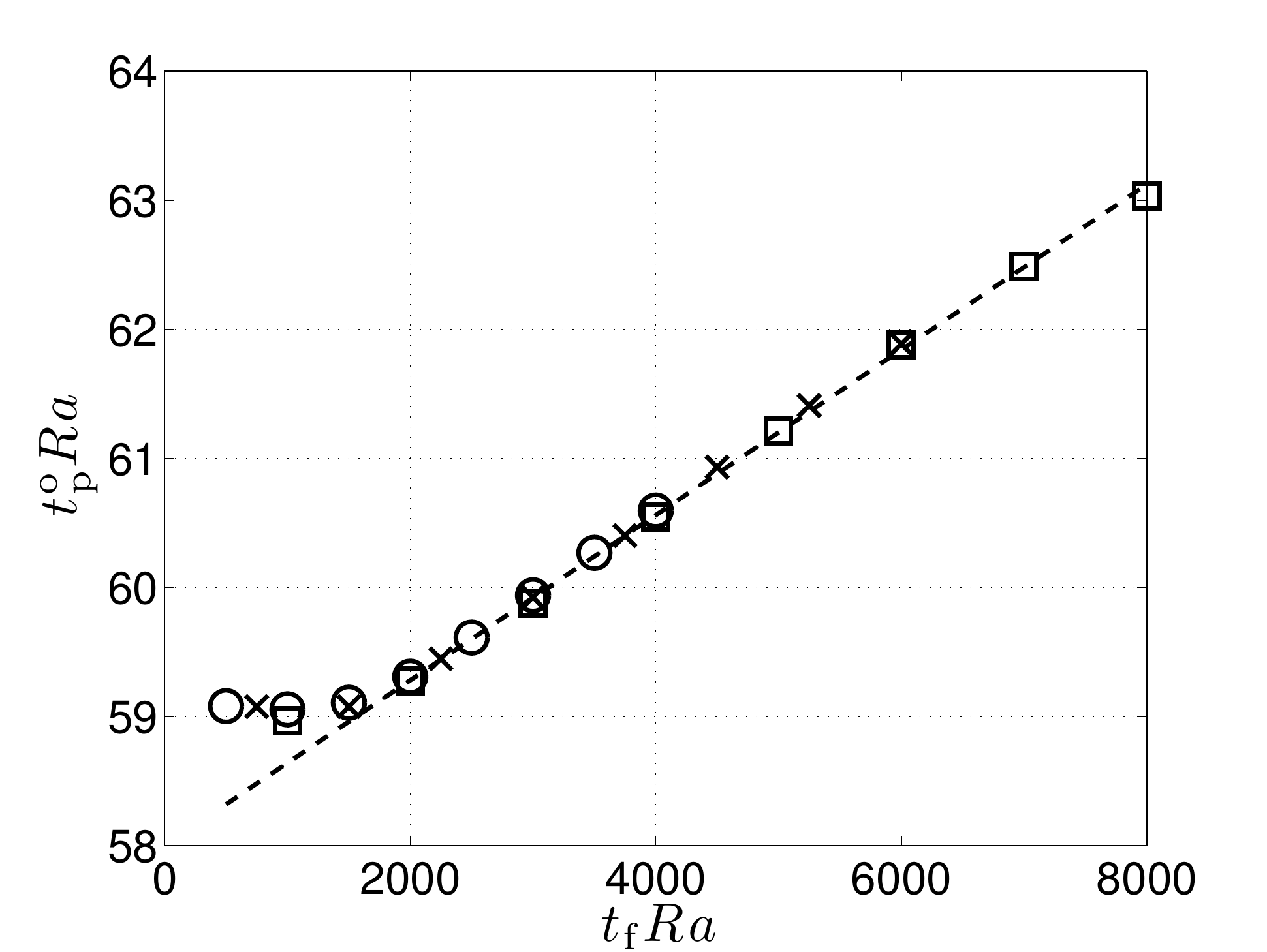}
    \caption{ The optimal point  $(\Phic^\mathrm{o}$, $\kw^\mathrm{o}$, $\ti^\mathrm{o})$ as a function of $\tf$ and $\Ra$. (\emph{a}) $\Phic$ vs. $\ti$ and $\kw$ for $\Ra=500$ and $\tf=1$. The solid dot marks $(\Phic^\mathrm{o}$, $\kw^\mathrm{o}$, $\ti^\mathrm{o})$.  (\emph{b})  $\Phic^\mathrm{o}$  vs. $\tf \Ra$  for $\Ra = 500$ (circles), $\Ra = 750$ (crosses), and $\Ra = 1000$ (squares). The dashed line shows relationship (\ref{eqphio}). (\emph{c}) $\kw^\mathrm{o}/\Ra$  vs. $\tf \Ra$ for $\Ra = 500$ (circles), $\Ra = 750$ (crosses), and $\Ra = 1000$ (squares). The dashed line shows relationship (\ref{eqko}) (\emph{d}) $\ti \Ra$  vs. $\tf \Ra$ for $\Ra = 500$ (circles), $\Ra = 750$ (crosses), and $\Ra = 1000$ (squares). The dashed line shows relationship (\ref{eqtio}).   } 
    \label{figure6}
\end{center}
\end{figure}

To explore the optimal initial perturbation time, we repeat the optimization procedure for a wide range of wavenumbers, initial times, and final times. Figure \ref{figure6}(\emph{a}) illustrates the optimal amplifications $\Phic$ for $\tf=1$, $\Ra=500$, $10 \le \kw \le 50$, and $0.01 \le \ti \le 0.5$. 
We define the maximum amplification, i.e. the peak of the $\Phic$ surface in figure \ref{figure6}(\emph{a}), as 
\begin{equation}
\Phic^\mathrm{o}(\tf) =  \sup_{\substack{0 \le \kw<\infty \\ 0 < \ti < \tf}} \{ \Phic(\tf, \kw, \ti) \},
\label{eq:ampo}
\end{equation}
and the optimal point $(\kw^\mathrm{o}, \ti^\mathrm{o})$ as the location in the $(\kw, \ti)$ plane where $\Phi = \Phic^\mathrm{o}$. 
To explore the dependence of the optimal point on $\tf$ and $\Ra$, we compute $(\Phic^\mathrm{o}$, $\kw^\mathrm{o}$, $\ti^\mathrm{o})$ for  $500 \le \Ra \le 1000$ and $1 \le \tf \le 8$. Figure \ref{figure6} demonstrates that the results collapse to three curves by plotting $\Phic^\mathrm{o}$ (panel \emph{b}), $\kw^\mathrm{o}/ \Ra$ (panel \emph{c}), and $\ti^\mathrm{o} \Ra$ (panel \emph{d}) as functions of $\tf \Ra$. This collapse occurs because the optimal perturbations are concentrated near $z=0$ and do not interact with the lower boundary at $z=1$. Consequently, the Rayleigh number dependence may be scaled out of the governing equations (\ref{eq:ge})--(\ref{eq:ge-bc}) by approximating the vertical depth as infinite, $H \rightarrow \infty$, and nondimensionalizing the problem with respect to the characteristic length $L=\phi D/ U$, and time, $T =\phi L / U$. From figure \ref{figure6}, we obtain the following relationships, 
 \begin{equation}
\log\Phic^\mathrm{o} =   - 4.458\! \times \! 10^{-8} (\tf \Ra)^2 + 0.001721 \tf \Ra - 0.05739, 
 \label{eqphio}
 \end{equation}
 \begin{equation}
\kw^\mathrm{o} = \Ra \big[  0.1152 -0.02023 \, \log (\tf Ra)  \big], 
 \label{eqko}
 \end{equation}
  \begin{equation}
 \ti^\mathrm{o} = 6.364\! \times \! 10^{-4} \,  \tf + 58.00/  \Ra,
 \label{eqtio}
 \end{equation}
For convenience, we also present relations (\ref{eqko}) and (\ref{eqtio}) in dimensional form, 
 \begin{equation}
\kw^* = \frac{U}{\phi D} \Big[  0.1152 -0.02023 \, \log \left(\frac{\tf^* U^2}{\phi^2 D} \right)  \Big],
\label{eqkod}
\end{equation}
  \begin{equation}
  \ti^* = 6.364\! \times \! 10^{-4} \,  \tf^* + 58.00 \frac{\phi^2 D}{ U^2},
  \label{eqtiod}
  \end{equation}
where $\kw^*$, $\ti^*$, and $\tf^*$ are the optimal wavenumber, initial time, and final time in dimensional form. Recall from \S2 that  $U= K \Delta \rho \, g / \mu$. These relations demonstrate that the optimal wavenumber and initial time are independent of the aquifer depth $H$. Note that when $\tf \Ra< 1500$, relations (\ref{eqphio}) and (\ref{eqko}) continue to provide accurate estimates, while relation (\ref{eqtio}) deviates significantly.

\cite{Ennis-King2005} report the following  typical parameter values for CO$_2$ sequestration: $\mu = 5 \! \times \! 10^{-4}$ Pa s, $\phi = 0.2$, $\Delta \rho = 10$ kg m$^{-3}$, $g = 9.81$ m s$^{-2}$, $D=10^{-9}$ m$^2$ s$^{-1}$, and $ 10^{-14} \le K \le 10^{-12}$ m$^2$. Using these values, figure \ref{figure6} predicts that  the optimal wavelength and initial time  for high permeability aquifers,  $K=10^{-12}$ m$^2$, vary in the range, $11 \, \mathrm{cm} \le 2 \pi / \kw^* \le  18 \, \mathrm{cm}$ and $17 \, \mathrm{hours} \le \ti^*\le 18 \, \mathrm{hours}$ as the final time varies between, $6 \, \mathrm{days} \le \tf^*\le 96 \, \mathrm{days}$. For low permeability aquifers, $K=10^{-14}$ m$^2$,  these parameters vary between $11 \, \mathrm{m} \le 2 \pi / \kw^* \le 18 \, \mathrm{m}$, $19 \, \mathrm{years} \le \ti^* \le 21 \, \mathrm{years}$, $165 \, \mathrm{years} \le \tf^*\le 2636 \, \mathrm{years}$. While these initial and final times for $K=10^{-14}$ m$^2$ aquifers may appear late, the optimal initial times are consistent with previous estimates of the critical time reported by \cite{Ennis-King2005} and \cite{Riaz2006JFM}. Furthermore, in \S6.2, we confirm that the range of final times are representative of actual onset times for nonlinear convection, $\ton$.   
 
The dependence of the optimal point $(\kw^\mathrm{o}, \ti^\mathrm{o})$ on $\tf$ indicates that the optimal initial perturbation depends on the initial perturbation amplitude and
consequently cannot be determined through purely linear analysis. Consider, for example, that direct numerical simulations show that the onset time for convection, $\ton$, decreases
with increasing initial perturbation amplitude \cite[]{Selim2007b, Rapaka2008}.  Consequently, figures \ref{figure6}(\emph{c}) and \ref{figure6}(\emph{d}) predict that large amplitude perturbations will have larger values of  $\kw^\mathrm{o}$ and smaller values of $\ti^\mathrm{o}$ than small amplitude perturbations. Their exact values, however, would require \emph{a priori} numerical or experimental results for the onset time $\ton$. 
 
\subsection{Influence of final time on initial perturbation profiles}
\begin{figure}
 \begin{center}
  \hspace{0.0cm}
    (\emph{a})
    \hspace{6.5cm}
    (\emph{b})
    \\
    \includegraphics[trim = 0 0 0 10, clip=true,width=7cm]{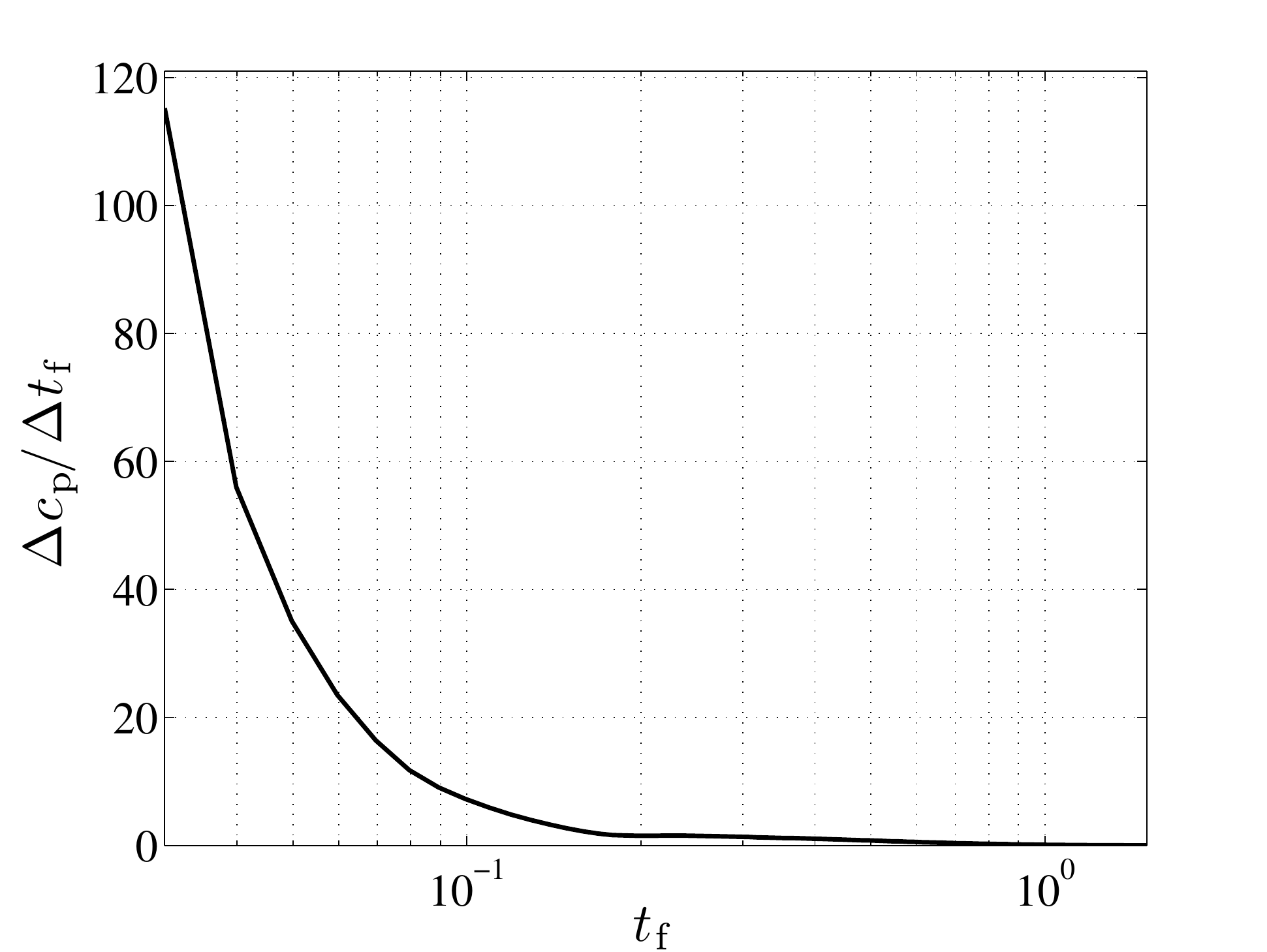}
    \includegraphics[trim = 0 0 0 10, clip=true,width=7cm]{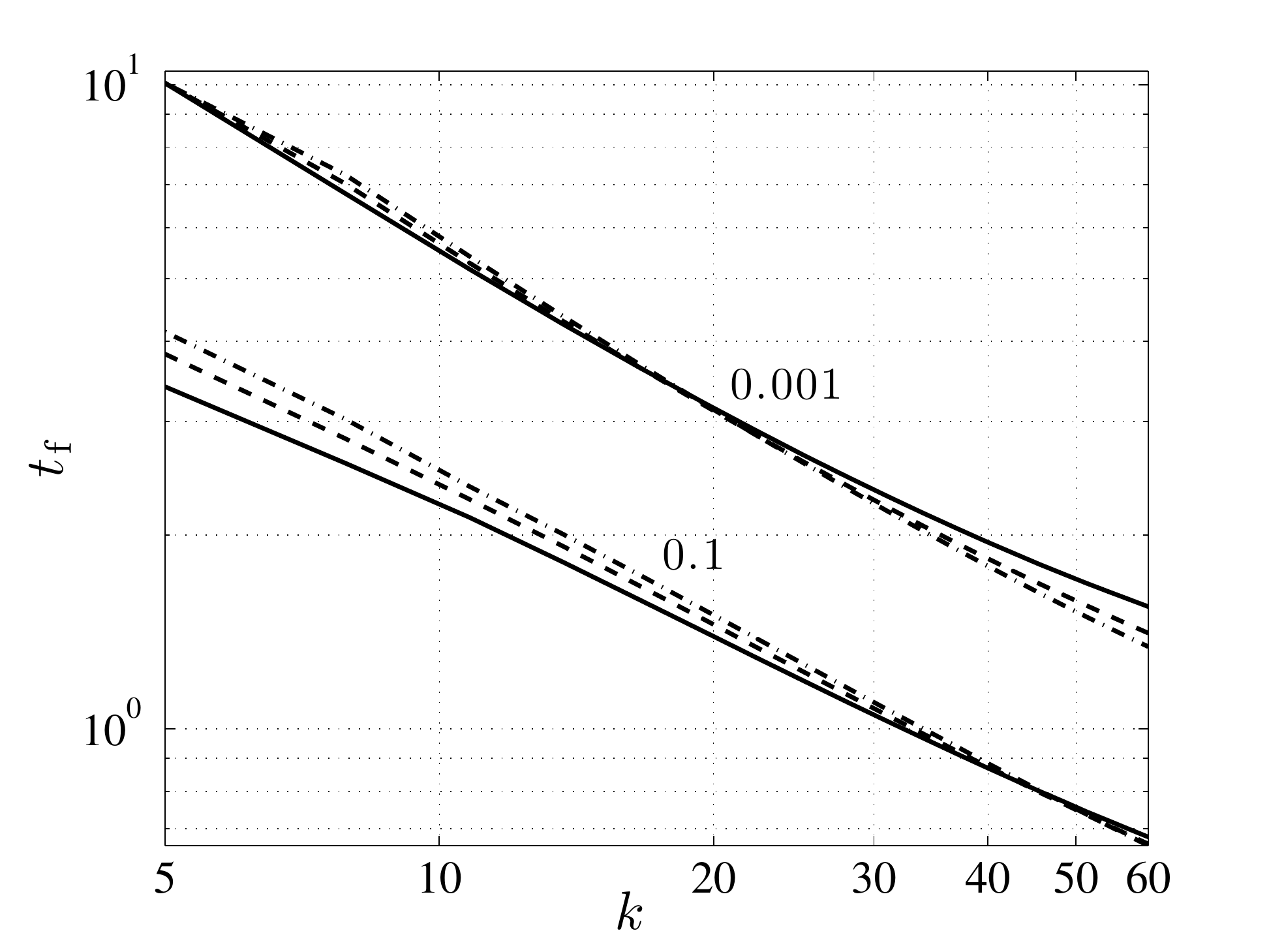}
        \caption{ Convergence of the optimal $\cip$ profiles for $\ti=0.01$. (\emph{a}) $\Delta \cip / \Delta \tf$ vs. $\tf$ for $\kw=30$ and $\Ra=500$ (\emph{b}) Isocontours of ${ \Delta \cip }/{\Delta \tf }$ in the $(\kw, \tf)$ plane for $Ra=500$ (solid line), $Ra=750$ (dashed line), and $Ra=1000$ (dash-dotted line).  } 
    \label{figure7}
\end{center}
\end{figure}
We observe that beyond a certain final time, the initial profiles, $\cip$ and $\wip$, are unaffected by further increases to $\tf$. 
To quantify the final time beyond which $\cip$ and $\wip$ do not depend on $\tf$, we measure the rate of change of $\cip$ with respect to $\tf$ as, 
\begin{equation}
\frac{ \Delta \cip }{\Delta \tf } =   \frac{  \| \cip (z;{\tf+\Delta t_\mathrm{f}})  -   \cip (z;{\tf} ) \|_\infty }{\Delta t_\mathrm{f}},
\label{eq:convergence}
\end{equation}
where ${\Delta t_\mathrm{f}}=0.01$ and the $\cip$ profiles are normalized with respect to their $L^2$ norms.
Figure \ref{figure7}(\emph{a}) illustrates ${ \Delta \cip }/{\Delta \tf }$ versus $\tf$ for $\ti=0.01$, $\kw=30$ and $\Ra=500$.  We observe that ${ \Delta \cip }/{\Delta \tf }$ is initially large but decreases rapidly to zero. Figure \ref{figure7}(\emph{b}) illustrates isocontours of ${ \Delta \cip }/{\Delta \tf }=0.1$ and 0.001 in the $(\kw, \tf)$ parameter plane for $\ti=0.01$ and $\Ra=500$ (solid line), $\Ra=750$ (dashed line), and $\Ra=1000$ (dash-dotted line). With increasing $\kw$, the final time after which $\cip$ and $\wip$ do not change decreases. We observe only a small influence of the Rayleigh number on the ${ \Delta \cip }/{\Delta \tf }$ isocontours.

\begin{figure}  
  \begin{center}  
    \hspace{0.0cm}
    (\emph{a})
    \hspace{6.5cm}
    (\emph{b})
    \\
    \includegraphics[trim = 0 0 0 10, clip=true,width=7cm]{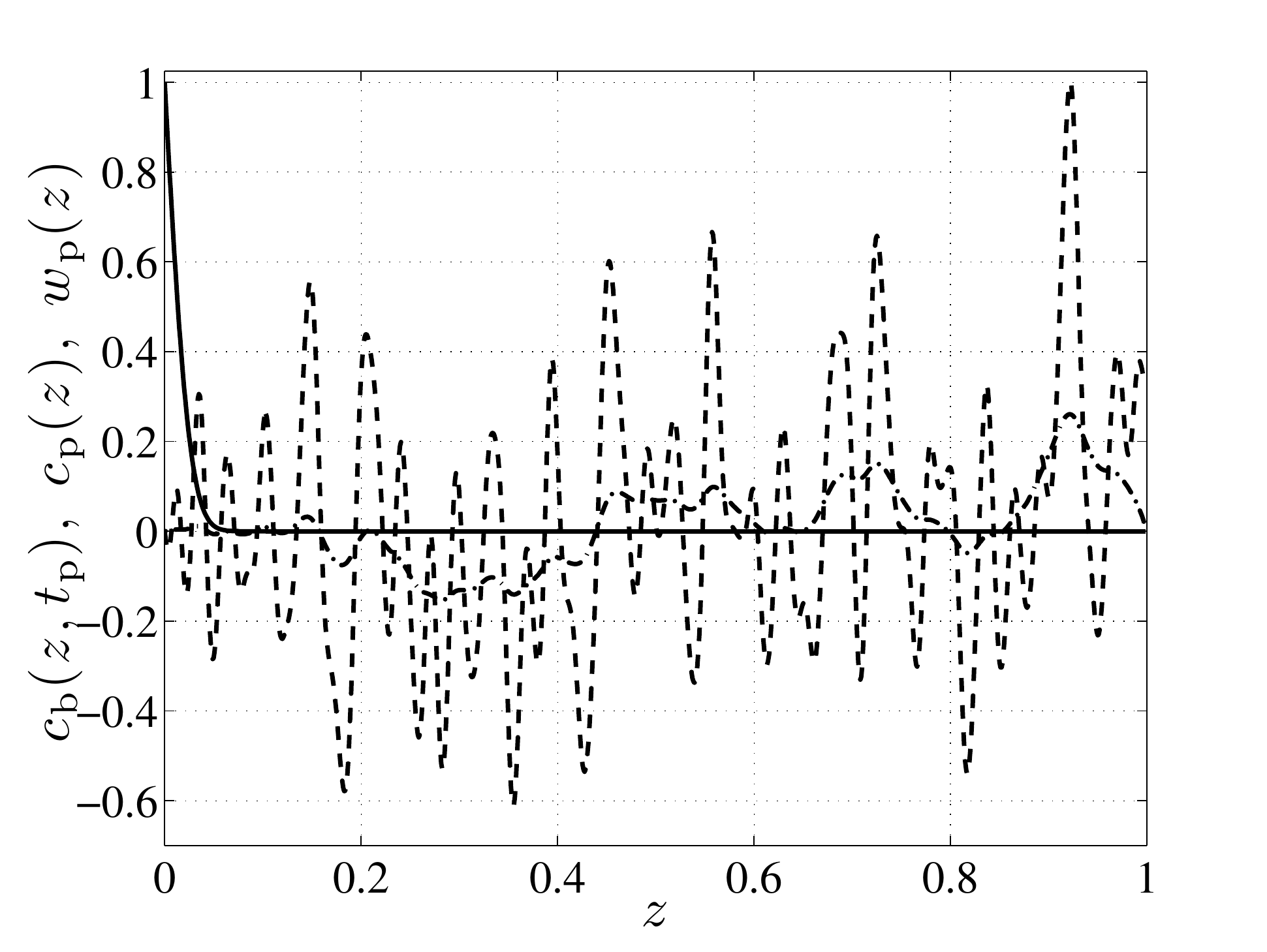}
    \includegraphics[trim = 0 0 0 10, clip=true,width=7cm]{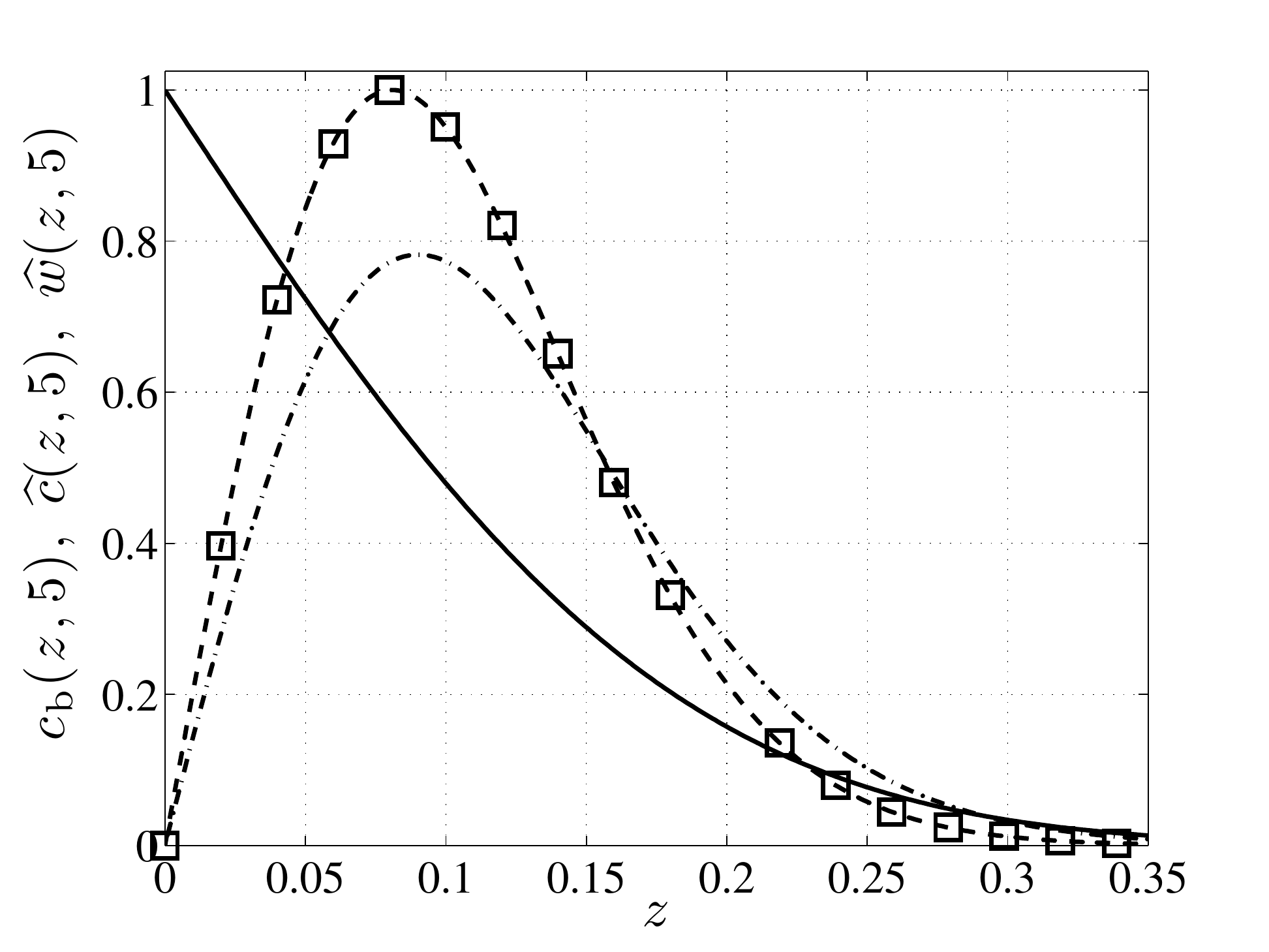}
    \caption{IVP results using random initial profiles, $\cip$ and $\wip$, for $\ti=0.01$, $\kw = 30$, and $\Ra = 500$. (\emph{a}) The base-state (solid line) and random initial profiles $\cip$ (dashed line) and $\wip$ (dash-dotted line) at $\ti=0.01$. (\emph{b}) Resulting perturbation profiles $\ch$ (dashed line) and $\wh$ (dash-dotted line) at $t=5$ The squares show the corresponding optimal perturbation $\ch$ when $\tf=5$. }
   \label{figure8}
 \end{center}
\end{figure}
The convergence of $\cip$ and $\wip$ beyond a certain $\tf$ may be explained by noting that the forward IVP (\ref{eq:linear})--(\ref{eq:ic}) always converges to the same dominant perturbations given sufficient time.  To demonstrate this behavior, figure \ref{figure8}(\emph{a}) illustrates random initial conditions for $\cip$ (dashed line) and $\wip$ (dash-dotted line) that span the entire vertical domain, $0 \le z \le 1$, at $\ti=0.01$ for $\kw=30$. Figure \ref{figure8}(\emph{b}) illustrates the resulting perturbation profiles, $\ch(z,5)$ and $\wh(z,5)$, generated by integrating the forward IVP to $t=5$. The final state of the forward IVP is identical to the corresponding optimal perturbation, shown using squares in figure \ref{figure8}(\emph{b}).  

\subsection{Comparison with QSSA modal analysis}

The convergence of the forward IVP and optimization procedure to identical dominant perturbations at late times may be explained by considering a quasi-steady modal analysis. For a prescribed final time $\tf$, this approach approximates the base-state, $\cb(z,t)$, as steady and decomposes perturbations into separable functions of $z$ and $t$,
\begin{equation}
\ch = c_\mathrm{e}(z;\tf) \mathrm{e}^{\sigma(\tf) t}, \qquad \wh = w_\mathrm{e}(z;\tf) \mathrm{e}^{\sigma(\tf) t},
\label{eq:shape}
\end{equation} 
where $\sigma(\tf)$ is the instantaneous growth rate at $t=\tf$. Substituting (\ref{eq:shape}) into (\ref{eq:linear})--(\ref{eq:bcs}) produces an eigenvalue problem for eigenvalues $\sigma$ and eigenfunctions $c_\mathrm{e}$ and $w_\mathrm{e}$.
We compare the optimal perturbations with the dominant QSSA modes by measuring,
\begin{equation}
\Delta \ch =  \int_0^1 \bigg|  \frac{ c_\mathrm{e}(z;\tf) }{ \| c_\mathrm{e}(z;\tf) \|_\infty}  - \frac{ \ch(z,\tf) }{ \| \ch(z,\tf) \|_\infty}    \bigg|  \mathrm{d}z.
\label{qssa}
\end{equation}
When $\Delta \ch=0$, the dominant QSSA mode and optimal  perturbation are identical. 

\begin{figure}
 \begin{center}
  \hspace{0.0cm}
    (\emph{a})
    \hspace{6.5cm}
    (\emph{b})
    \\
    \includegraphics[trim = 0 0 0 10, clip=true,width=7cm]{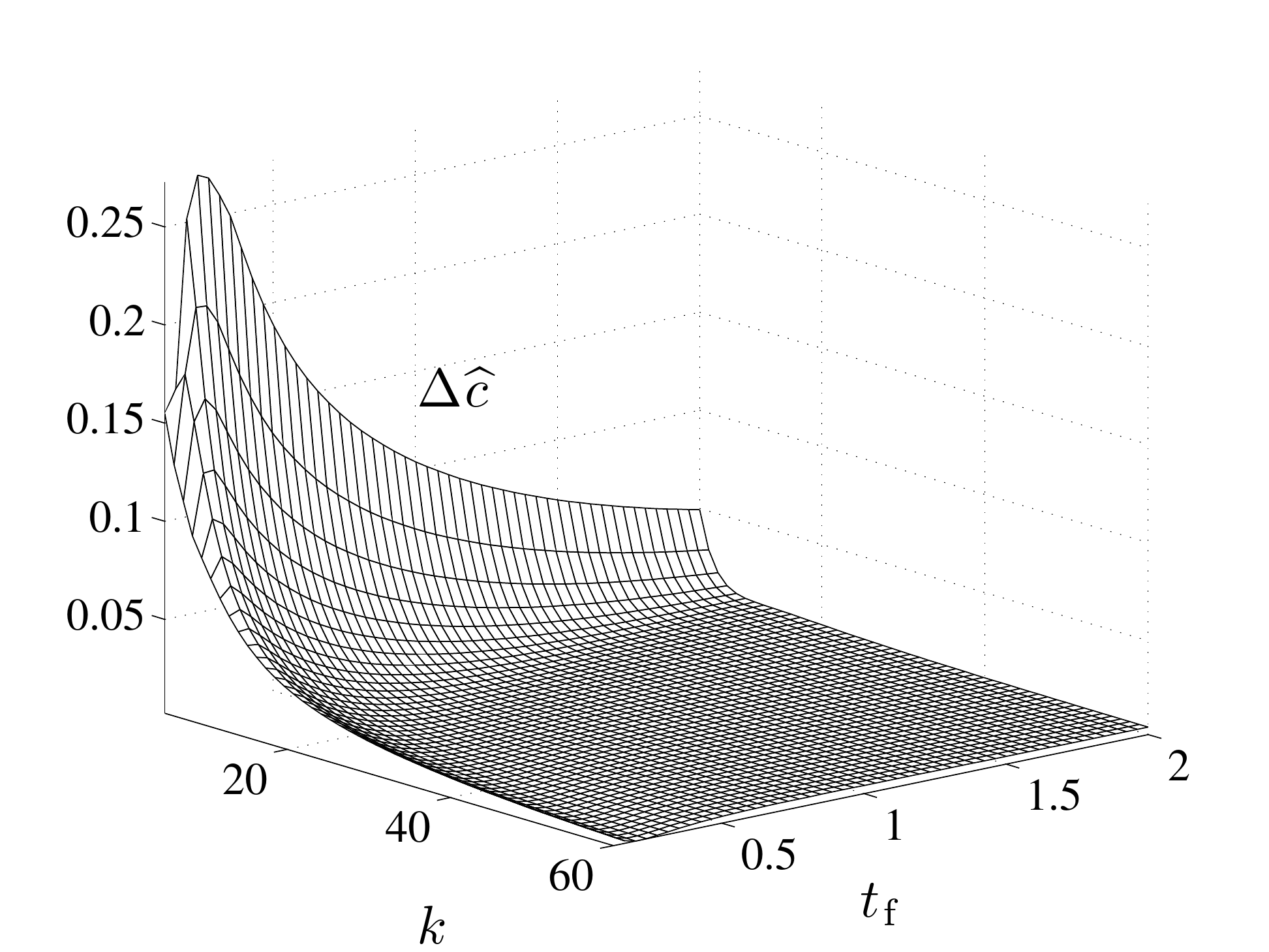}
     \includegraphics[trim = 0 0 0 10, clip=true,width=7cm]{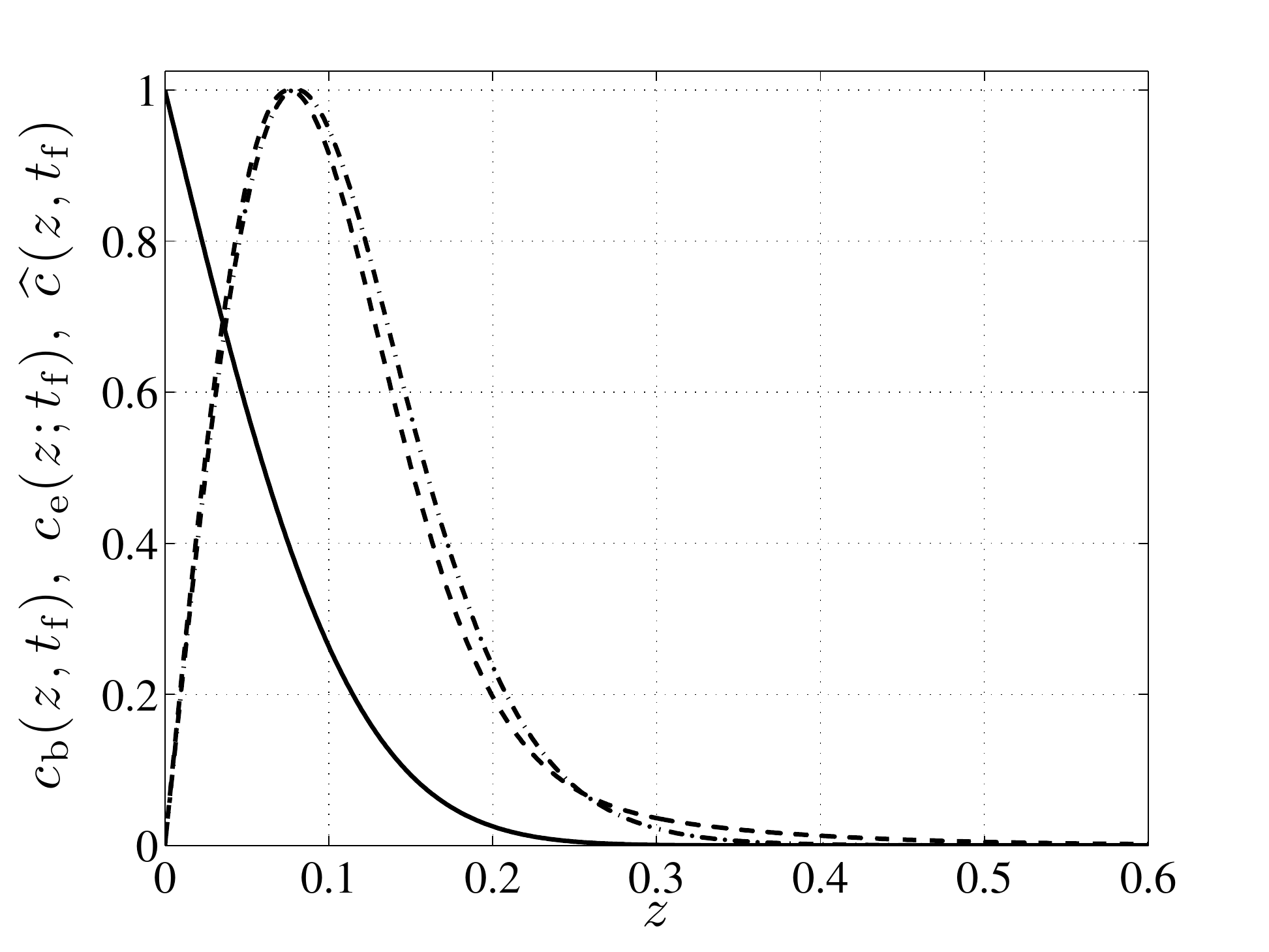}
     \\
     \hspace{0.0cm}
     (\emph{c})
     \hspace{6.5cm}
     (\emph{d})
     \\
   \includegraphics[trim = 0 0 0 10, clip=true,width=7cm]{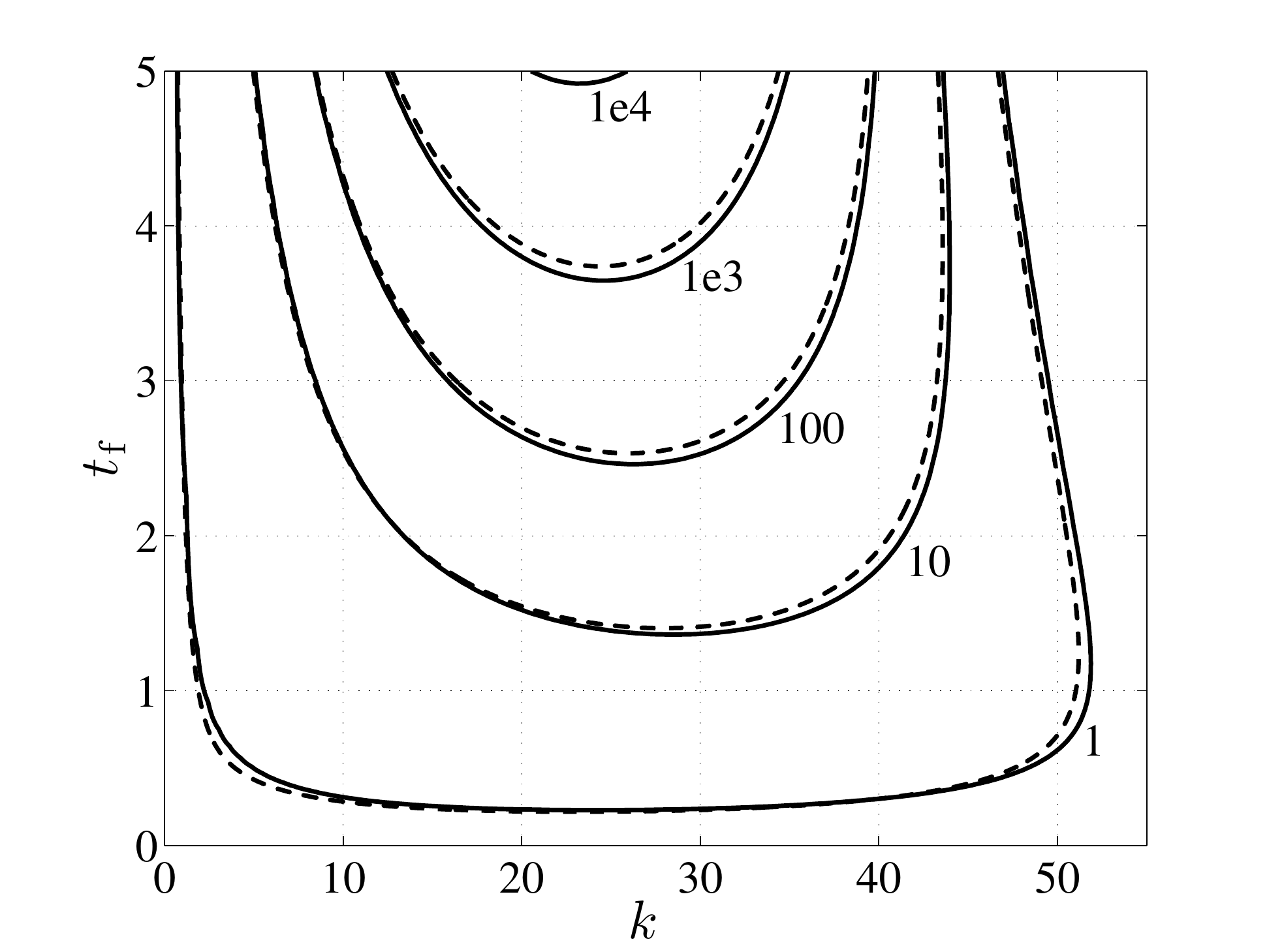}
    \includegraphics[trim = 0 0 0 10, clip=true,width=7cm]{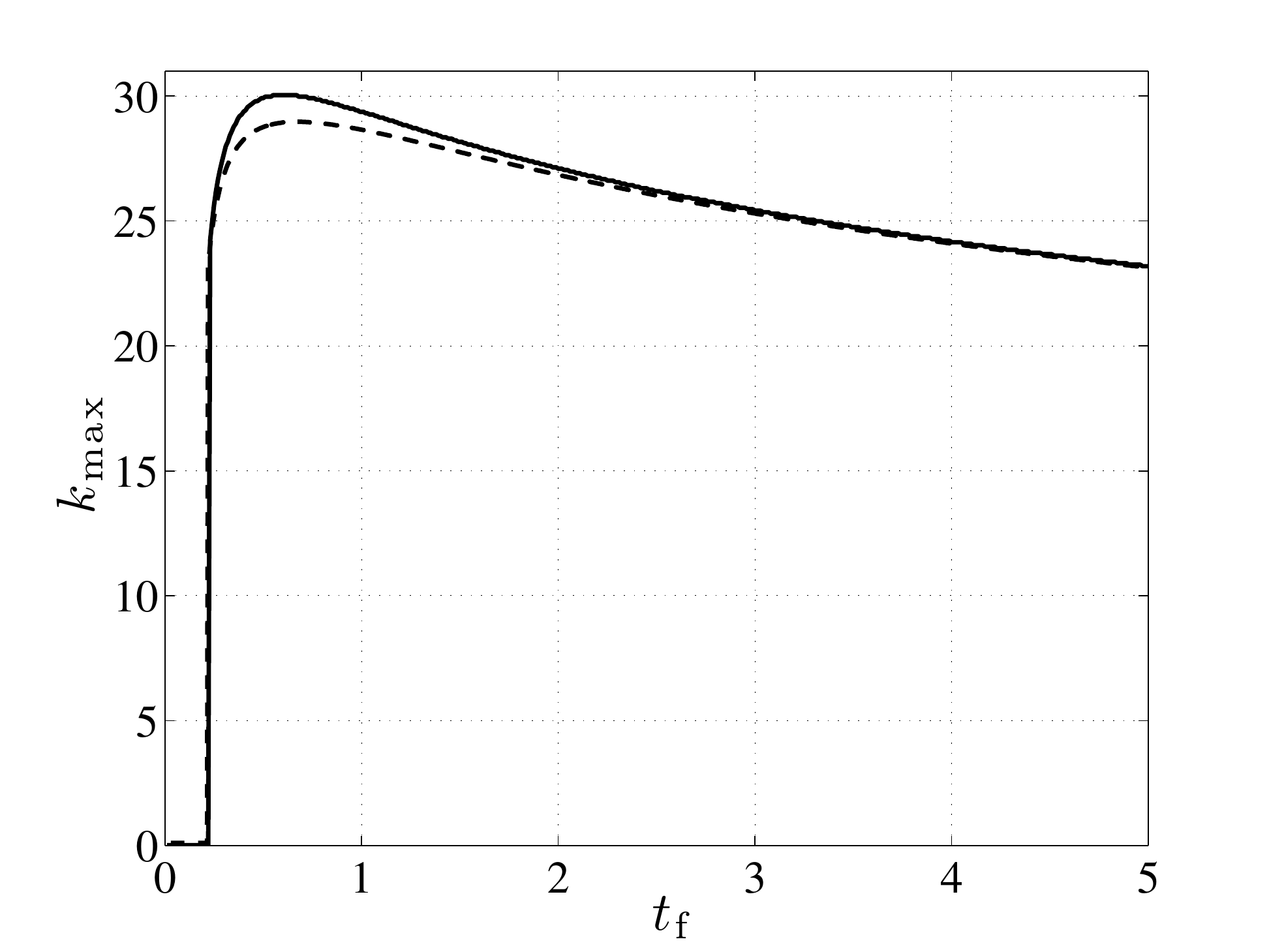}
        \caption{ Comparison of optimal perturbations with the least stable QSSA mode for $\Ra = 500$. (\emph{a}) $\Delta \ch$ in the $(k, \tf)$ plane for $\ti=0.1$. (\emph{b}) illustrates base-state, $\cb(z,\tf)$ (solid line), optimal perturbation $\ch(z,\tf)$ (dashed line) and least stable eigenmode, $c_\mathrm{e}(z;\tf) $ (dash-dotted line) at $\tf$ for $\ti=0.1$, $\kw=10$, and $\tf=2$. (\emph{c}) Isocontours of amplification in the ($\kw$, $\tf$) plane when $\ti=0.01$ for QSSA (solid lines)  and optimization (dashed lines). (\emph{d}) The dominant wavenumbers $\kw_\mathrm{max}$ vs. $\tf$  when $\ti=0.01$ for QSSA (solid lines)  and optimization (dashed lines) } 
    \label{figure9}
  \end{center}
\end{figure}

Figure \ref{figure9} compares optimal perturbations with dominant QSSA modes for $\ti=0.1$ and $\Ra=500$. Note that $\ti$ is chosen to be close to the optimal initial time $\ti^\mathrm{o}$. Figure \ref{figure9}(\emph{a}) illustrates the variation of $\Delta \ch$ for wavenumbers $5 \le k \le 60$ and final times $0.12 \le \tf \le 2$. 
We observe large values of $\Delta \ch$ at small wavenumbers and final times.  In the limit of $\kw \rightarrow 0$, however, $\Delta \ch$ tends to zero because the optimal perturbation and dominant QSSA mode both tend to $\ch = \sin{(\pi z /2)}\exp{(-\pi^2 \Ra^{-1} t/4)}$, see discussion of equation (\ref{eq:diff}) in \S4.2.  With increasing wavenumber and final time, $\Delta \ch$ becomes small, indicating that the optimal perturbations essentially recover the dominant QSSA modes.  This behavior is  confirmed in figure \ref{figure9}(\emph{b}) which illustrates the base-state (solid line), optimal perturbation (dashed line), and dominant QSSA eigenmode (dash-dotted line) at $\tf=2$ and $\kw=10$. Note that $\Delta \ch$ remains small for the optimal perturbations with wavenumber, $\kw_\mathrm{max}$, illustrated in figure \ref{figure5}(\emph{a}). 

The amplification produced by temporal integration of the dominant QSSA growth rate, $\sigma$, can be computed through the relation
\begin{equation}
\label{eq:sigma}
\Phiq(t)=\mathrm{e}^{g(t)}, \qquad g(t)= \int_{\ti}^{t} \sigma(\tf) \, \mathrm{d}\tf.
\end{equation}
Figure \ref{figure9}(\emph{c}) compares isocontours of $\Phiq$ (solid line) with optimal results for $\Phic$ (dashed line) in the $(\kw, \tf)$ parameter plane for $\Ra=500$ and $\ti=0.01$. We observe excellent agreement between the amplifications produced by optimal perturbations and dominant QSSA eigenmodes. Counterintuitively, for much of the $(\kw, \tf)$ plane, $\Phiq$ is marginally greater than $\Phic$. This occurs for the following reasons. At late times, the boundary layer grows slowly and the optimal perturbations tend to the dominant QSSA eigenmodes. At small times, however, the boundary layer varies rapidly and the optimal perturbations cannot continuously adhere to the quasi-steady eigenmodes.  Consequently, temporal integration of the dominant QSSA growth rates produces marginally larger amplifications than $\Phic$. This also helps explain why dominant perturbations tend to differ from the dominant eigenmode at small times. 

Figure \ref{figure9}(\emph{d}) illustrates the dominant wavenumbers, $\kw_\mathrm{max}$, that maximize $\Phiq$ (solid line) and $\Phic$ (dashed line) for $0.03 \le \tf \le 5$. We repeat the optimization procedure for different $\ti$ and observe similar agreement between the QSSA and optimization results.  This suggests that optimal perturbations are primarily composed of the dominant QSSA mode. In contrast, nonmodal stability analyses of steady wall-bounded shear flows, such as channel flows and flat plate boundary layers, typically produce optimal perturbations that are qualitatively very different from the corresponding dominant eigenmodes. 
This suggests that for the current study, the deviation of the optimal perturbations from the dominant eigenmodes at small times is primarily due to the transient base-state, rather than the nonorthogonality of the quasi-steady eigenmodes.

\section{Modified Optimization Procedure}

\begin{table}
\begin{minipage}[b]{0.3\linewidth} \centering 
\begin{tabular}{c c} 
 \multicolumn{2}{c}{$\ti=0.01$} \\
\hline\hline 
$\M$ & $  c_\mathrm{net}^\mathrm{min}$  \\ 	
\hline 
$10^{-02}$ & $-10^{-02}$   \\
$10^{-05}$ & $-10^{-05}$ \\
$10^{-10}$ & $-10^{-10}$ \\ 
\hline 
\end{tabular}
\end{minipage}
\hspace{0.5cm}
\begin{minipage}[b]{0.3\linewidth} \centering 
\begin{tabular}{c c} 
  \multicolumn{2}{c}{$\ti=0.1$} \\
\hline\hline 
$\M$ & $  c_\mathrm{net}^\mathrm{min}$  \\ 	
\hline 
$10^{-02}$ & $-8.0 \times 10^{-03}$   \\
$10^{-05}$ & $-4.9 \times 10^{-06}$ \\
$10^{-10}$ & $-2.5 \times 10^{-11}$ \\ 
\hline 
\end{tabular}
\end{minipage}
\hspace{0.5cm}
\begin{minipage}[b]{0.3\linewidth} \centering 
\begin{tabular}{c c} 
\multicolumn{2}{c}{$\ti=1$} \\
\hline\hline 
$\M$ & $  c_\mathrm{net}^\mathrm{min}$  \\ 
\hline 
$10^{-02}$ & $-2.1 \times 10^{-03}$   \\
$10^{-05}$ & $-3.3 \times 10^{-08}$ \\
$10^{-10}$ & $-5.1 \times 10^{-15}$ \\ 
\hline 
\end{tabular}
\end{minipage}
\caption{ Minimum net concentrations $ c_\mathrm{net}^\mathrm{min}$ produced by the classical optimal $\cip$ profiles when $\kw=30$, $\Ra=500$, $\tf=5$, $\ti=0.01$, 0.1, 1, and $\M= 10^{-2},$ $10^{-5},$ and $10^{-10}$. At $\tp=0.01$, the negative concentration is of the same order as $\M$. As $\ti$ increases, the perturbation profiles become increasingly concentrated within the boundary layer and consequently the magnitude of the negative concentrations $ c_\mathrm{net}^\mathrm{min}$ diminish. } 
\label{table:lin} 
\end{table}

Experimental studies observe that perturbations are initially localized within the boundary layer  \cite[]{Spangenberg1961, Elder1968, Blair1969, Green1975,Wooding1997}.
 To determine whether the optimal perturbations obtained in \S4 reflect those observed experimentally, we consider the following argument. If the optimal perturbation is observed experimentally, the net concentration can be expressed as the sum of the base-state and perturbation through the relation  
\begin{equation}
 c_\mathrm{net}(x,z,\ti) =  c_\mathrm{b}(z,\ti) + \M \cos( \kw x) \frac{\cip(z) }{|| \cip ||_\infty} ,
\label{netconc}
\end{equation}
where $\cip$ is the optimal initial profile and $\M$ is the perturbation amplitude measured using the $L^\infty$ norm.
Table \ref{table:lin} lists the minimum net concentrations, $ c_\mathrm{net}^\mathrm{min}$,  for various $\M$ and $\ti$ when $\tf=5$, $\kw=30$, and $\Ra=500$. For $\ti=0.01$, we observe unphysical negative net concentrations equal to $\M$. This occurs because the maxima of the optimal $\cip$ profiles are located outside the boundary layer, see figure \ref{figure5}(\emph{b}). For $\ti=0.1$ and 1, the magnitude of the negative concentrations become increasingly smaller because the optimal $\cip$ profiles  become increasingly concentrated within the boundary layer, see figures \ref{figure5}(\emph{c})--\ref{figure5}(\emph{d}).  

Direct numerical simulations show that the onset time for convection decreases with increasing initial perturbation amplitude $\M$\cite[]{Rapaka2008}. Consequently, though the classical optimal perturbations are mathematically valid optimal solutions, onset of convection in physical systems may more likely be triggered by suboptimal perturbations concentrated within boundary layer. Those perturbations support finite initial amplitudes, and consequently require less time to grow sufficiently for onset of convection.
To investigate this alternate path to onset of convection, we propose a modified optimization procedure that constrains the initial concentration fields  of the perturbations to be within the boundary layer.

\subsection{Methodology}

The classical optimization procedure described in \S 3 is modified by replacing the constraint $E(\ti)=1$ with the modified constraint $E_\Psi (\ti) = 1$, where
\begin{equation}
E_\Psi(\ti) = \int_0^1  { \Psi(z) \, \ch(z,\ti)^2}  \mathrm{d} z,
\label{eq:epsi}
\end{equation}
where $\Psi(z)$ is a filter function that tends to  infinity, $\Psi \rightarrow \infty$, outside the boundary layer. We then maximize $\Phi_\Psi=\sqrt{E(\tf)/\Ep}$.The filter function assures that $\Ep = \infty$, when $\cip$ extends beyond the boundary layer. This forces $\Phi_\Psi$ to zero and effectively filters such perturbations from the optimization procedure. In practice, the infinite values of $\Psi$ are approximated numerically using a large finite value.

Following an analogous procedure to that in \S 3, we formulate the Lagrangian,
\begin{eqnarray}
\nonumber \mathcal{L}(\ch,\cs ,\wh,\ws,\mathrm{s}) &=& E(\tf) -  \mathrm{s} \big[ \Ep - 1 \big]  - \int_{\ti}^{\tf} \int_0^1 \ws \left( \D \wh + k^2 \ch \right)  \,\mathrm  {d}z \, \mathrm{d}t  \\
&& -    \int_{\ti}^{\tf} \int_0^1 \cs \left( \frac{\partial \ch}{\partial t}   -  \frac{1}{Ra}  \D \ch + \wh \frac{\partial \cb}{\partial z}  \right) \,\mathrm{d}z \,\mathrm{d}t,
\label{eq:L_w}
\end{eqnarray}
and obtain the following coupling conditions between physical and adjoint variables,
\begin{equation}
2 \mathrm{s} \ch \, \big|_{\ti}  = \Psi^{-1}\cs \big|_{\ti} ,   \qquad  2 \ch \, \big|_{\tf}   = \cs \big|_{\tf} .
\label{eq:rel_weight}
\end{equation}
The adjoint IVP (\ref{eq:adj1})--(\ref{eq:bc_adj}) remains unchanged. After convergence of the iterative procedure for the optimal profile that maximizes $\Phi_\Psi$, we compute the final amplification using the traditional definition of  $\Phic = \sqrt{ {E(\tf)}/{E(\ti)} }$. This allows us to compare results of the modified optimization procedure with those of the classical procedure. 

\subsection{Filter Functions}

\begin{figure}   
   \begin{center} 
     \hspace{0.0cm}
     (\emph{a})
     \hspace{6.5cm}
     (\emph{b})
     \\
     \includegraphics[trim = 0 0 0 10, clip=true,width=7cm]{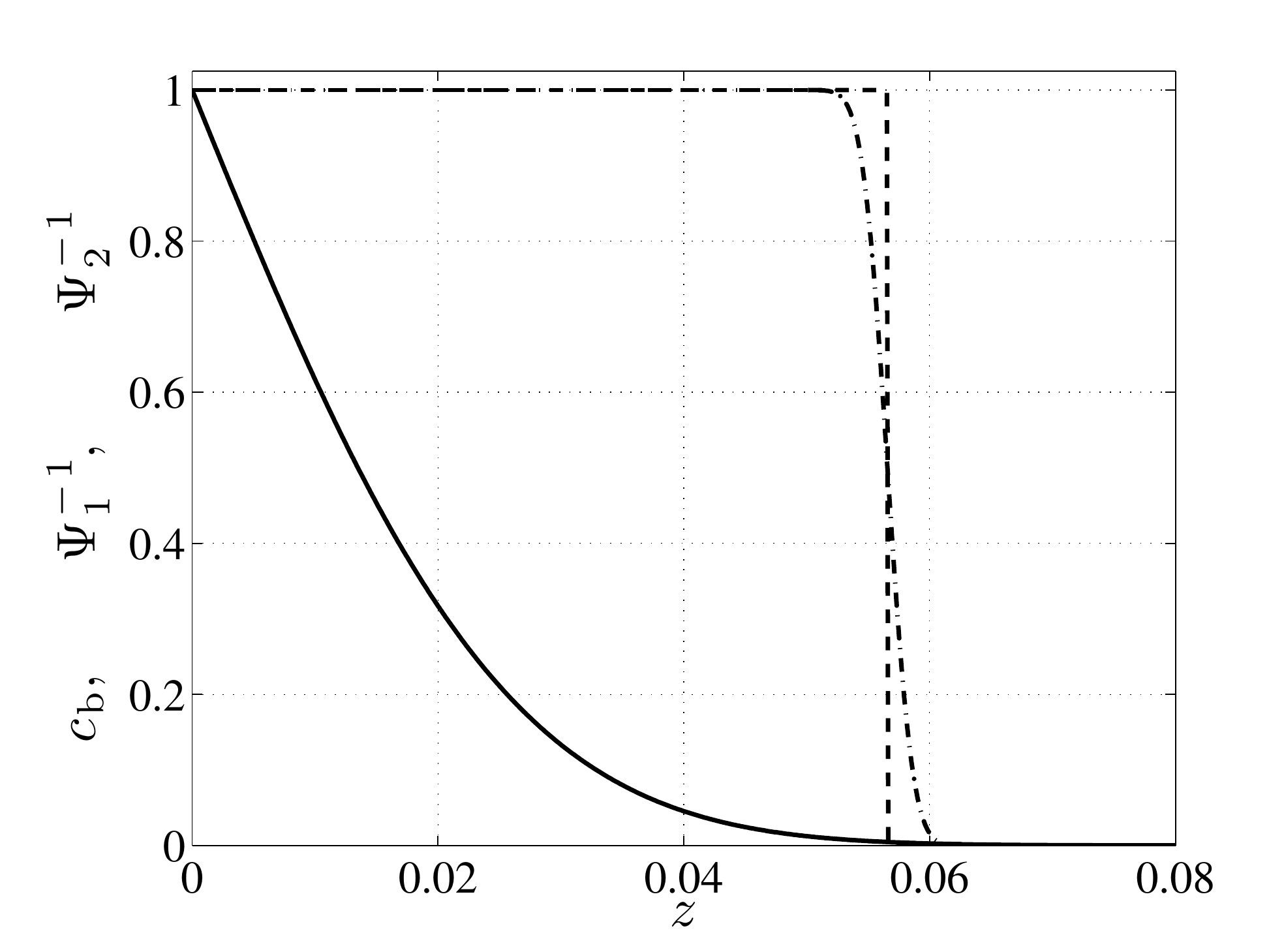}
     \includegraphics[trim = 0 0 0 10, clip=true,width=7cm]{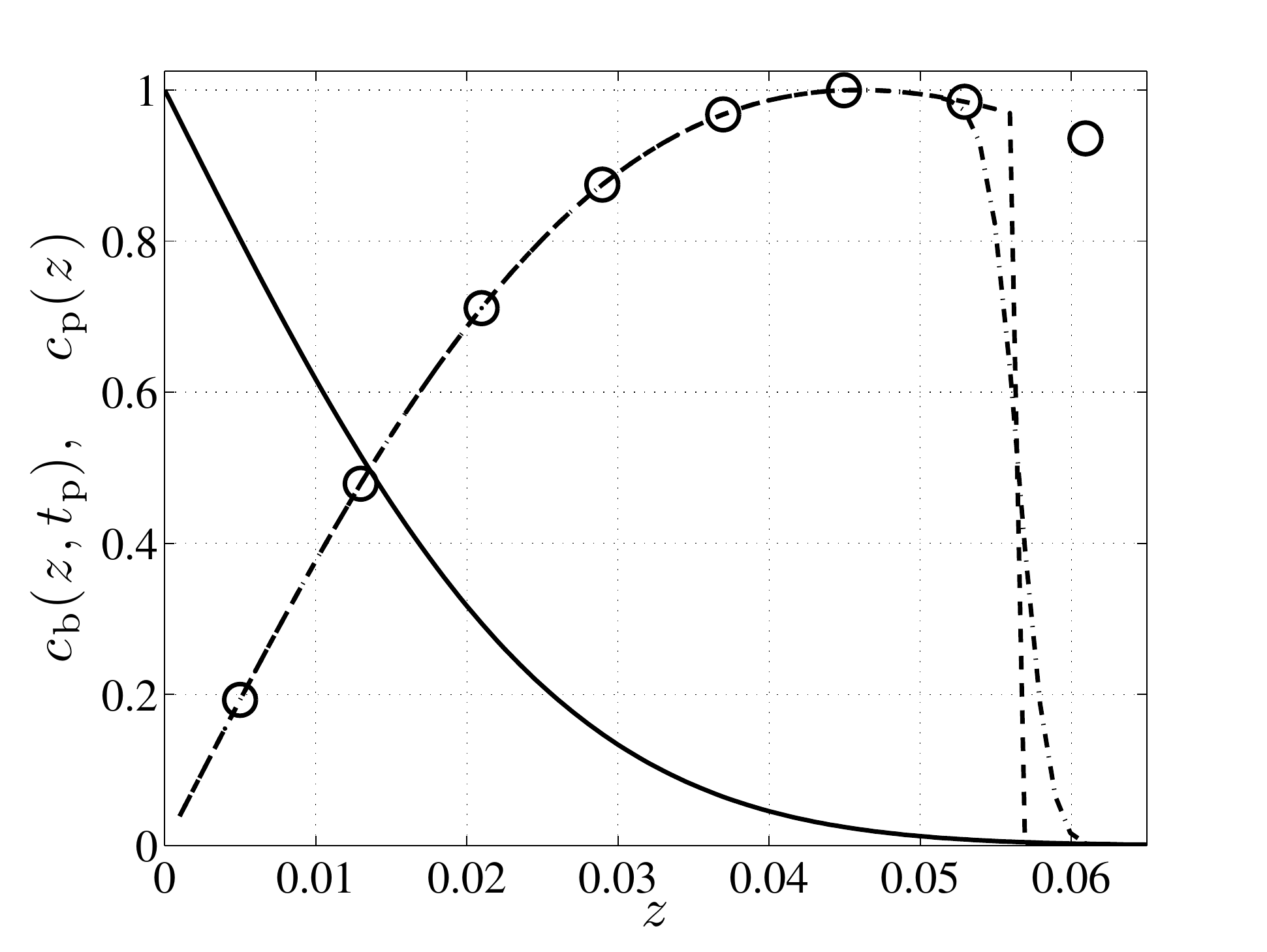}
    \caption{Optimization using $\Psi_1$ and $\Psi_2$ for $\kw = 30$, $\Ra = 500$, $\ti=0.1$, and $\tf=3$. (\emph{a}) Base-state (solid line), $\Psi^{-1}_1$ (dashed line), and $\Psi^{-1}_2$ (dash-dotted line). (\emph{b}) Base-state (solid line), classical $\cip$ (circles), and modified $\cip$ profiles using $\Psi_1$ (dashed line) and $\Psi_2$ (dash-dotted line). } 
    \label{figure10}
   \end{center}
\end{figure}

We first consider a filter function whose inverse is a step function of the form,
\begin{equation} 
\Psi^{-1}_1(z)  = \begin{cases}
1 & \text{if $z \le \delta$,}\\ 
0 & \text{if $\delta<z \le1 $,}
\end{cases} 
\end{equation}
where $\delta$ is the boundary layer depth defined as $\cb(\delta,\ti) =0.005$. 
Figure \ref{figure10}(\emph{a}) illustrates $\Psi^{-1}_1$ as a dashed line for $\ti=0.1$ and $\Ra=500$. The base-state is shown as a solid line. Figure \ref{figure10}(\emph{b}) illustrates the corresponding optimal $\cip$ profile (dashed line) for $\Ra=500$, $\kw=30$, $\ti=0.1$, and $\tf=3$.  The base-state is shown as a solid line and the classical optimal $\cip$ profile is shown using circles. Within the boundary layer, the modified profile follows the classical profile and then vanishes discontinuously at  $z=\delta$. 
Consequently, though concentrated within the boundary layer, the perturbations generated by $\Psi_1$  are unlikely to arise in nature. 

To produce continuously differentiable perturbations, we introduce the following
filter function that is equal to unity in most of the boundary layer,
but varies smoothly to zero beyond the boundary layer depth,
\begin{equation}
\Psi^{-1}_2(z) = \frac{1}{2}  \mathrm{erfc} \left(  \frac{25 \left( z-\delta \right) }{ \delta } \right).
\end{equation} 
Figure \ref{figure10}(\emph{a}) illustrates $\Psi^{-1}_2$ using a dash-dotted line.  Figure \ref{figure10}(\emph{b}) illustrates that the corresponding optimal modified $\cip$ profile  (dash-dotted line) decreases rapidly, but smoothly, to zero outside the boundary layer, but is otherwise similar to that produced by $\Psi_1$. The modified profiles illustrated in figure \ref{figure10}(\emph{b}) produce  physical initial conditions, $c^\mathrm{min}_\mathrm{net}=0 $, when $\M < 10^{-3}$.

Though $\Psi_2$ produces physically realizable optimal perturbations,
the perturbations have maxima near the boundary layer depth,
$z=\delta$, where the base-state concentration is very small. This
limits the maximum allowable initial amplitude of these perturbations.
In contrast, perturbations with maxima near $z=0$ can support larger
initial amplitudes and may consequently trigger onset of convection
sooner. Furthermore, one may expect that perturbations would naturally
tend to have maxima near the upper boundary, $z = 0$, where the
base-state has a maximum and there is consequently more solute to
perturb. To investigate this possibility, we first note that the
inverse filter functions may be interpreted as weight functions.
Because $\Psi_1^{-1}$ and $\Psi_2^{-1}$ are equal to unity in most of
the boundary layer, they give equal weight to most of the boundary layer.
Optimal perturbations with maxima near $z=0$ can be obtained using an
inverse filter function that decreases with the base-state
concentration. A natural candidate is $\Psi_3^{-1}=\cb$ because this
naturally weighs regions of high base-state concentration over those
with low base-state concentration.

\begin{figure}  
   \begin{center}  
    \hspace{0.0cm}
    (\emph{a})
    \hspace{6.5cm}
    (\emph{b})
    \\
    \includegraphics[trim = 0 0 0 10, clip=true,width=7cm]{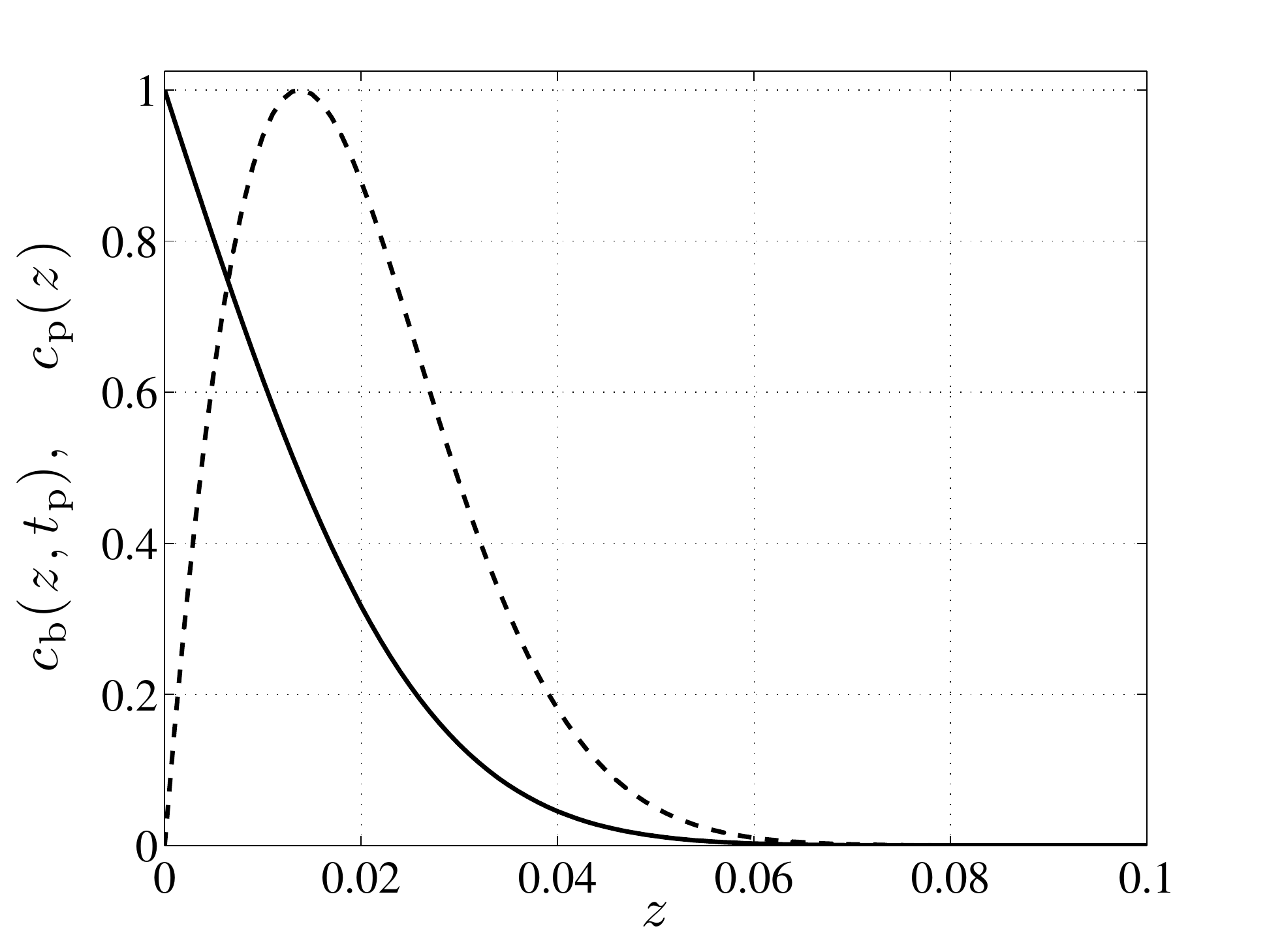}
    \includegraphics[trim = 0 0 0 10, clip=true,width=7cm]{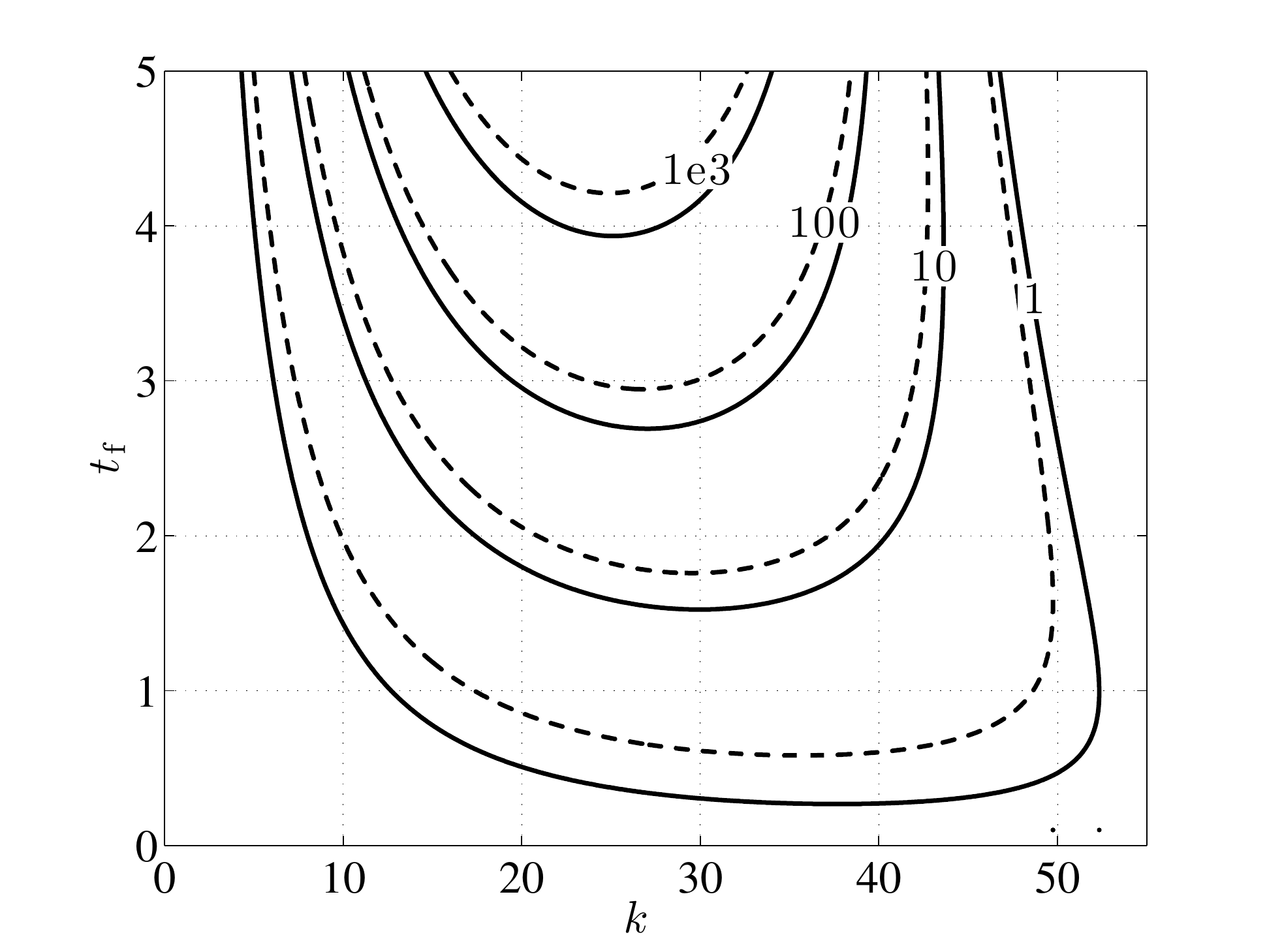}
    \caption{ Optimization results for $\ti=0.01$ and $\Ra=500$. (\emph{a}) The base-state $\cb(z,\ti)$ (solid line) and  optimal $\cip(z)$ profile using $\Psi_3$ (dashed line) for $\tf=5$ and $\kw=30$. Note that the initial $\cip(z)$ profile obtained using $\Psi_2$ is shown in figure \ref{figure10}(\emph{b}). (\emph{b}) Isocontours of $\Phic$ in $(k,\tf)$ plane using $\Psi_2$ (solid line) and $\Psi_3$ (dashed line).  }
    \label{figure11}
   \end{center}
\end{figure}

Figure \ref{figure11}(\emph{a})  illustrates the base-state (solid line) and $\cip$ profile generated using $\Psi_3$ (dashed line) for $\ti=0.1$, $\tf=5$, $\kw=30$, and $\Ra=500$.  As expected, $\Psi_3$
produces a profile with a maximum closer to $z=0$ than $z=\delta$.   Consequently, the profile shown in  figure \ref{figure11}(\emph{a}) supports initial amplitudes as large as $\M = 10^{-1}$ without producing negative values of $c_\mathrm{net}$.
Figure \ref{figure11}(\emph{b}) illustrates optimal isocontours of $\Phic$ in the $(\kw,\tf
)$ parameter plane using $\Psi_2$ (solid lines) and $\Psi_3$ (dashed lines).
As expected, though $\Psi_3$ supports larger initial amplitudes, $\Psi_2$
produces greater amplifications. This raises the possibility that
there exists an optimal filter function, $\Psi_{\mathrm{opt}}$, that
balances the tradeoff between the initial amplitude and subsequent
amplification in order to minimize the onset time for convection. This
is beyond the scope of the current study, however, because it requires
a nonlinear analysis. Therefore, for brevity, we focus on the
perturbations produced by $\Psi_3$ because these support large initial
amplitudes.
 
\subsection{Comparison with classical optimization scheme}

\begin{figure}
   \begin{center}    
    \hspace{0.0cm}
    (\emph{a})
    \hspace{6.5cm}
    (\emph{b})
    \\
        \includegraphics[trim = 0 0 0 10, clip=true,width=7cm]{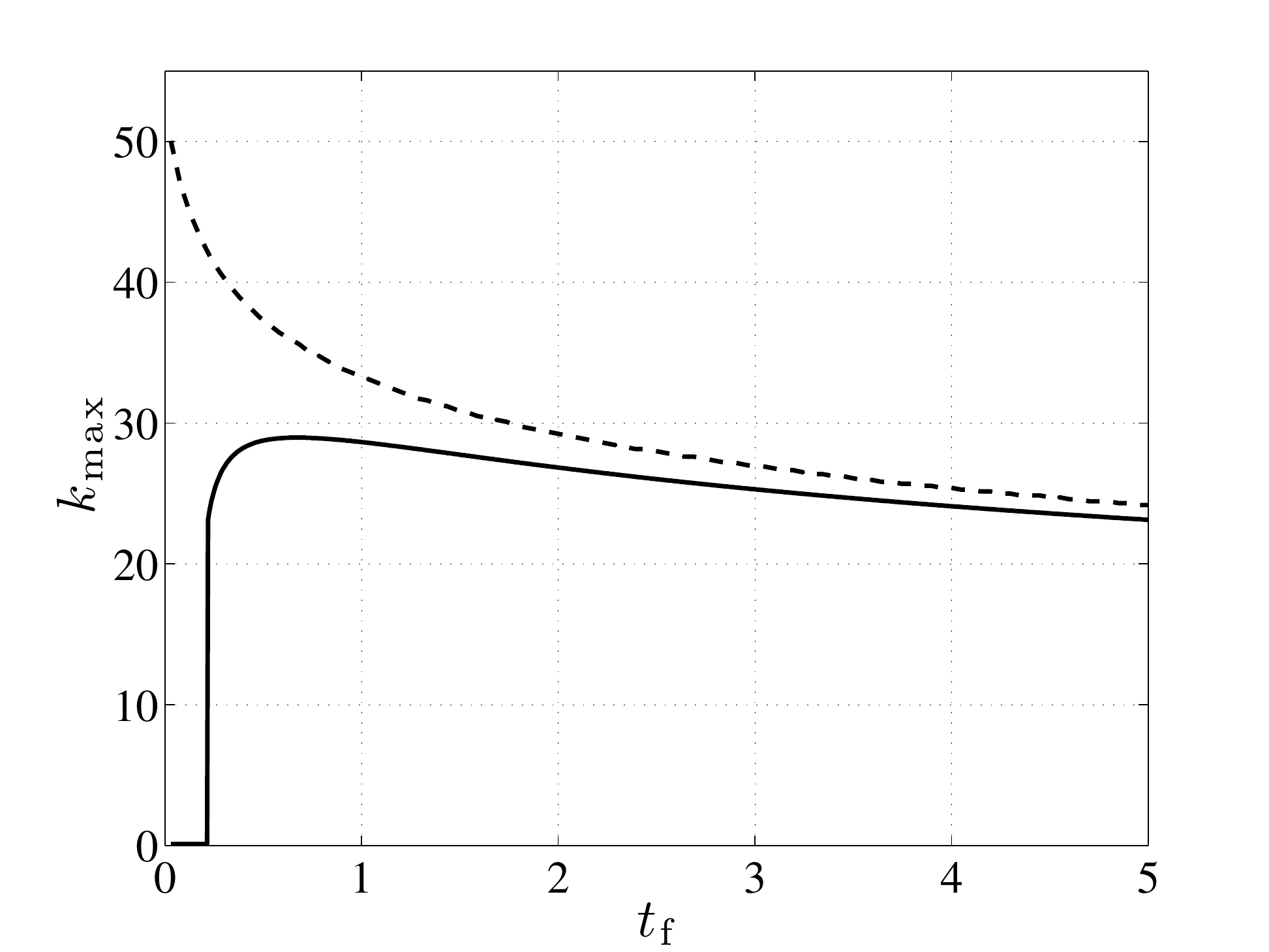} 
     \includegraphics[trim = 0 0 0 10, clip=true,width=7cm]{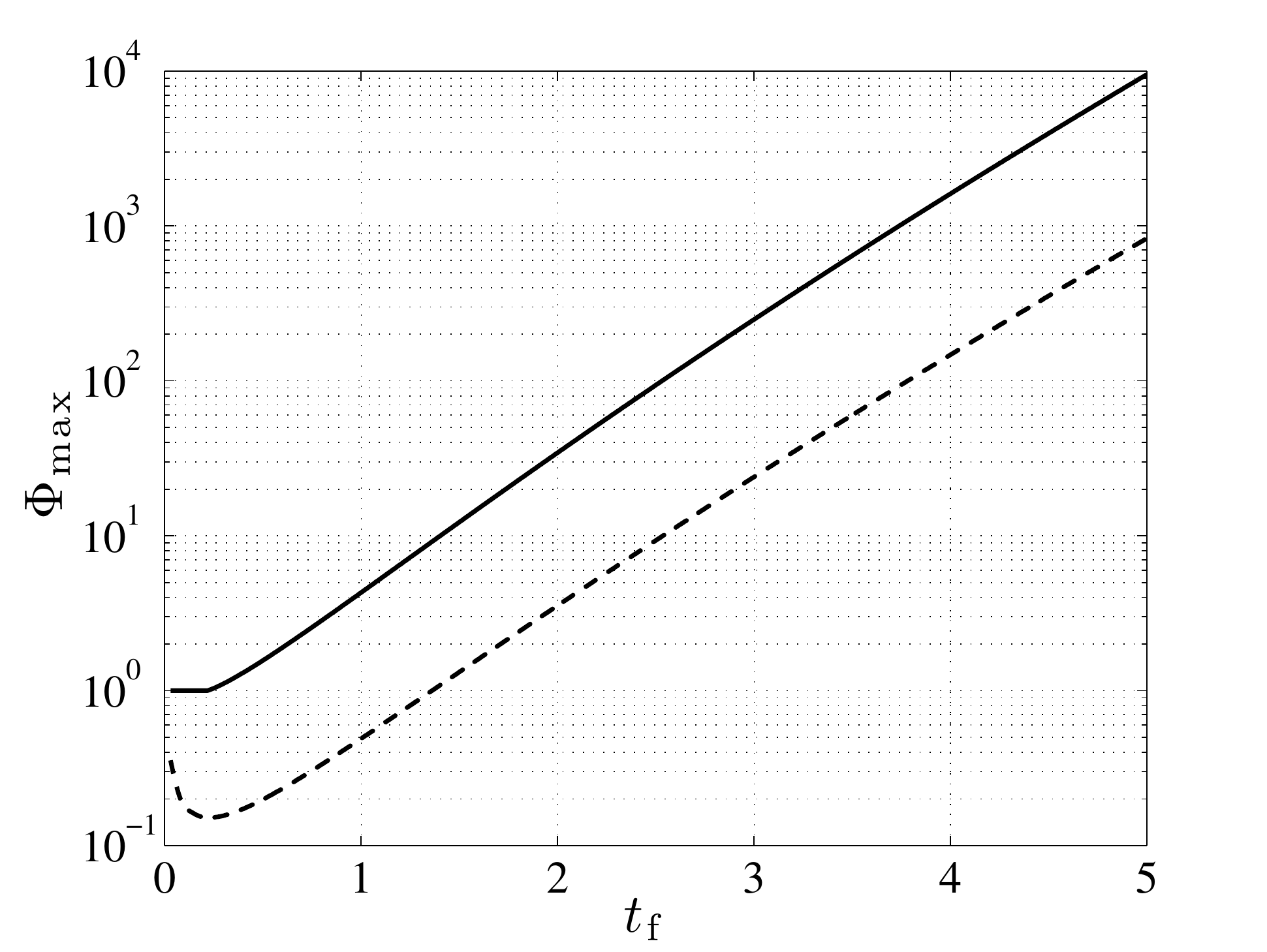}
    \\
    \hspace{0.0cm}
    (\emph{c})
    \hspace{6.5cm}
    (\emph{d})
    \\
      \includegraphics[trim = 0 0 0 10, clip=true,width=7cm]{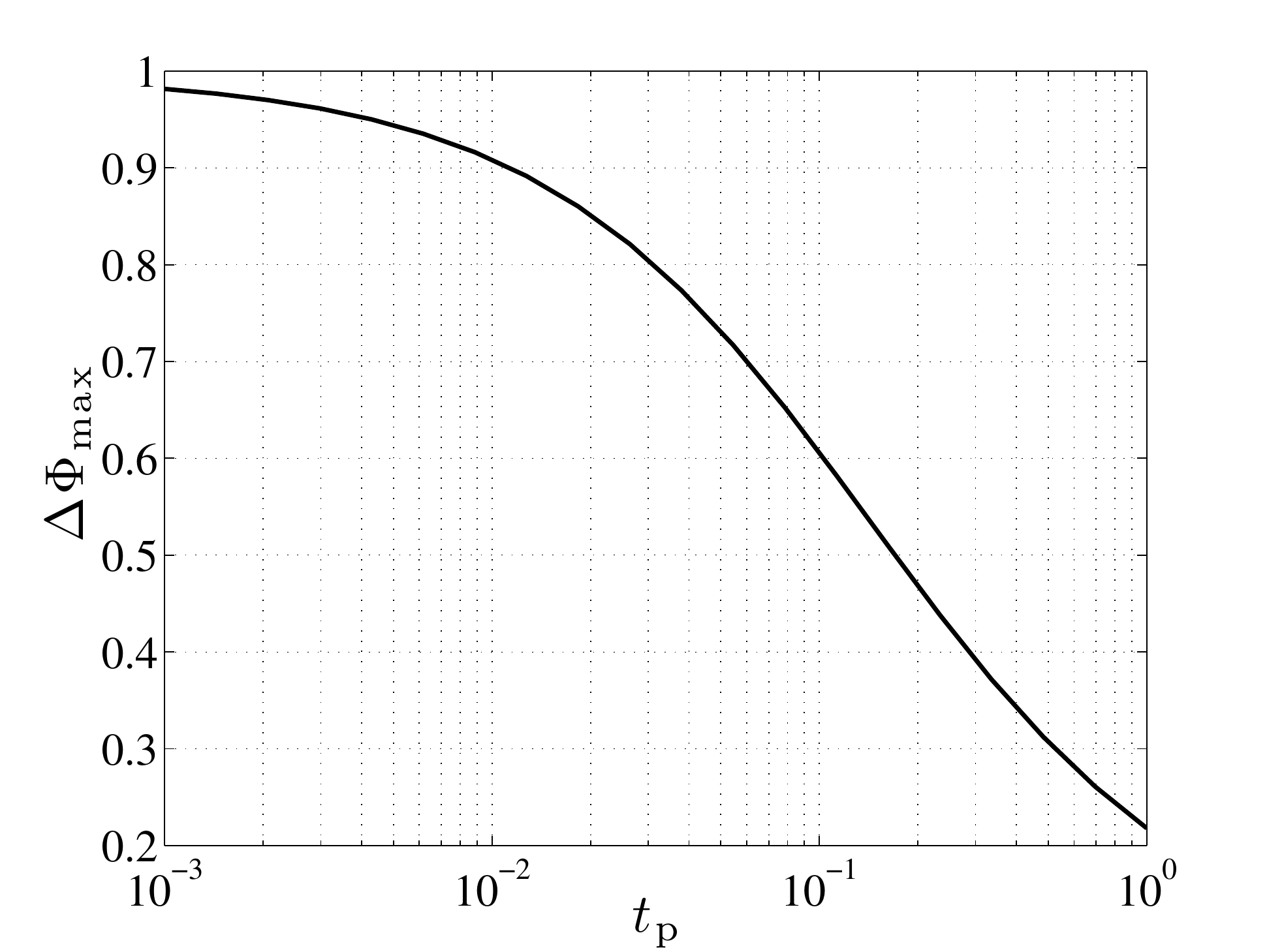}
      \includegraphics[trim = 0 0 0 10, clip=true,width=7cm]{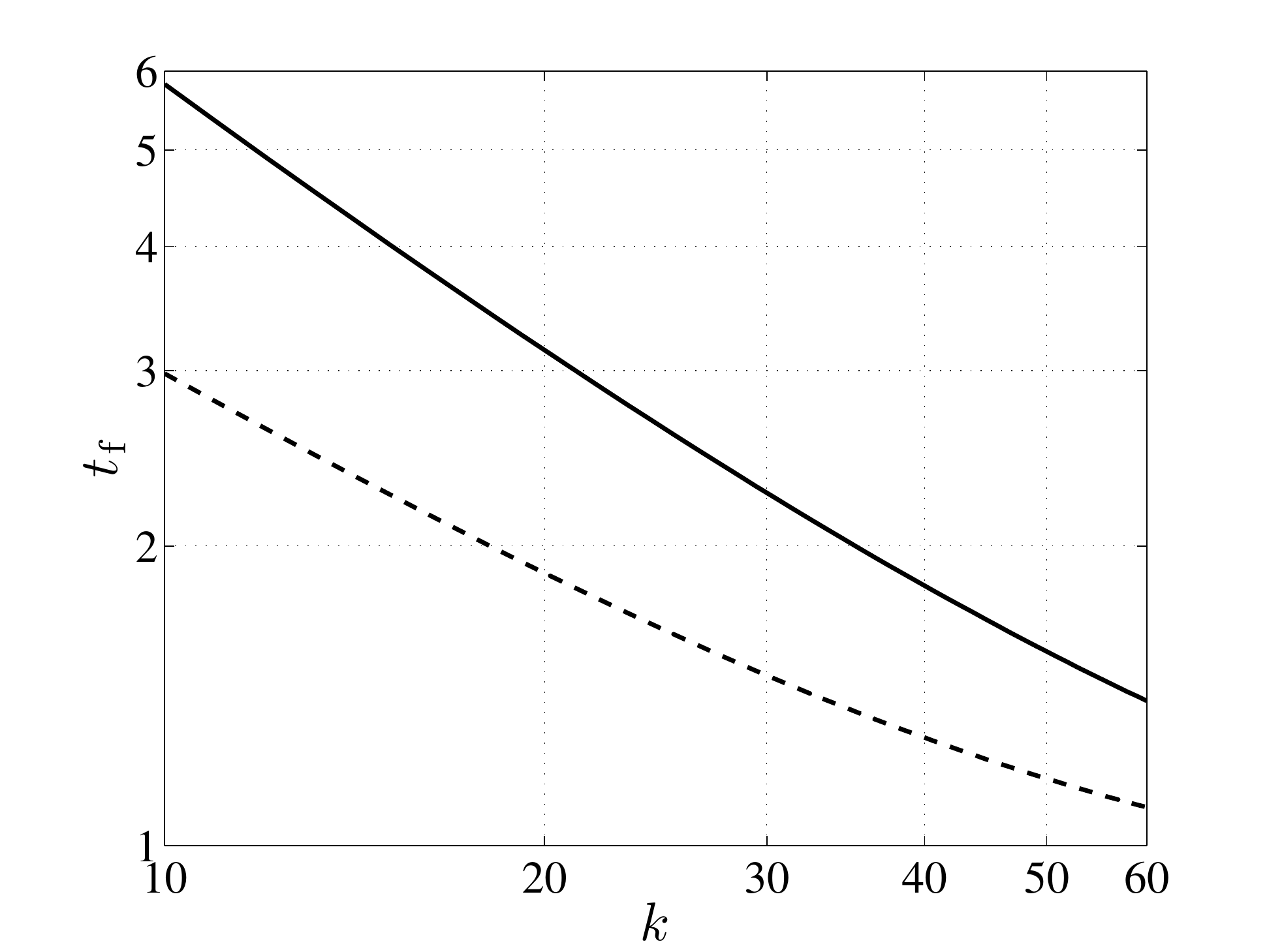}
      \caption{ Comparison of the $\opa$ and $\opb$ schemes for $\Ra=500$. (\emph{a}) $\kw_\mathrm{max}$ vs. $\tf$ for $\ti=0.01$ for $\opa$ (solid line) and $\opb$ (dashed line). (\emph{b}) $\Phi_\mathrm{max}$ vs. $\tf$ for $\ti=0.01$ for $\opa$ (solid line) and $\opb$ (dashed line). (\emph{c}) $\Delta \Phi_\mathrm{max}$ vs. $\ti$ for $\tf=4$.
(\emph{d}) Isocontours of $\Delta \cip / \Delta \tf = 0.001$  in the $(\kw, \tf)$ plane for $\ti=0.1$ for $\opa$ (solid line) and $\opb$ (dashed line). }  
    \label{figure12}
  \end{center}
\end{figure}

Hereinafter, we refer to the classical optimization procedure as $\opa$ and the modified optimization procedure using $\Psi_3$ as $\opb$. 
Figure \ref{figure12}(\emph{a}) illustrates the temporal evolution of the dominant wavenumbers, $\kw_\mathrm{max}$, produced by the $\opa$ (solid line) and $\opb$ (dashed line) schemes for $\ti=0.01$ and $\Ra=500$.
For early final times, $\tf < 0.21$, the $\opb$ scheme produces nonzero dominant wavenumbers,
$\kw_\mathrm{max} \ne 0$, while the $\opa$ scheme predicts $\kw_\mathrm{max} = 0$. The large difference in dominant wavenumbers at small times occurs
because the zero-wavenumber perturbations produced by the $\opa$ scheme
span the entire vertical domain, $\ch = \sin{(\pi z /2)}\exp{(-\pi^2 \Ra^{-1} t/4)} $, as
discussed in \S4.2. Using the $\opb$ scheme, these perturbations are
filtered by $\Psi_3$. At late $\tf$, the $\opb$ dominant wavenumbers tend towards those predicted by the $\opa$.
Figure \ref{figure12}(\emph{b}) illustrates the corresponding maximum amplifications, $\Phi_\mathrm{max}$, see equation (\ref{phimax}), produced by the $\opa$ (solid line) and $\opb$ (dashed line) schemes. For final times, $\tf < 0.21$, the $\opa$
amplifications are close to unity because the zero-wavenumber
perturbations have a small constant decay rate, see discussion in
\S4.2. The $\opb$ amplifications are an order-of-magnitude smaller
because the perturbations are constrained to the boundary layer and
undergo substantial damping up to the critical time for instability,
$\tc$. For $\tf > 1$, the amplifications produced by $\opa$ and $\opb$ in
figure  \ref{figure12}(\emph{b}) have identical slopes.  This occurs because of similar growth rates between the final states of the dominant wavenumber perturbations obtained using the $\opa$ and $\opb$ schemes. 

The difference between the $\opa$ and $\opb$ amplifications depends on the
initial time, $\ti$. To explore this, we measure
\begin{equation}
\Delta \Phi_\mathrm{max} =   \frac{\Phi_{\opa} - \Phi_{\opb}}{\Phi_{\opa}},
\end{equation}
where $\Phi_{\opa}$ and $\Phi_{\opb}$  are the maximum amplifications, $\Phi_\mathrm{max}$, obtained using $\opa$ and $\opb$, respectively. Figure \ref{figure12}(\emph{c}) illustrates $\Delta \Phi_\mathrm{max}$ for $\tf=4$ as the initial perturbation time varies from $\ti=10^{-3}$ to $\ti=1$.  Note that the results are independent of final time $\tf$ when $\tf>1$.
$\Delta \Phi_\mathrm{max}$ tends to a maximum as $\ti \rightarrow 0$ because $\Phi_{\opb} \rightarrow 0$, while $\Phi_{\opa}$ converges to finite values, see figures \ref{figure4}(\emph{c})--\ref{figure4}(\emph{d}).  With increasing $\ti$, $\Delta \Phi_\mathrm{max}$ decreases indicating better agreement between the $\opa$ and $\opb$ amplifications. Note that the  maxima of the optimal initial $\opb$ profiles are always closer to the top boundary, $z=0$, compared to the initial $\opa$ profiles. 

Recall from \S4.4, that beyond certain final times, the initial
$\cip$ profiles generated by the $\opa$ scheme are insensitive to further
increases in $\tf$. To investigate this behavior for the $\opb$
scheme, figure \ref{figure12}(\emph{d}) illustrates the isocontours $\Delta \cip / \Delta \tf=0.001$, see equation (\ref{eq:convergence}), in the $(\kw,  \tf)$ parameter plane generated
using the $\opa$ (solid line) and $\opb$ (dashed line) schemes for $\ti = 0.1$
and $\Ra = 500$. The final times beyond which the initial $\opb$ profiles do not change shape are much smaller than the
$\opa$ profiles. This suggests that initial perturbations confined within
the boundary layer rapidly converge to a common shape.

\begin{figure}    
   \begin{center}
    \hspace{0.0cm}
    (\emph{a})
    \hspace{6.5cm}
    (\emph{b})
    \\
    \includegraphics[trim = 0 0 0 10, clip=true,width=7cm]{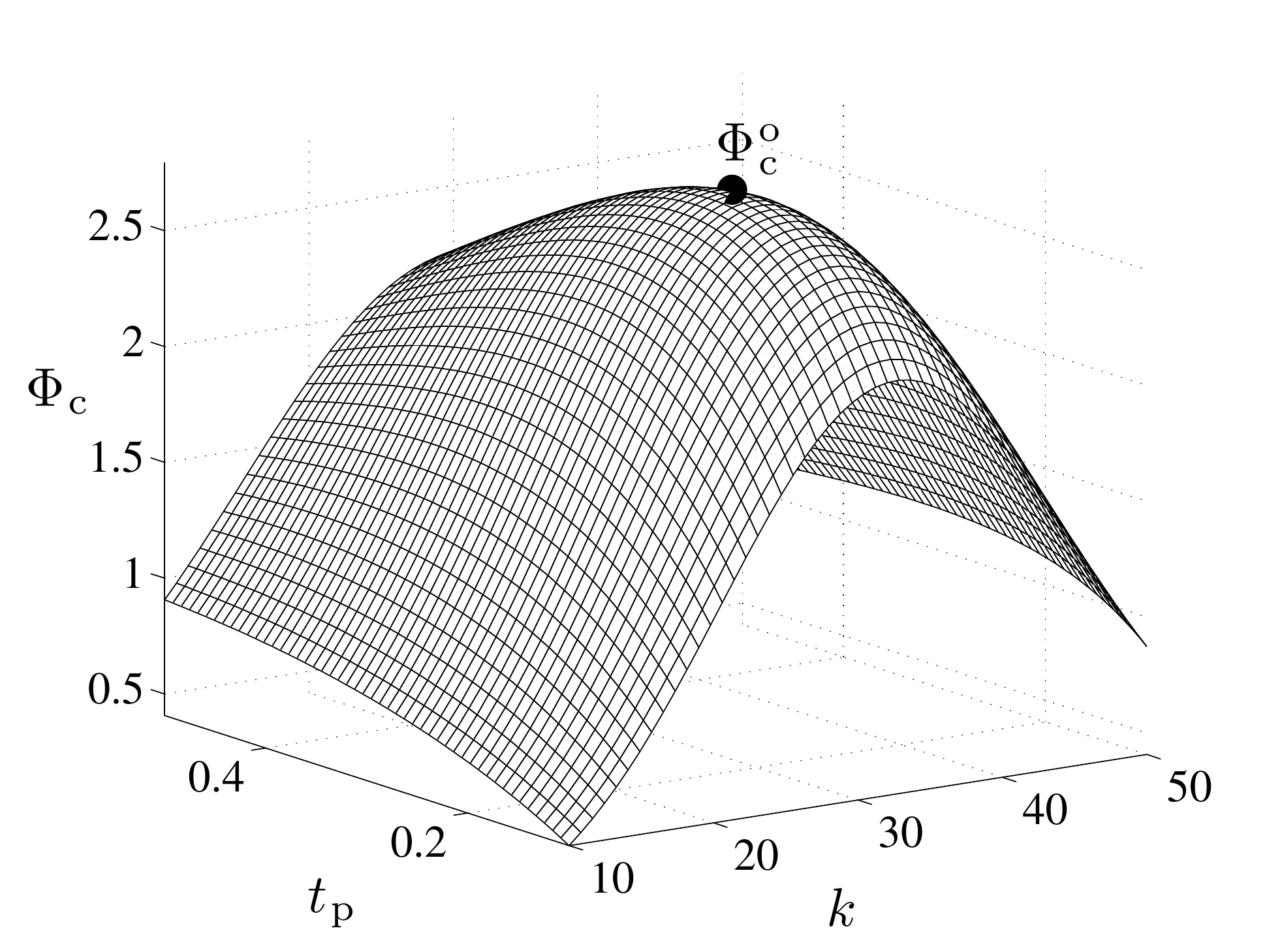}
    \includegraphics[trim = 0 0 0 10, clip=true,width=7cm]{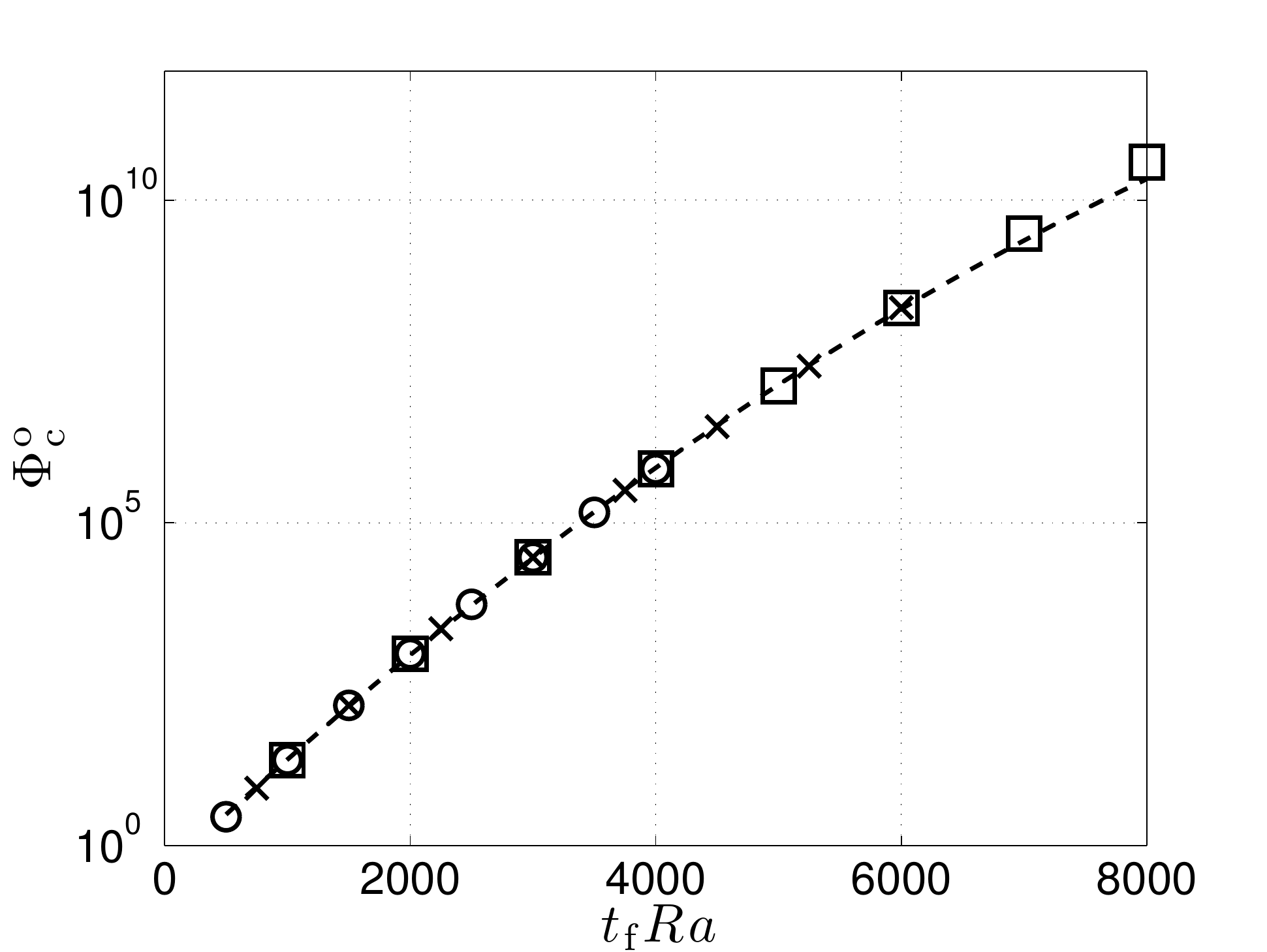}
    \\
     \hspace{0.0cm}
    (\emph{c})
    \hspace{6.5cm}
    (\emph{d})
    \\
    \includegraphics[trim = 0 0 0 10, clip=true,width=7cm]{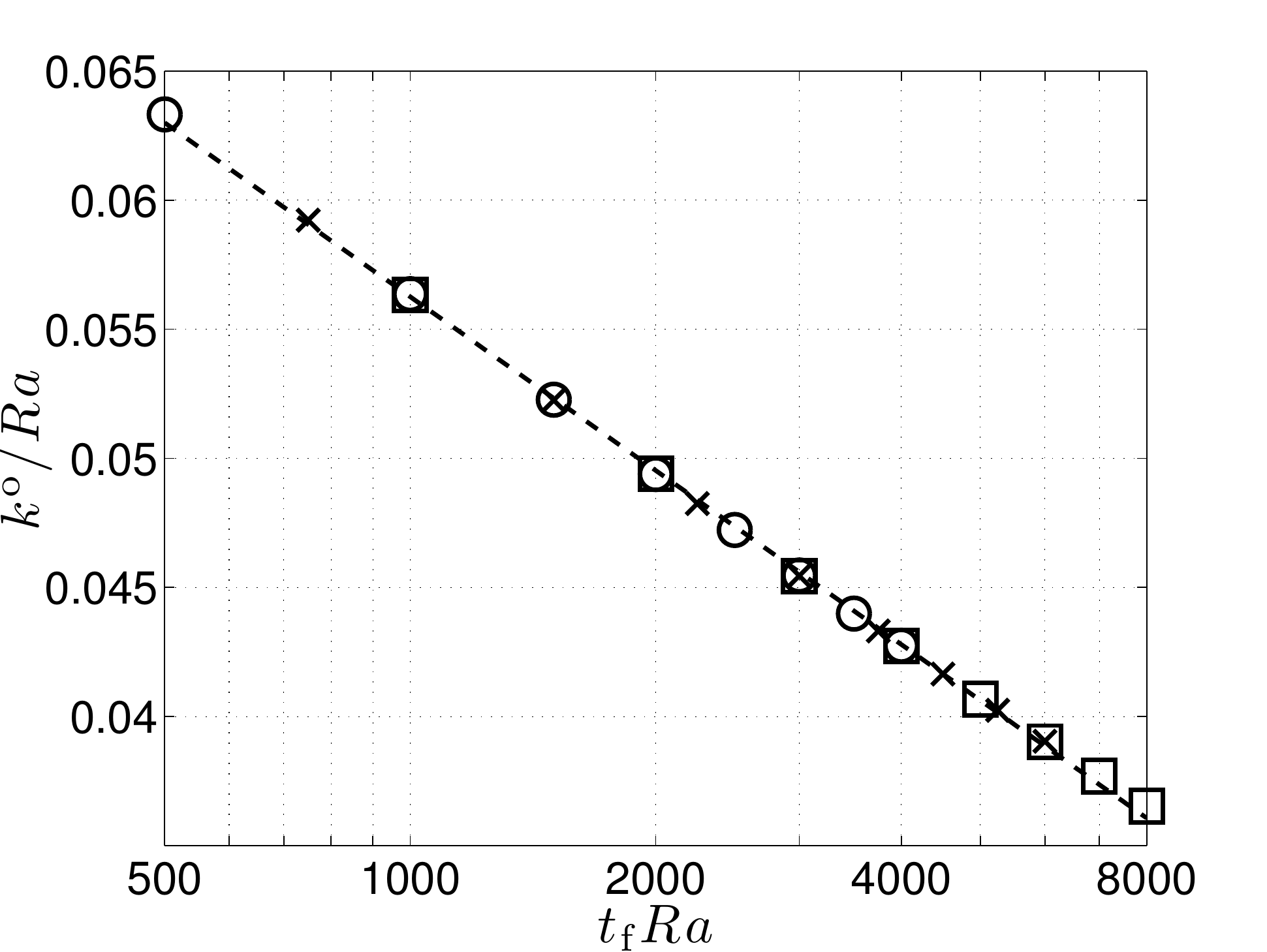}
    \includegraphics[trim = 0 0 0 10, clip=true,width=7cm]{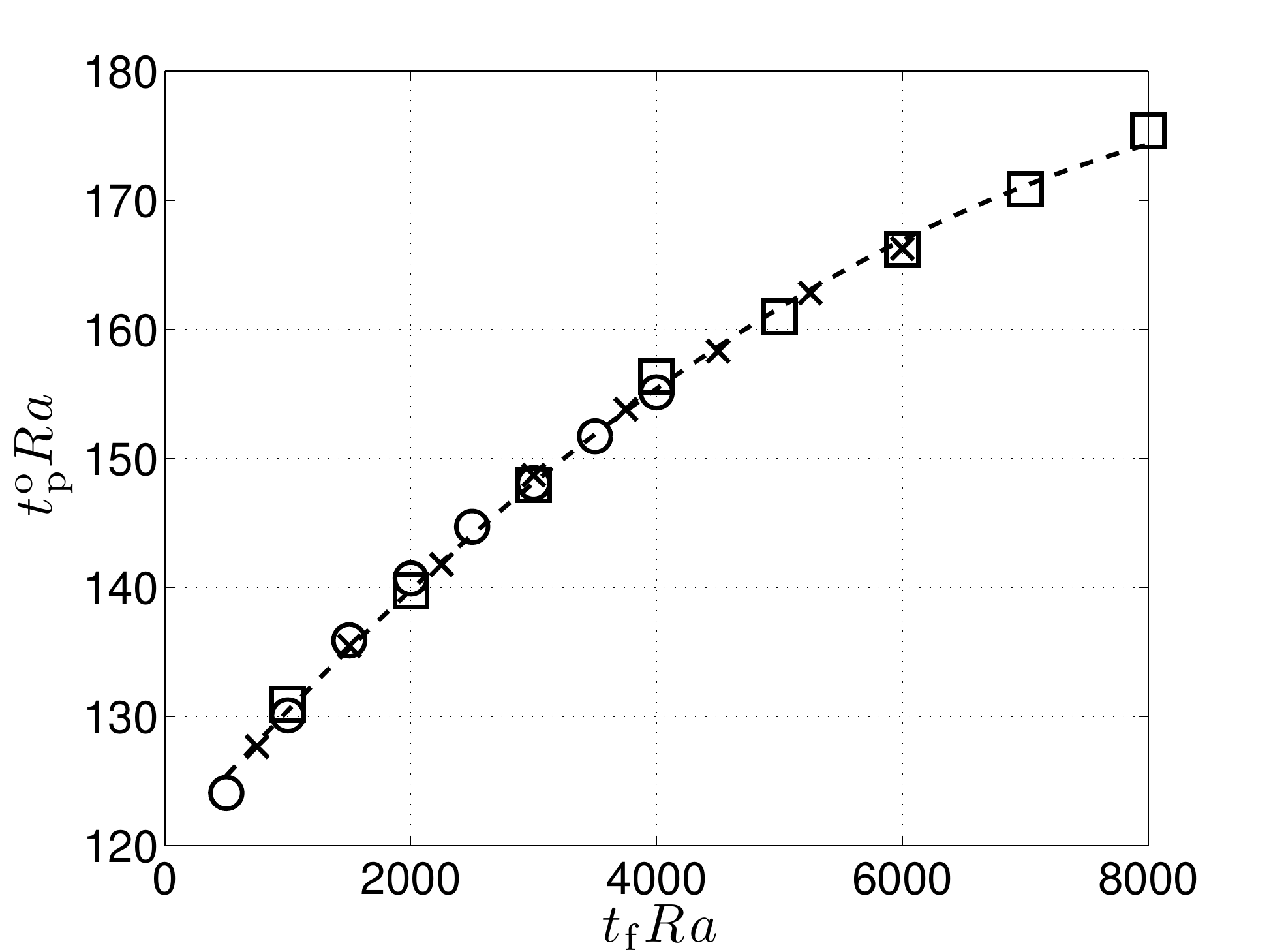}
    \caption{The optimal $\opb$ point  $(\Phic^\mathrm{o}$, $\kw^\mathrm{o}$, $\ti^\mathrm{o})$ as a function of $\tf$ and $\Ra$. (\emph{a}) $\Phic$ vs. $\ti$ and $\kw$ for $\Ra=500$ and $\tf=1$. The solid dot marks $(\Phic^\mathrm{o}$, $\kw^\mathrm{o}$, $\ti^\mathrm{o})$.  (\emph{b})  $\Phic^\mathrm{o}$  vs. $\tf \Ra$  for $\Ra = 500$ (circles), $\Ra = 750$ (crosses), and $\Ra = 1000$ (squares). The dashed line shows relationship (\ref{eqmphio}). (\emph{c}) $\kw^\mathrm{o}/\Ra$  vs. $\tf \Ra$ for $\Ra = 500$ (circles), $\Ra = 750$ (crosses), and $\Ra = 1000$ (squares). The dashed line shows relationship (\ref{eqmko}). (\emph{d}) $\ti \Ra$  vs. $\tf \Ra$ for $\Ra = 500$ (circles), $\Ra = 750$ (crosses), and $\Ra = 1000$ (squares). The dashed line shows relationship (\ref{eqmtio}).   }
    \label{figure13}
   \end{center}
\end{figure}

As discussed in \S4.3, due to the transient growth of the base-state, there exists an optimal combination of initial time and wavenumber, $\ti^\mathrm{o}$ and $\kw^\mathrm{o}$, that produces the subsequent optimal amplification $\Phic^\mathrm{o}$. Figure \ref{figure13}(\emph{a}) illustrates the $\opb$ amplifications,
$\Phic$, for $10 \le \kw \le 50$,  $0.1 \le \tp \le 0.5$, $\tf = 1$, and $\Ra = 500$.
The solid dot marks the peak of the surface, $\Phic^\mathrm{o}$.  
Figure \ref{figure13} demonstrates $\Phic^\mathrm{o}$ (panel \emph{b}), $\kw^\mathrm{o}/ \Ra$ (panel \emph{c}), and $\ti^\mathrm{o} \Ra$ (panel \emph{d}) as functions of $\tf \Ra$. The results for different $\Ra$ collapse as previously demonstrated for the $\opa$ scheme in figure \ref{figure6}.  We obtain the following relationships for $\Phic^\mathrm{o}$ and the dimensional forms of wavenumber, $\kw^*$, and the initial time, $\ti^*$, 
\begin{equation}
\log\Phic^\mathrm{o} =   -5.550 \! \times \! 10^{-8} {\tf^*}^2 \left(\frac{U^2}{\phi^2 D} \right)^2 + 0.001785 \tf^*  \, \frac{U^2}{\phi^2 D}  - 0.3967, 
\label{eqmphio}
 \end{equation}
 \begin{equation}
\kw^* = \frac{U}{\phi D} \Big[  0.1234 -0.02237 \, \log \left(\frac{\tf^* U^2}{\phi^2 D} \right)  \Big],
 \label{eqmko}
\end{equation}
\begin{equation}
 \ti^* = -5.107 \! \times\! 10^{-7} {\tf^*}^2 \frac{U^2}{\phi^2 D} +0.01086 \tf^* + 120.1 \frac{\phi^2 D}{ U^2}.
 \label{eqmtio}
\end{equation} 
 
For high permeability aquifers, $K=10^{-12}$ m$^2$ (see \S4.3), figure \ref{figure13} predicts that the optimal perturbation wavelength and initial perturbation time vary in the range, $10 \, \mathrm{cm} \le 2 \pi / \kw^* \le  18 \, \mathrm{cm}$ and $36 \, \mathrm{hours} \le \ti^*\le 51 \, \mathrm{hours}$ as the final time varies between, $6 \, \mathrm{days} \le \tf^*\le 96 \, \mathrm{days}.$ For low permeability aquifers, $K=10^{-14}$ m$^2$,  these parameters vary in the range $10 \, \mathrm{m} \le 2 \pi / \kw^* \le 18 \, \mathrm{m}$, $41 \, \mathrm{years} \le \ti^* \le 58 \, \mathrm{years}$, $165 \, \mathrm{years} \le \tf^*\le 2636 \, \mathrm{years}$.
The optimal amplifications,  $\Phic^\mathrm{o}$, are approximately 50 \% those produced by the $\opa$ scheme, see figure \ref{figure6}.  We observe that $\kw^\mathrm{o}$ agrees closely with those produced, using the $\opa$ scheme. The optimal initial perturbation times, $\ti^\mathrm{o}$, however,  are roughly twice as large as those for the $\opa$ scheme due to the large
initial damping periods experienced by the $\opb$ perturbations. The optimal initial time, $\ti^\mathrm{o}$,  is also more sensitive to $\tf$ than the $\opa$ scheme. Recall from \S4.3, that the optimal initial perturbation time would require \emph{a priori} knowledge of the onset time of convection, i.e. $\tf =\ton$. Because of the increased sensitivity of $\ti^\mathrm{o}$ to $\tf$, we expect the optimal $\opb$ perturbations to be more sensitive to initial perturbation amplitude, $\M$, than the $\opa$ perturbations.

\subsection{Comparison with $\mathrm{QSSA}_\xi$ eigenvalue and IVP problems}

\begin{figure}
    \begin{center}    
    \hspace{0.0cm}
    (\emph{a})
    \hspace{6.5cm}
    (\emph{b})
    \\
     \includegraphics[trim = 0 0 0 10, clip=true,width=7cm]{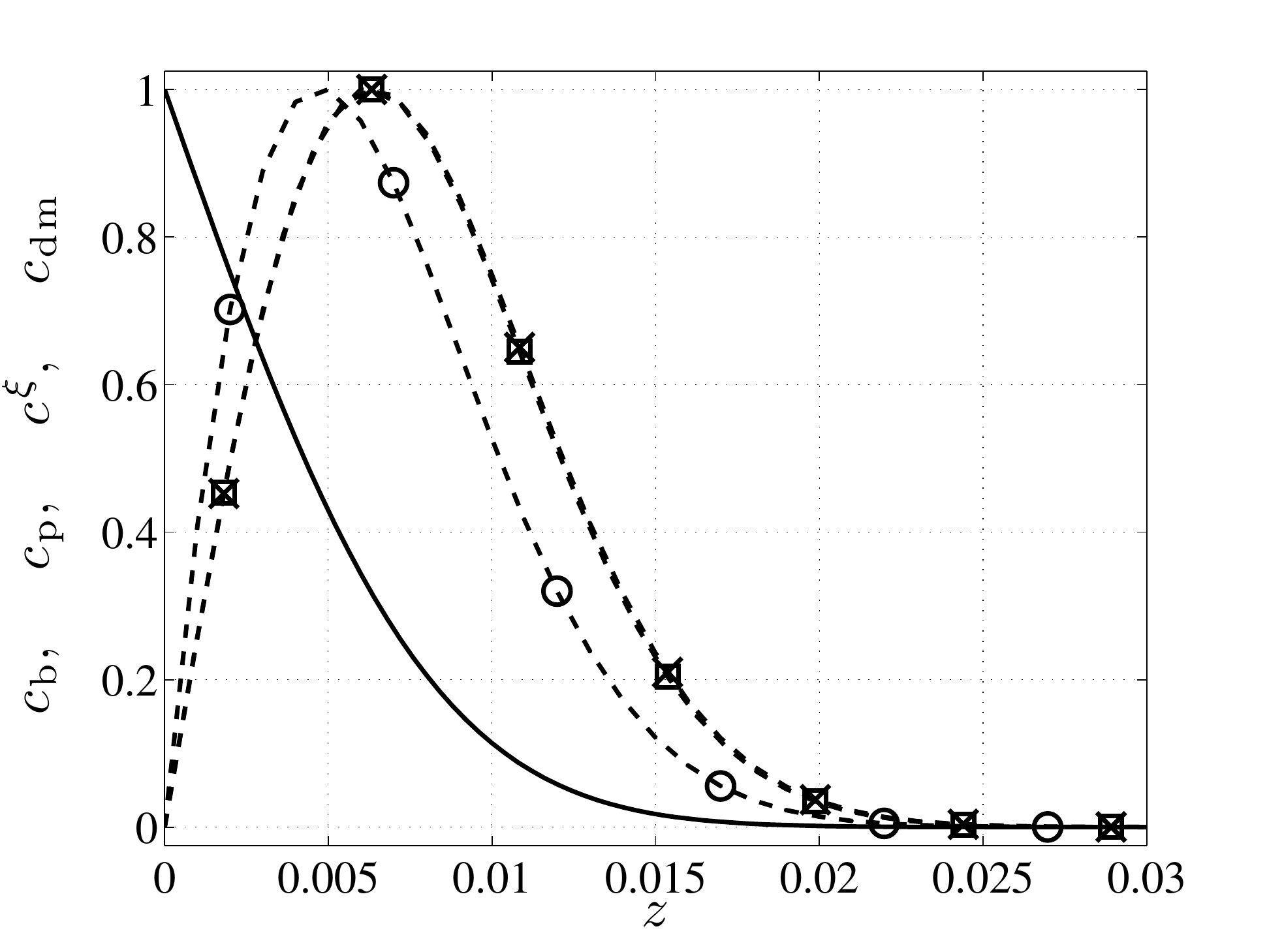}
    \includegraphics[trim = 0 0 0 10, clip=true,width=7cm]{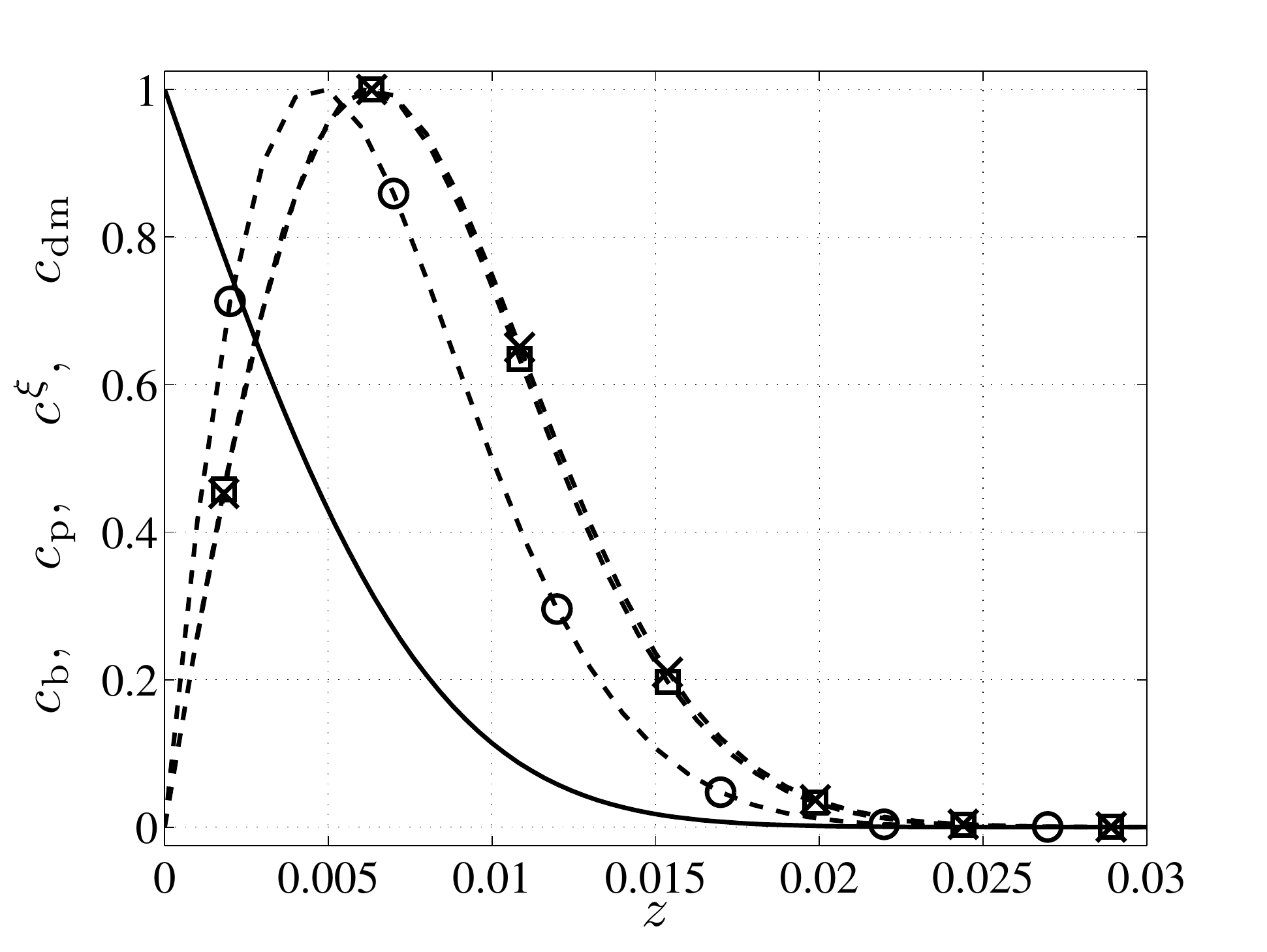}
        \\
    \hspace{0.0cm}
    (\emph{c})
    \hspace{6.5cm}
    (\emph{d})
    \\
      \includegraphics[trim = 0 0 0 10, clip=true,width=7cm]{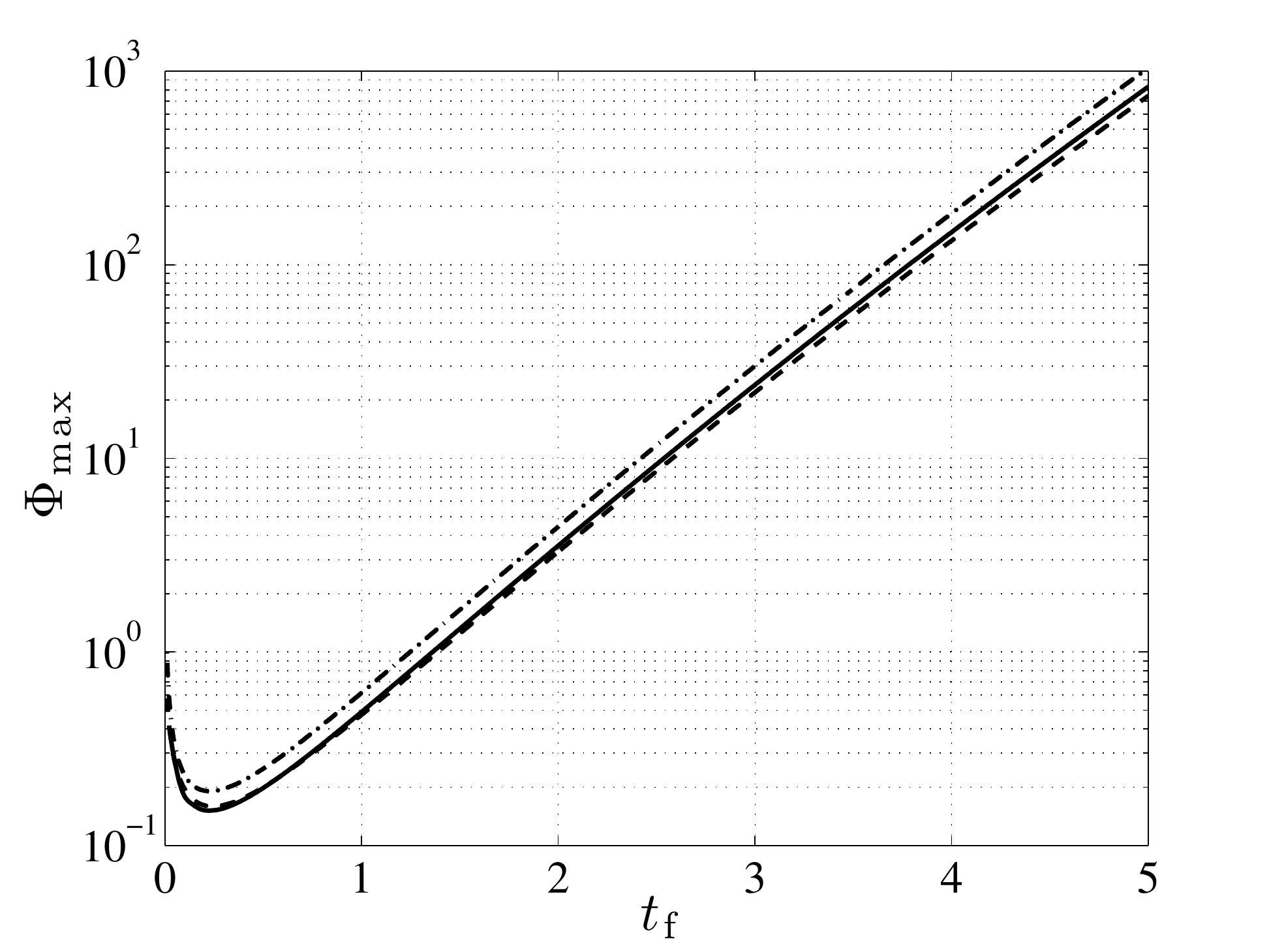}
       \includegraphics[trim = 0 0 0 10, clip=true,width=7cm]{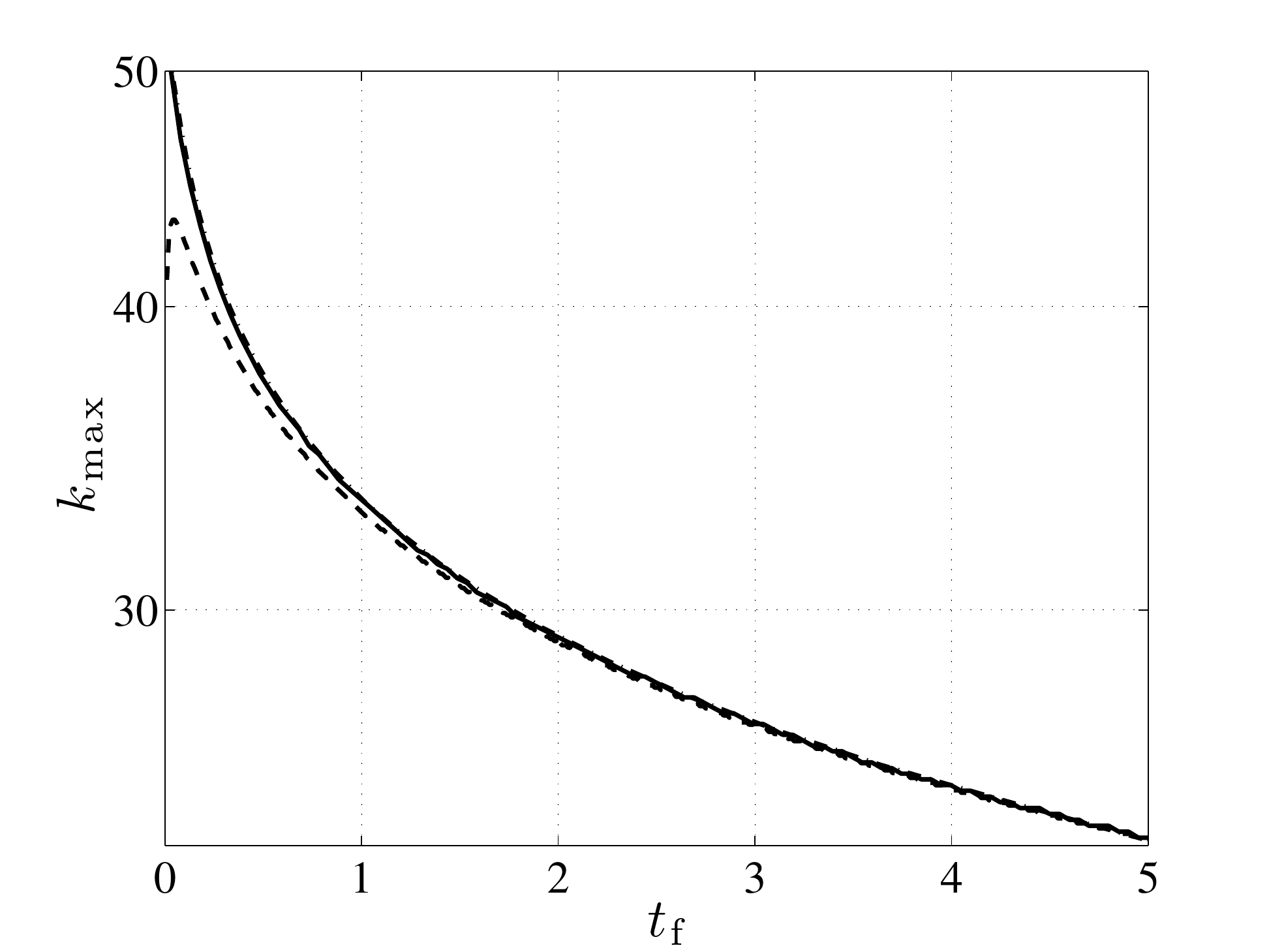}
      \caption{ Comparison of modified optimization with  QSSA in self-similar space and IVP with $\cip=c_\mathrm{dm}$ for $\Ra=500$. (\emph{a}) Base-state, $\cb$ (solid line),  initial $\opb$ profile (circles), dominant $\mathrm{QSSA}_\xi$ eigenmode, $c^\xi$ (squares), and $c_\mathrm{dm}$  perturbation (\ref{eq:dm}) (crosses) at $\ti=0.01$ for $\kw=10$. (\emph{b}) Same as in panel (\emph{a}) for $k=50$. (\emph{c})--(\emph{d}) Temporal evolution of $\Phi_\mathrm{max}$ and $\kw_\mathrm{max}$ for $\ti=0.01$ using $\opb$ (solid line),  $\mathrm{QSSA}_\xi$ (dashed line), and initial condition (\ref{eq:dm}) (dash-dotted line). }  
    \label{figure14}
   \end{center}
\end{figure}

In this section, we compare the modified optimization procedure to
previously published linear stability methods that ensure
perturbations are localized within the boundary layer. The first
approach approximates the vertical domain as semi-infinite. In this
case there is a similarity solution for the base-state,
$\cb=1-\mathrm{erf}(\xi)$,  where $\xi(z,t)=z \sqrt{\Ra/(4t)}$ is the similarity
variable. \cite{Riaz2006JFM} demonstrated that a quasi-steady modal analysis
with respect to the $(\xi, t)$ space produces eigenmodes concentrated in
the boundary layer. For convenience of notation, we refer to this as
the QSSA$_\xi$ problem. We refer to the eigenvectors of the QSSA$_\xi$ problem
as $c^{\xi}$ and $w^{\xi}$.
The second procedure we consider is the solution of the forward IVP
 (\ref{eq:linear})--(\ref{eq:ic})  using $\cip = c_\mathrm{dm}$, where $ c_\mathrm{dm}$ is the ``dominant mode'' of \cite{Riaz2006JFM} given by,
\begin{equation}
   c_\mathrm{dm}(z) = \xi \mathrm{e}^{-\xi^2}.
 \label{eq:dm}
\end{equation}
Initial condition (\ref{eq:dm}) is the leading-order term of a Hermite
polynomial expansion in the ($\xi,t)$ space and has been used in
numerous previous studies \cite[]{ben2002, Pritchard2004, Riaz2006JFM, Selim2007a, Wessel2009, Kim2011, Kim2012, Elenius2012}.

Figure \ref{figure14}(\emph{a}) illustrates the initial perturbation concentration profiles,
$\ch(z,\ti)$, produced by the $\opb$ (circles), dominant QSSA$_\xi$
eigenmode (squares), and initial condition (\ref{eq:dm})
(crosses),  for $\ti = 0.01$, $\Ra = 500$, and
$\kw = 10$. Figure \ref{figure14}(\emph{b}) repeats figure \ref{figure14}(\emph{a}) for the larger wavenumber, $\kw= 50$. In both figures, the base-state is shown as a solid line. For
both wavenumbers, the three methodologies produce qualitatively
similar profiles.  The profiles produced by QSSA$_\xi$  and $c_\mathrm{dm}$ are
indistinguishable, while the $\opb$ profiles have maxima closer to $z = 0$. Note that the $\opb$ profiles support slightly larger initial magnitudes, $\M$,  than the QSSA$_\xi$  and $c_\mathrm{dm}$ profiles, without producing negative net concentrations, $c_\mathrm{net}$. 

Figure \ref{figure14}(\emph{c}) illustrates results for $\Phi_\mathrm{max}$ versus $\tf$ obtained using  the $\opb$ (solid line), $\mathrm{QSSA}_\xi$ (dashed line), and  initial
condition (\ref{eq:dm})  for $\ti=0.01$, $0.03 \le \tf \le 5$, and $\Ra=500$. 
The three procedures again produce
similar results, though initial condition (\ref{eq:dm}) produces marginally larger amplifications. 
Note that the $\mathrm{QSSA}_\xi$ amplifications are obtained by first transforming the dominant $\mathrm{QSSA}_\xi$ growth rates to the $(z, t)$ coordinates using a $L^2$ norm, before integrating equation (\ref{eq:sigma}). 
Figure  \ref{figure14}(\emph{d}) illustrates the corresponding dominant
wavenumbers, $\kw_\mathrm{max}$, of the three procedures. The results produced by
initial condition (\ref{eq:dm}) and the $\opb$ are indistinguishable.
 
\section{Direct Numerical Simulations}

We perform two-dimensional direct numerical simulations (DNS) of the nonlinear governing equations (\ref{eq:ge})--(\ref{eq:ge-bc}) using a traditional pseudospectral method with spectral spatial accuracy \cite[]{Peyret2002}.  The horizontal domain is truncated to $x \in [0, L]$ with periodic boundary conditions on $x = 0$ and $x = L$. Equations (\ref{eq:ge})--(\ref{eq:ge-bc}) are then discretized spatially using Chebyshev polynomials in the vertical $z$ direction and a Fourier expansion in the horizontal $x$ direction.
The advection-diffusion equation is discretized temporally using a third-order, semi-implicit, backwards-difference scheme \cite[]{Peyret2002}. This temporal discretization is chosen for its favorable stability and allows us to investigate small initial times, $\ti \rightarrow 0$, for which the DNS scheme of \cite{Rapaka2008} was numerically unstable.   
 The initial concentration field is prescribed at $t=\ti$ as
\begin{equation}
c_\mathrm{dns}(z,x) = \cb(z) + \M \frac{c_\mathrm{i}(x,z)}{\| c_\mathrm{i}\|_\infty},
\end{equation}
where $\M$ is the initial perturbation magnitude measured with respect to the infinity norm of the perturbation concentration field, $c_\mathrm{i}$. 

\subsection{DNS of physical systems}

To emulate physical experiments, we perform DNS in which the boundary layer is simultaneously perturbed with all wavenumbers resolved numerically, 
\begin{equation}
c_\mathrm{i}(x,z) = \sum_{m=0}^{N/2-1}  a_m \mathrm{cos} \left( \frac{2 \pi m} {\mathrm{L}} x \right) G(z) F(z), 
\label{eq:rdm}
\end{equation}
where $N$ is the number of collocation points in the $x$ direction, and $-1 \le F(z) \le 1$ is a random function generated using Fortran's random number generator. The coefficients $a_m$ are computed to ensure that each horizontal Fourier mode is perturbed with equal energy. We set $L = 4 \pi$ and $N=1024$ in order to resolve wavenumbers, $\kw = 0,0.5,1,...\, , 255$.
To ensure that $c_\mathrm{i}$ satisfies the boundary condition at $z=0$ and remains concentrated within the boundary layer,  we introduce the Gaussian function, 
\begin{equation} 
G(z)   = \begin{cases}
0 & \text{if $z = 0 $,}\\ 
\mathrm{exp} \left( -\frac{1}{2} \left(  \frac{\zeta-\zeta_c}{\sigma} \right)^2 \right) & \text{if $ 0 < z \le \delta    $,}\\ 
0 & \text{if $ \delta  < z \le 1   $,}
\end{cases} 
\end{equation}
where $\zeta = z / \delta $,  $\zeta_c$ is the mean and $\sigma$ is the standard deviation.
For example, when $\zeta_c=0.5$, the peak of the Gaussian function is located midway between $z=0$ and $z=\delta$. We vary the peak location, $\zeta_c$, and the width, $\sigma$, to recreate several experimental possibilities listed in table \ref{table:initial}. 

\begin{table}
\centering
\begin{tabular}{c c c c}
\hline\hline 
$\mathrm{Case}$ & $\zeta_c$ & $\sigma$ & $\mathrm{Symbol}$   \\ 	
\hline 
$1$ & $0.50$ & $0.05$ & Circle    \\
$2$ & $0.50$ & $0.10$ & Square \\
$3$ & $0.50$ & $0.15$ & Cross \\
$4$ & $0.25$ & $0.10$ & Diamond \\
$5$ & $0.75$ & $0.10$ & Plus \\
\hline 
\end{tabular}
\caption{ The parameters used for the Gaussian, $G(z)$.} 
\label{table:initial} 
\end{table}

\begin{figure}  
  \begin{center}  
  \hspace{0.0cm}
    (\emph{a})
    \hspace{6.5cm}
    (\emph{b})
    \\
    \includegraphics[trim = 0 0 0 10, clip=true,width=7cm]{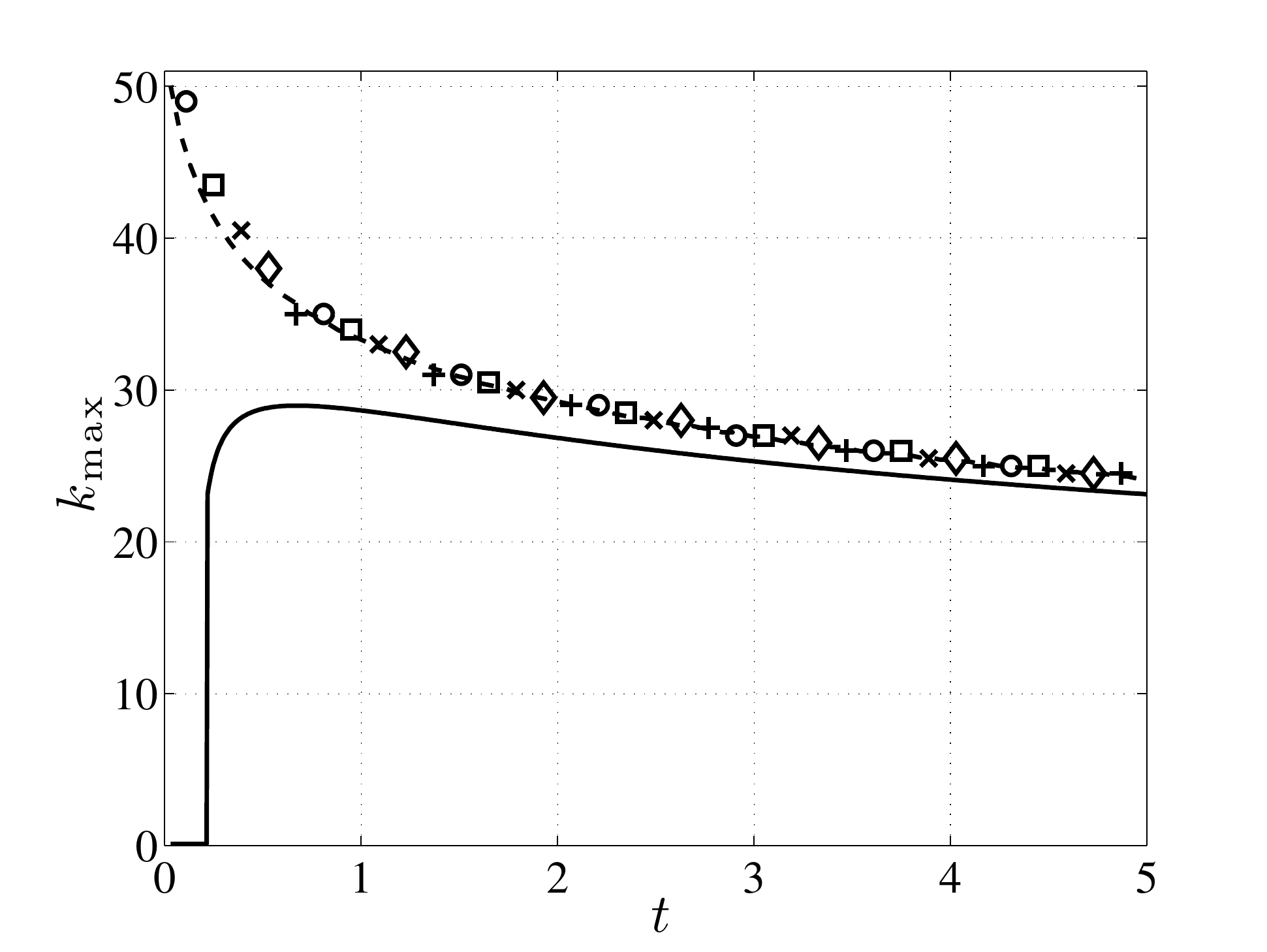}
     \includegraphics[trim = 0 0 0 10, clip=true,width=7cm]{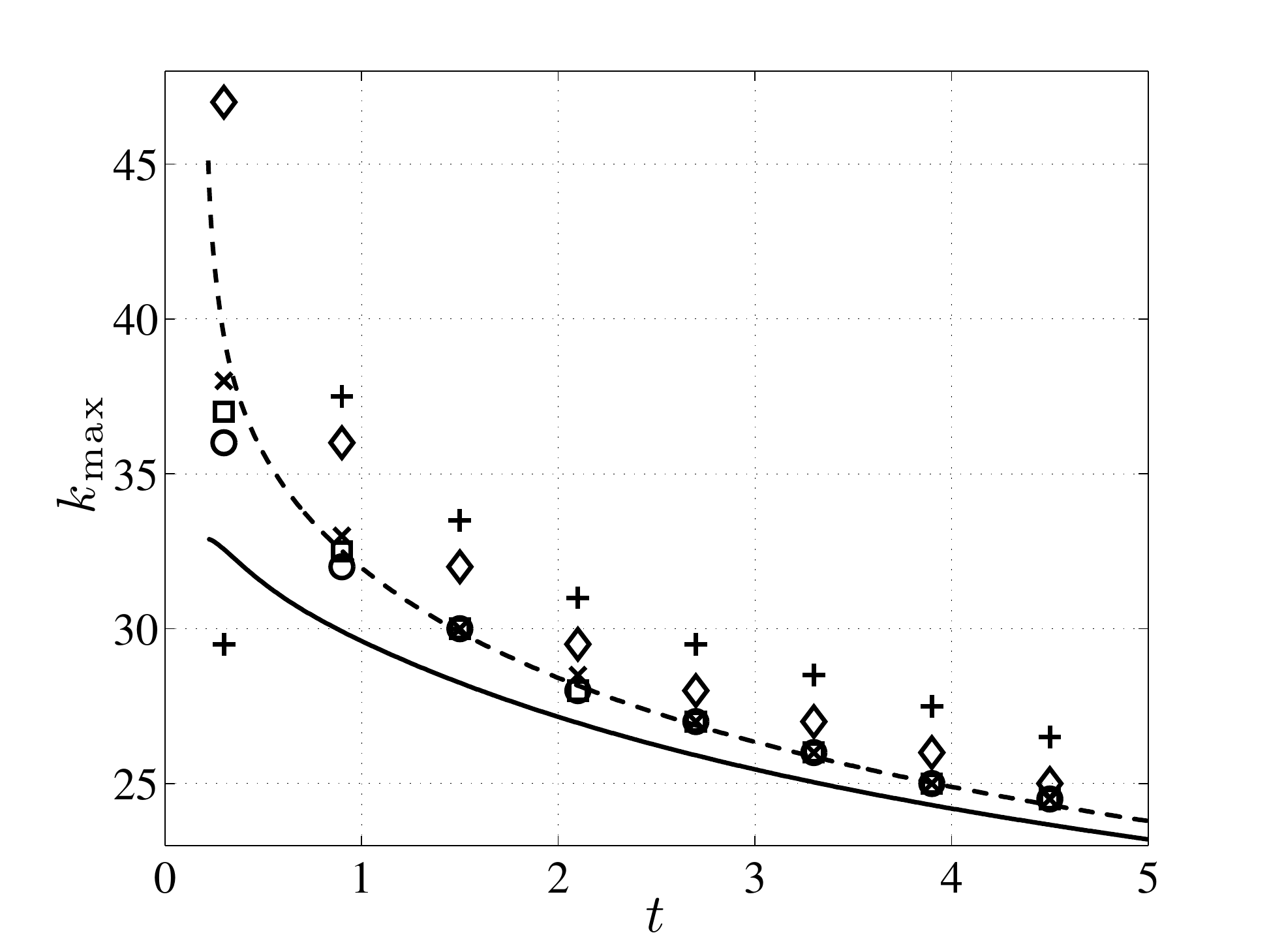}
    \caption{ (\emph{a}) Temporal evolution of dominant wavenumbers, $\kw_\mathrm{max}$, produced by DNS (symbols, see table \ref{table:initial}), $\opa$ (solid line), and $\opb$ (dashed line) for $\Ra=500$ and $\ti=0.01$ (\emph{b}) Same as in panel (\emph{a}) for $\ti=0.2$.   }
   \label{figure15}
   \end{center}
\end{figure}

Figure \ref{figure15}(\emph{a}) illustrates the temporal evolution of the dominant wavenumbers,  $\kw_\mathrm{max}$, produced by $\opa$ (solid line), $\opb$ (dashed line), and five DNS recreating the experimental conditions in table \ref{table:initial},  for $\Ra=500$ and $\ti=0.01$. 
All simulations are run using the initial amplitude $\M=10^{-4}$ to produce a long linear regime, $\ton > 5$, to
facilitate comparison of the dominant wavenumbers predicted by $\opa$,
$\opb$ and DNS.  
We observe excellent agreement between the dominant wavenumbers
produced by the $\opb$ and DNS, while those predicted  by the $\opa$ show poor agreement. 

Figure \ref{figure15}(\emph{b}) repeats figure \ref{figure15}(\emph{a}) for the initial perturbation time $\ti=0.2$, chosen to be near the optimal perturbation time, $\ti^\mathrm{o}$.
We first note that the DNS results for $\kw_\mathrm{max}$ have a much wider spread than those for $\ti=0.01$. This likely occurs because the initial damping period is much shorter for $\ti=0.2$. Overall, we observe that $\opb$ shows much better agreement with DNS than $\opa$. For cases 1, 2, and 3 (see table \ref{table:initial}) the agreement between $\opb$ and DNS is excellent. For cases 4 and 5, $\opb$ underpredicts $\kw_\mathrm{max}$, though it still outperforms $\opa$.  The improved agreement for cases 1, 2, and 3 may stem from the fact that the boundary layer was perturbed near $z=0.5\delta$ in these cases. In cases 4 and 5, the layer was perturbed near $z=0.25\delta$ and $z=0.75\delta$, respectively.

\begin{figure} 
  \begin{center}   
   \hspace*{-5mm}
    \includegraphics[trim = 35 0 0 10, clip=true,width=15.3cm]{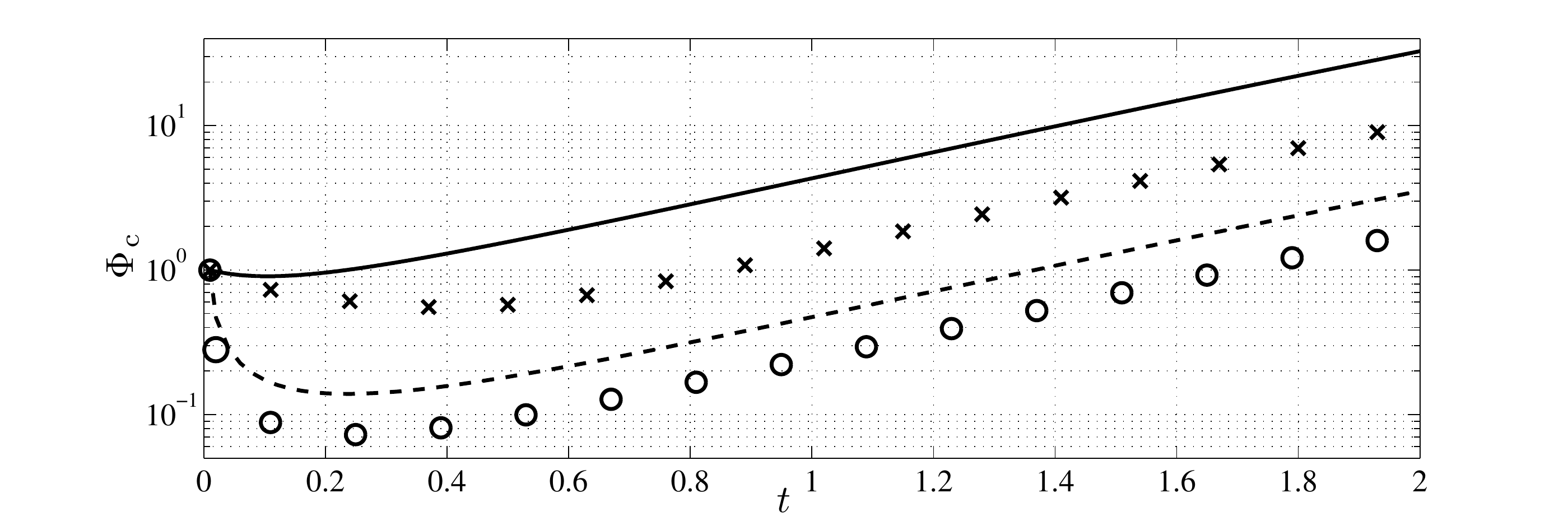}
    \caption{  DNS result (circles) for the temporal amplification, $\Phic$, of the $\kw=30$ mode when the boundary layer is perturbed with initial condition (\ref{eq:rdm}) at $\ti=0.01$ for case 1 (see table 2) and $\Ra=500$.  For comparison, we also show the corresponding optimal amplifications produced by $\opa$ (solid line) and $\opb$ (dashed line)  for $\kw=30$, $\Ra=500$, and $\ti=0.01$. The crosses illustrate the amplifications produced when the forward IVP is integrated using the unphysical random initial condition illustrated in figure \ref{figure8}(\emph{a}).}
   \label{figure16}
   \end{center}
\end{figure}

Figure \ref{figure16} illustrates the DNS result (circles) for the temporal amplification, $\Phic$, of the $\kw=30$ mode when the boundary layer is perturbed with initial condition (\ref{eq:rdm}) at $\ti=0.01$ for case 1 (see table 2) and $\Ra=500$.  For comparison,  the figure also illustrates the  corresponding optimal amplifications produced by $\opa$ (solid line) and $\opb$ (dashed line) for $\kw=30$, $\Ra=500$, and $\ti=0.01$. As expected, the unphysical $\opa$ perturbation has the smallest initial damping period and largest amplifications because it is not constrained to the boundary layer region. The DNS and $\opb$  perturbations both experience considerable damping; however, the DNS perturbation experiences greater damping because condition (\ref{eq:rdm}) initially excites heavily damped modes.  Following the initial damping period, $t>0.4$, the DNS, $\opb$, and $\opa$ perturbations experience identical growth rates such that the amplifications in figure \ref{figure16} have identical slopes. Previously, \cite{Rapaka2009} interpreted these identical slopes as confirmation of their nonmodal stability analysis. We find, however, that the identical slopes are due to the fact that all initial perturbations, optimal or suboptimal, rapidly converge to the dominant QSSA eigenmode (see \S 4.5). To illustrate this point, the crosses in figure \ref{figure16} show the amplifications produced when the forward IVP is integrated for a random initial condition, illustrated in figure \ref{figure8}(\emph{a}), that spans the entire domain.  The random initial condition produces identical slopes for $t>0.6$.

\subsection{Extent of linear regime and onset of convection}

\begin{figure}
    \begin{center}    
    \hspace{0.0cm}
    (\emph{a})
    \hspace{6.5cm}
    (\emph{b})
    \\
    \includegraphics[trim = 0 0 0 10, clip=true,width=7cm]{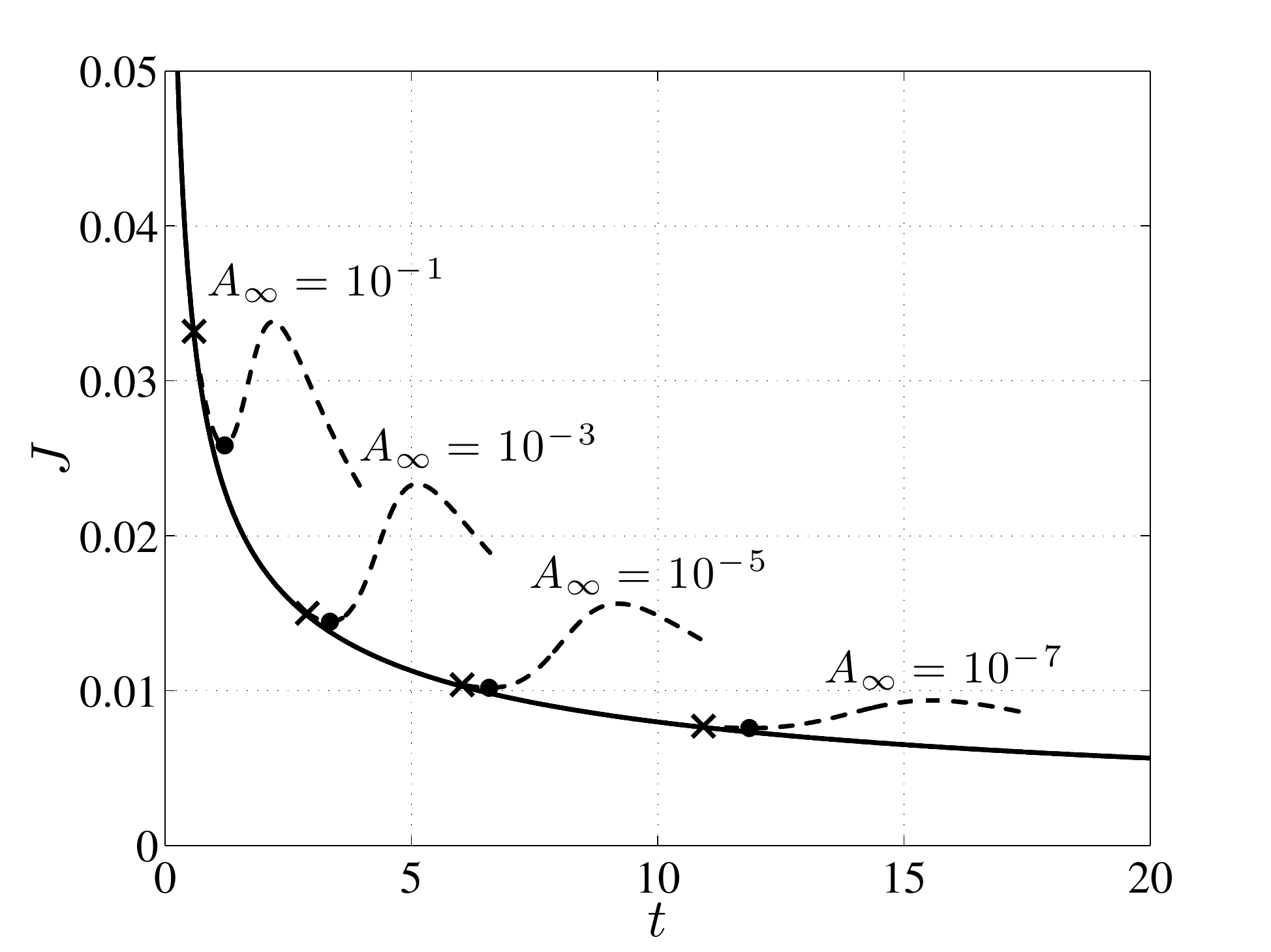}
    \includegraphics[trim = 0 0 0 10, clip=true,width=7cm]{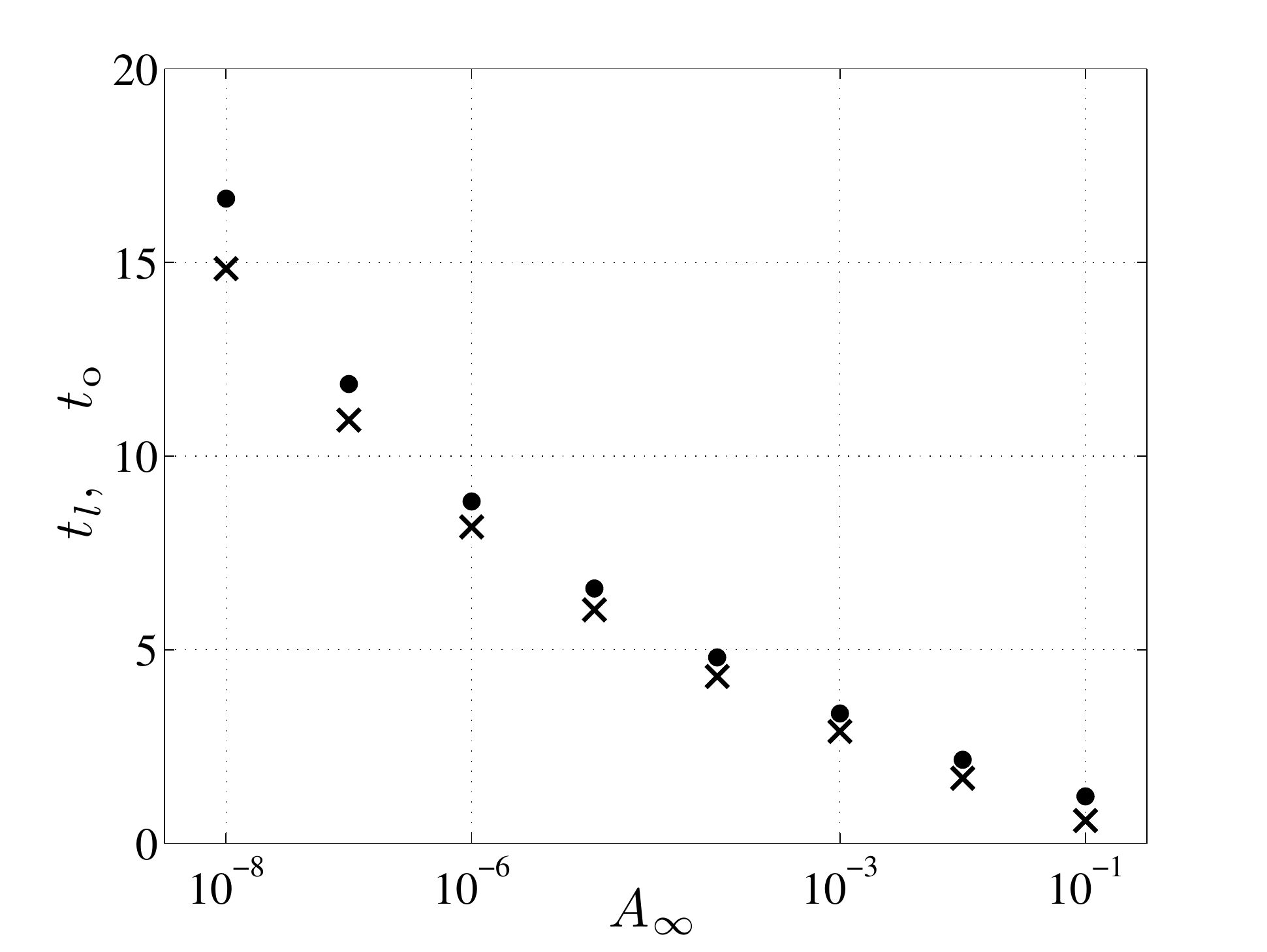}
    \\
      \hspace{0.0cm}
    (\emph{c})
    \hspace{6.5cm}
    (\emph{d})
    \\
    \includegraphics[trim = 0 0 0 10, clip=true,width=7cm]{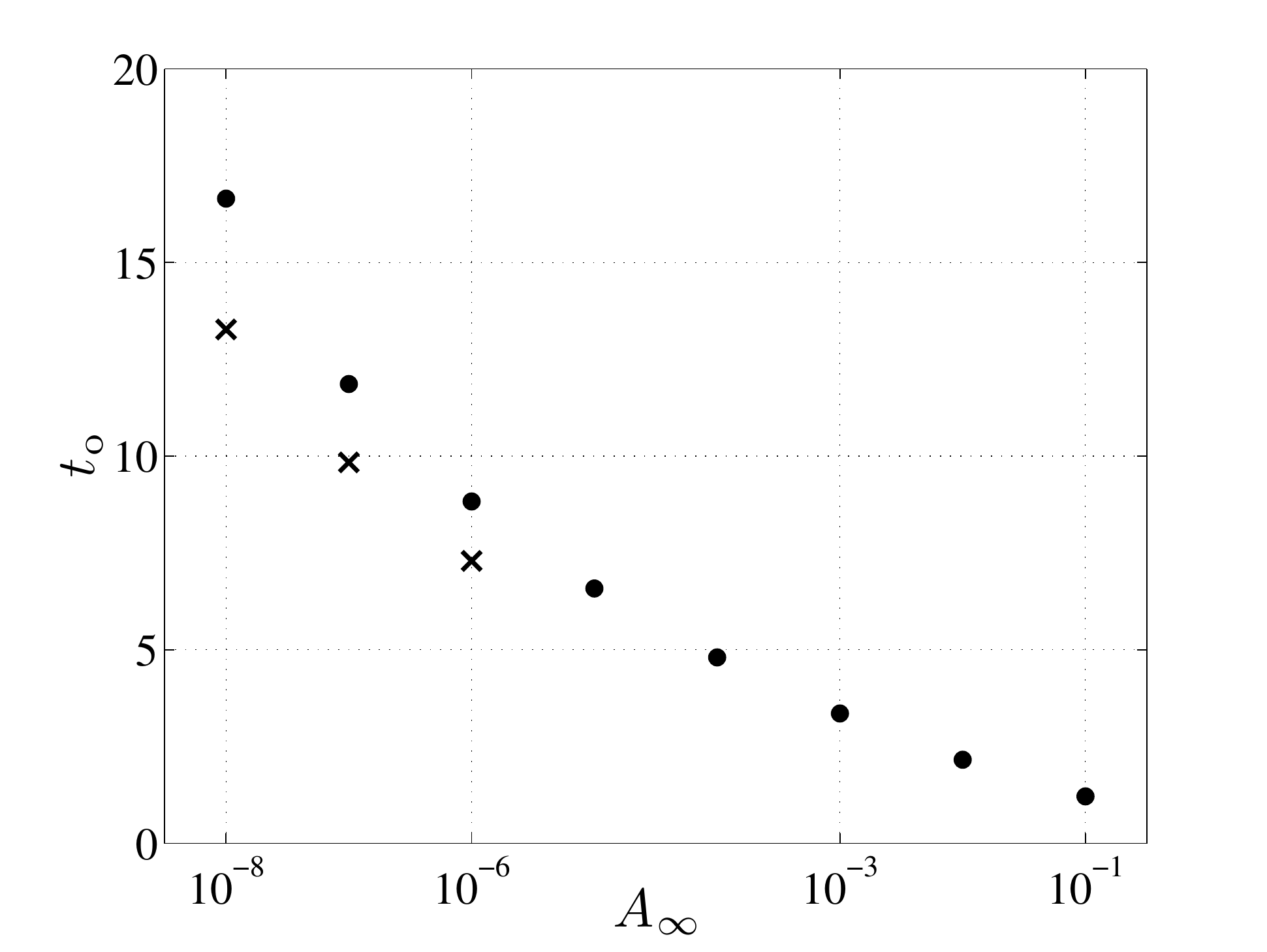}
    \includegraphics[trim = 0 0 0 10, clip=true,width=7cm]{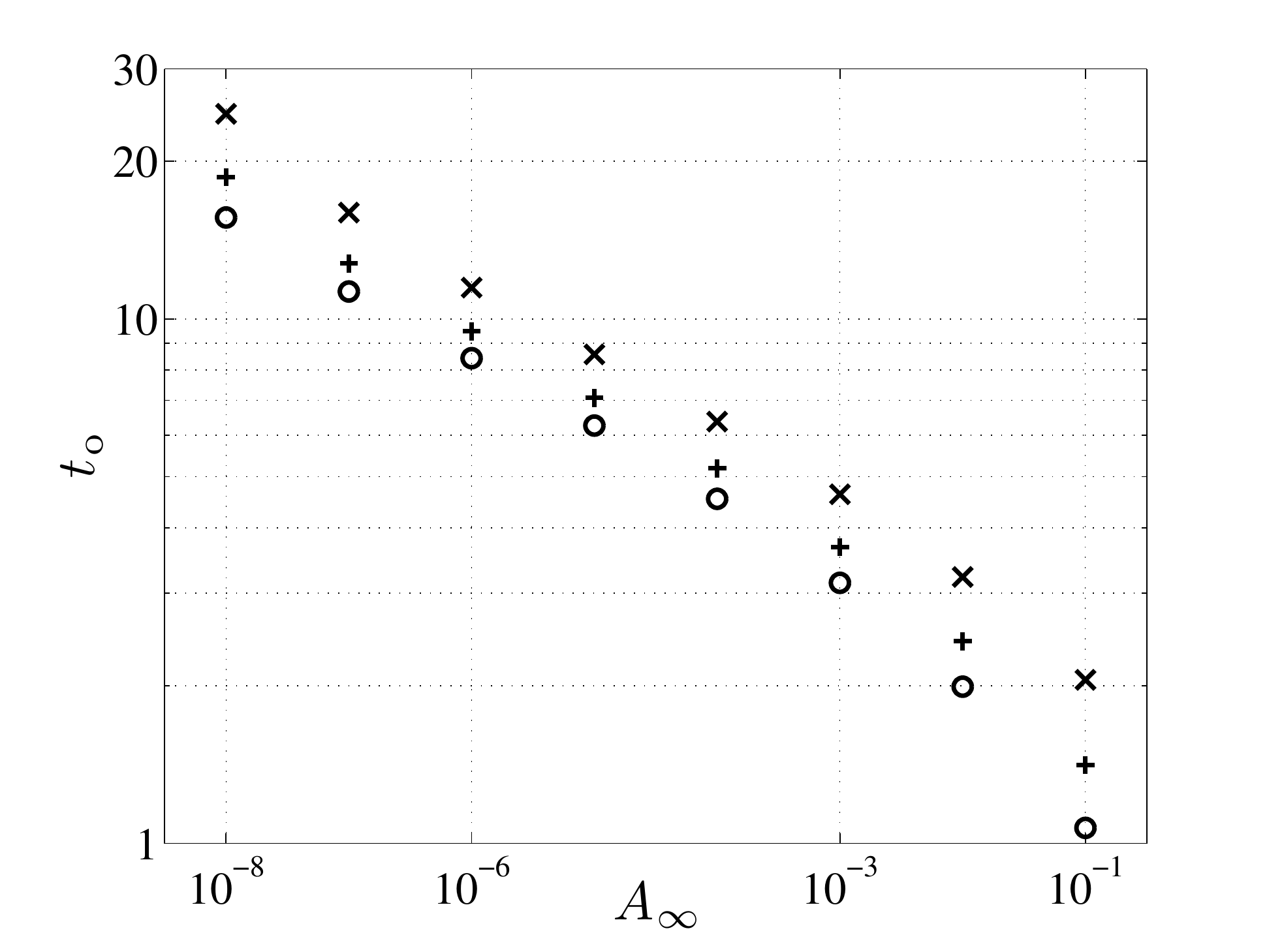}
    \caption{ DNS results for $\Ra=500$ and $\kw=30$ (\emph{a}) the flux due to base-state, $J_\mathrm{b}$ (solid lines), and the flux from DNS, $J$,  (dashed lines) using the $\opb$ $\cip$ profile at $\ti=0.1$. The crosses denote  $t_l$ while the solid dots denote $\ton$. (\emph{b}) $t_l$ (crosses) and $\ton$ (solid dots) vs. $\M$ using the $\opb$ $\cip$ profile at $\ti=0.1$. (\emph{c}) $\ton$ vs. $\M$ using the $\opa$ (crosses) and $\opb$ (solid dots)  $\cip$ profiles at $\ti=0.1$. (\emph{d}) $\ton$ vs. $\M$ using the $\opb$  $\cip$ profiles at $\ti=0.01$ (crosses), $\ti=0.05$ (plus signs), and $\ti=0.2$ (circles). Note that a log scale has been used for $\ton$. } 
    \label{figure17}
    \end{center}
\end{figure}

We now demonstrate the existence of a well-defined linear regime preceding onset of convection, and we compare the onset times, $\ton$, produced by the $\opb$ and $\opa$ schemes. We measure $\ton$ for different values of $\M$ and $\ti$ by specifying the following initial concentration field,
\begin{equation}
 c_\mathrm{dns}(x,z) =  c_\mathrm{b}(z) + \M \cos( \kw x) \frac{\cip(z) }{|| \cip ||_\infty},
\label{netconc1}
\end{equation}
where  $\cip$ are the optimal initial profiles determined by $\opa$ or $\opb$. Motivated by experiments \cite[]{Blair1969, Kaviany1984}, we define $\ton$ as the time at which $ dJ / dt=0$, where  $J$ is the mean flux of $\mathrm{CO}_2$ into the brine given by,
\begin{equation}
J(t) =  -\frac{1}{L} \int_0^L \frac{1}{\Ra} \frac{\partial c_\mathrm{dns}}{\partial z} \Big|_{z=0} \, \mathrm{d} x.
\label{eq:flux}
\end{equation}
Note from (\ref{eq:flux}) that perturbations
oscillating sinusoidally in the horizontal direction have no net
effect on $J$. Consequently, during the linear regime, the net flux is
due to pure diffusion of the base-state, i.e. $J = J_\mathrm{b}$. The deviation of the DNS results for $J$ from $J_\mathrm{b}$ is due to the growth of a zero-wavenumber mode, $\kw=0$, due to nonlinear interactions \cite[]{Jhaveri1982}. To further quantify the
duration of the linear regime, we also measure the time, $t = t_l$, for
which $J/J_\mathrm{b}=1.01$.

Figure \ref{figure17}(\emph{a}) presents DNS results for $J$ using the optimal $\cip$ profile produced by $\opb$ for $\ti=0.1$, $\tf=5$, $\kw=30$, and $\Ra=500$. Note that the $\opb$ $\cip$ profiles are  insensitive to the final time when $\tf>1$. The solid line shows the temporal evolution of the flux due to the base-state, $J_\mathrm{b}$, while the dashed lines show DNS results for $J$ when $\M = 10^{-1}$, $10^{-3},$ $10^{-5},$ and $10^{-7}$. The times, $t_l$ and $\ton$, are marked with solid dots and crosses respectively.  The flux $J$ initially agrees with $J_\mathrm{b}$ and then deviates after $t=t_l$ due to nonlinear effects. The initial linear regime exists even in the case of large initial amplitude $\M=10^{-1}$.
Figure \ref{figure17}(\emph{b}) illustrates $t_l$ (crosses) and $\ton$ (solid dots) for various perturbation amplitudes $\M$. 

Figure \ref{figure17}(\emph{c})  illustrates $\ton$ versus $\M$ using the optimal profiles produced by $\opa$ (crosses) and $\opb$ (solid dots) for $\ti=0.1$, $\tf=5$, $\kw=30$, and $\Ra=500$.
The $\opa$ scheme produces negative net concentration fields,  $c_\mathrm{net}$, for all finite perturbation amplitudes,  see table \ref{table:lin}. 
For illustration purposes, we arbitrarily set the maximum amplitude for $\opa$ to $\M=10^{-6}$ for which $c_\mathrm{net}^\mathrm{min}=-4.1 \times 10^{-7}$. In this case, $\opa$ produces onset times as low as $\ton=7.29$ for $\M=10^{-6}$. We stress, however, that the onset times predicted by $\opa$ cannot be realized in physical systems because of $c_\mathrm{net}^\mathrm{min}<0$, and are shown for illustration purposes only. In comparison, the $\opb$ supports finite initial amplitudes as large as $\M=10^{-1}$ for which $\ton=1.21$. We conclude that the perturbations produced by the $\opb$ are more likely to trigger onset of convection in physical systems.

Figure \ref{figure17}(\emph{d})  illustrates $\ton$ versus $\M$ using the $\opb$ $\cip$ profiles at $\ti=0.01$ (crosses), $\ti=0.05$ (plus signs), and $\ti=0.2$ (circles) for $\tf=5$, $\Ra=500$, and $\kw=30$. Onset of convection occurs later for smaller $\ti$ due to the strong initial damping periods. Note that a log scale has also been used for $\ton$ to highlight the difference for larger $\M$. For large amplitude perturbations, we observe that onset of convection can occur around $\ton \approx 1$. For typical aquifer conditions (see \S 4.3), with permeability  $K=10^{-14}$ m$^2$ and height, $H=51$ m, this corresponds to a dimensional onset time of $\ton^* \approx $ 165 years. 

\section{Conclusions and summary} 

We investigated the linear stability of gravitationally unstable,
transient, diffusive boundary layers in isotropic, homogeneous porous media. 
 We began by performing a classical optimization procedure ($\opa$) to determine optimal
perturbations with maximum amplifications. Previous studies \cite[]{Tan1986, Doumenc2010} have observed that perturbation amplification is sensitive to the perturbation flow field used to measure perturbation magnitude. Because this sensitivity has not been addressed for applications to CO$_2$ sequestration, we compared three different measures of perturbation amplitude that
maximize either the perturbation concentration field, vertical velocity field,
or the sum of the perturbation velocity and concentration fields,
which we refer to as the total energy.  We determined that maximizing
the perturbation concentration field naturally maximizes the total
energy. Maximizing the perturbation velocity field, however, does so
at the expense of the concentration field and total energy.
Consequently, we focus our study on perturbations that maximize the
concentration field because we expect these to be the dominant trigger for
onset of nonlinear convection.

Due to the transient nature of the base-state, optimal perturbations are sensitive to the initial time, $t=\ti$, at which the boundary layer is perturbed. Moreover, for a given final time, $t=\tf$, there is a unique initial perturbation time, $\ti^\mathrm{o}$, and wavenumber, $\kw^\mathrm{o}$, that maximize perturbation growth. By rescaling the problem, we obtained approximate analytical relationships, see equations (\ref{eqphio})--(\ref{eqtio}), for the optimal amplification, wavenumber, and initial perturbation time. These relationships show that $\ti^\mathrm{o}$ and $\kw^\mathrm{o}$ are independent of the aquifer height, $H$, but sensitive to the final time $\tf$. This indicates that the optimal initial perturbation depends on the onset time for nonlinear convection, $\ton$, and consequently the initial perturbation amplitude. Relationships (\ref{eqphio})--(\ref{eqtio}) also  predict that large amplitude perturbations with small onset times will have larger optimal wavenumbers, $\kw^\mathrm{o}$, and smaller optimal initial perturbation times, $\ti^\mathrm{o}$, than small amplitude perturbations with late onset times.

As the final time, $\tf$, increases, the optimal initial perturbations eventually converge to a fixed shape and cease to vary with increasing $\tf$. This occurs because the final perturbations at $t=\tf$ rapidly tend to the dominant quasi-steady eigenmode.
In fact, we demonstrate that for the current problem, the quasi-steady modal analysis is a good approximation
to the $\opa$. Both methods produce nearly identical amplifications and dominant wavenumbers. 
 This suggests that the deviation of the optimal perturbations from the dominant eigenmodes at small times may be primarily due to the transient base-state, rather than the nonorthogonality of the quasi-steady eigenmodes. This is in stark contrast to wall-bounded shear flows for which non-orthogonal eigenmodes often play a dominant role.

To judge the relevance of optimal perturbations to physical systems, we demonstrate that every perturbation has a maximum allowable initial amplitude above which the sum of the base-state and perturbation produces unphysical negative concentrations. We demonstrate that the optimal initial perturbations predicted by the $\opa$ produce unphysical
negative concentrations for all finite initial amplitudes. 
Consequently, onset of convection in physical systems is more likely triggered by suboptimal perturbations that support finite amplitudes.
To explore this
alternate path to onset of convection, we developed a modified
optimization procedure ($\opb$) that constrains the initial perturbations
to be concentrated within the boundary layer.

An integral characteristic of the MOP is the concept of a filter function, $\Psi(z)$, that effectively filters out perturbations with concentration fields extending beyond the boundary layer, see equation (\ref{eq:epsi}). The choice of filter function is not unique, and determines both the maximum allowable initial perturbation amplitude as well as the subsequent perturbation amplification.  Filter functions that concentrate the initial perturbation close to $z=0$ support large initial amplitudes, but produce small subsequent amplifications. Filter functions that concentrate the perturbations near the boundary layer depth support small initial amplitudes, but produce large subsequent amplifications. This raises the possibility that there exists an optimal filter function that balances the effects of the initial amplitude and subsequent amplification in order to minimize the onset time for convection. Because this requires a nonlinear analysis, we leave its consideration to future work. Rather, we focussed on perturbations produced by $\Psi=\cb^{-1}$ because this naturally concentrates perturbations in regions of large base-state concentration, and because it shows good agreement with corresponding DNS of physical systems.

The  alternate path to onset of convection taken by the $\opb$ features smaller amplifications and larger dominant wavenumbers than the $\opa$, especially at small initial perturbation times, $\ti \ll \ti^\mathrm{o}$. 
This occurs because the dominant $\opb$ perturbations are concentrated within the boundary layer, and consequently experience more initial damping than the $\opa$ perturbations. We obtained approximate analytical relationships (\ref{eqmphio})--(\ref{eqmtio}) for the optimal amplification, wavenumber, and initial perturbation time. The optimal initial times produced by the $\opb$ are roughly twice those produced by the $\opa$. We also observed that $\opb$ perturbations are more sensitive to variations in $\tf$, and consequently more sensitive to the initial perturbation amplitude, than the $\opa$ perturbations. We demonstrated that the results produced by $\opb$ agree well with the ``dominant mode'' approach of \cite{Riaz2006JFM} as well as quasi-steady modal analyses performed in the  similarity space of the base-state  \cite[]{Riaz2006JFM, Selim2007a, Wessel2009}.

To emulate physical experiments, we performed DNS in which the boundary layer is simultaneously perturbed with all wavenumbers resolved by the simulations. The perturbations have a random structure but are concentrated within the boundary layer. The DNS results confirm that physical systems follow the alternate path to convection predicted by  the $\opb$ scheme and show poor agreement with $\opa$. Furthermore, the $\opb$ perturbations support large initial amplitudes, $\M \sim 10^{-1}$, and produce early onset times for nonlinear convection. In contrast, the $\opa$ perturbations support neither finite amplitudes nor finite onset times. 
In an ongoing study, we are comprehensively exploring the effects of wavenumber, initial amplitude, initial time, and Rayleigh number on the onset of nonlinear convection. This is being performed using a weakly nonlinear expansion that is beyond the scope of the current study. \\

\section{Acknowledgements}
DD gratefully acknowledges all researchers in this field, without their tireless efforts, this research would not have come into
fruition. 
DD also thanks  co-authors NT and AR for guidance and inspiration that aided him in completing his PhD at the University of Maryland, College Park.  
This research was supported through a research grant from the Petroleum Institute, Abu Dhabi.

\bibliographystyle{unsrt}
\bibliography{referencesjfm}

\end{document}